\documentclass[letterpaper,11pt,fleqn]{article}
\usepackage{jheppub}
\usepackage{bm,amsmath,amssymb}

\long\def\comment#1{ }

\newcommand{\eqn}[1]{Eq.~\eqref{#1}}

\newcommand{\beq}{\begin{equation}}
\newcommand{\eeq}{\end{equation}}
\newcommand{\nn}{\nonumber\\}

\newcommand{\dif}{{\rm d}}
\newcommand{\rmd}{{\rm d}}
\newcommand{\rme}{{\rm e}}
\newcommand{\rmi}{i}

\newcommand{\rmH}{{\rm H}}

\newcommand{\rmJ}{{\rm J}}
\newcommand{\rmK}{{\rm K}}

\newcommand{\del}{\partial}

\newcommand{\order}[1]{\mcal{O}{(#1)}}
\newcommand{\mcal}{\mathcal}

\newcommand{\bmp}{\bm{p}}
\newcommand{\bmk}{\bm{k}}
\newcommand{\bmx}{\bm{x}}

\newcommand{\bmeps}{\bm{\varepsilon}}

\newcommand{\tp}{\acute{t}}
\newcommand{\xp}{\acute{x}}

\newcommand{\rp}{\acute{r}}

\title{\Large Jet evolution from weak to strong coupling}

\author[a]{Y.~Hatta,}
\author[b]{E.~Iancu,}
\author[c]{A.H.~Mueller}
\author[d]{and D.N.~Triantafyllopoulos}

\affiliation[a]{Graduate School of Pure and Applied Sciences,
University of Tsukuba,
Tsukuba, Ibaraki 305-8571, Japan}
\affiliation[b]{Institut de Physique Th\'{e}orique de Saclay,
F-91191 Gif-sur-Yvette, France}
\affiliation[c]{Department of Physics, Columbia University, New York,
NY 10027, U.S.A.}
\affiliation[d]{ECT*, European Centre for Theoretical Studies in Nuclear Physics and Related Areas, Strada delle Tabarelle 286, I-38123 Villazzano (TN), Italy}

\emailAdd{hatta@het.ph.tsukuba.ac.jp}
\emailAdd{edmond.iancu@cea.fr}
\emailAdd{amh@phys.columbia.edu}
\emailAdd{trianta@ectstar.eu}

\abstract{Recent studies, using the AdS/CFT correspondence, of the radiation produced
by a decaying system or by an accelerated charge in the $\mcal{N}=4$ supersymmetric 
Yang--Mills theory, led to a striking result: the `supergravity
backreaction', which is supposed to describe the energy density at infinitely strong
coupling, yields exactly the same result as at zero coupling, 
that is, it shows no trace of quantum broadening.
We argue that this is not a real property of the radiation at strong coupling,
but an artifact of the backreaction calculation, which is unable to faithfully capture
the space--time distribution of the radiation. This becomes obvious in the case
of a decaying system (`virtual photon'), 
for which the backreaction is tantamount to computing 
a three--point function in the conformal gauge theory, which is 
independent of the coupling since protected by symmetries. 
Whereas this non--renormalization property is specific to the conformal
$\mcal{N}=4$ SYM theory, we argue that the failure of the three--point function 
to provide a local measurement is in fact generic: it holds in any field theory with
non--trivial interactions. To properly study a localized distribution, one should rather 
compute a four--point function, as standard in deep inelastic scattering. 
We substantiate 
these considerations with studies of the radiation produced by the decay
of a time--like photon at both weak and strong coupling. 
We show that by computing four--point functions, in perturbation theory at 
weak coupling and, respectively, from Witten diagrams at strong coupling, 
one can follow the quantum evolution and thus demonstrate the broadening
of the energy distribution. This broadening is slow when the coupling is weak
but it proceeds as fast as possible in the limit of a strong coupling.
}
\vfill
\begin{document}
\maketitle

\section{Introduction}
\label{sect:intro}

One topic which has received much attention over the last few years within the context
of the gauge/string duality is the space--time distribution of the radiation in the strong
coupling limit of the ${\mathcal N}=4$ supersymmetric Yang--Mills (SYM) theory.
Originally motivated by studies of strongly coupled plasmas in relation with
the energy loss by an energetic parton \cite{Herzog:2006gh,
Gubser:2006bz,CasalderreySolana:2006rq,Liu:2006ug,CaronHuot:2006te,Chernicoff:2006yp,
HIM3,Gubser:2008as,Dominguez:2008vd,
Guijosa:2011hf,Fadafan:2012qu}, 
this problem turned out to be interesting and intriguing
for the {\em vacuum} case as well, because of a surprising result. AdS/CFT calculations
of the radiated energy density at {\em infinitely} strong coupling, using the method of the 
backreaction within the supergravity approximation to the dual string theory, led to results which 
exhibit the same space--time pattern as in the corresponding problems at {\em zero} coupling:
the radiation appears to propagate at the speed of light, without any trace of quantum 
broadening. Originally identified for the case of the synchrotron radiation by a
heavy quark \cite{Athanasiou:2010pv}, 
this property has subsequently been shown to extend to
more general sources of radiation \cite{Hatta:2010dz,Hatta:2011gh,Baier:2011dh,Hubeny:2010bq,
Chernicoff:2011vn,Fiol:2011zg,Fiol:2012sg,Correa:2012at,Agon:2012rz}, 
like an accelerated heavy quark which follows an 
arbitrary trajectory or the decay of a virtual photon.
%with time--like virtuality. 

The lack of broadening is surprising in that
it contradicts our general expectations for a quantum theory of interacting fields and,
in particular, the experience that we have with perturbative studies at weak, but non--zero,
coupling. Indeed, in a gauge field theory like ${\mathcal N}=4$ SYM, one expects the 
radiation to involve a superposition of quanta with various virtualities, including time--like 
quanta which propagate at subluminal velocity. With increasing time, such quanta will
separate from each other and also dissociate into other quanta with lower virtualities,
leading to a spread in the energy distribution along the direction of motion which
increases with time. At weak coupling, this evolution is well known to lead to
parton cascades, in which the original virtuality gets evacuated via successive branchings.
The associated spreading of the parton distribution turns out to be quite slow,
because the rate for branching is proportional to the strength of the coupling
(say, the 't Hooft coupling $\lambda=g^2 N_c$ in the case of the ${\mathcal N}=4$ SYM
theory at large $N_c$). With increasing coupling, the branching becomes more
and more effective, and the spreading goes faster and faster. In particular in
the strong coupling limit $\lambda\to\infty$ one expects this spreading to proceed
as fast as possible and to occupy the whole region in space and time which is allowed
by causality and special relativity. 

The following example, to be discussed at length in this work, 
should illustrate the situation. Consider the decay of a `heavy'
photon (an off--shell photon with time--like virtuality) in its rest frame. 
More precisely, the photon is in a localized state represented 
by a wave--packet centered at $t=0$ and $\bm{x}=0$
and which carries a typical 4--momentum $p^\mu=(Q,0,0,0)+\order{1/\sigma}$,
with $\sigma$ the width of the wave--packet, assumed to be large: 
$\sigma Q \gg 1$ (see Sect.~\ref{sec:WP} for details). The photon 
splits into a pair of electrically--charged, massless, partons (`quarks'),  which can subsequently 
evolve via `colour' interactions, that is, by emitting other `quarks' and `gluons'. We 
shall follow this evolution to leading order in the electromagnetic coupling,  but by letting
the strength $\lambda$ of the colour interactions to vary from weak to strong. 

When
$\lambda\to 0$, there is no further evolution, so the final state consists in two on--shell
quarks propagating back--to--back (by momentum conservation) at the speed of light.
The direction of propagation of the two quarks is arbitrary, so if one averages over many
events one finds an energy distribution in the form of a thin spherical shell\footnote{In QCD,
the average distribution has no spherical symmetry because of the bias introduced
by the polarization vector of the virtual photon. But in ${\mathcal N}=4$ SYM, the
anisotropy exactly cancels between the (adjoint) fermion and scalar contributions,
so the ensuing distribution is isotropic indeed.} of essentially
zero width which radially expands at the speed of light: $r=t$. More precisely, 
this energy shell has a small width $t-r\sim \sigma$, which however
can be neglected at large times $t\gg \sigma$. 

If the coupling is non--zero but weak
($\lambda\ll 1$), the original quarks will be generally off--shell, but their virtualities will 
typically be much smaller than the respective energies. Hence, the quarks will 
propagate with a large boost factor $\gamma\gg 1$
before eventually decaying into massless quanta. 
Their radiation will be collimated within an angle 
$\sim 1/\gamma$ around their direction of propagation, leading to
a pair of jets in the final state. %(This is the typical situation in QCD at high energy.)
After averaging over many events, the energy distribution has spherical symmetry
and a radial spreading $t-r$ which increases with time, because of
the virtuality distribution of the quanta within the jets. By the uncertainty principle, 
it takes a time $t\sim xQ/\mu^2$ to emit a quantum with energy fraction $x$ and
virtuality $\mu^2$. Then, to leading order in perturbation theory, the radial spreading can
be estimated as
 \beq\label{spread0}
 t-r\,\sim\,\frac{\lambda}{Q}\,\ln\frac{Q^2}{\mu^2}\,
 \,\sim\,\frac{\lambda}{Q}\,\ln(Q t)\,,\qquad\mbox{with}\qquad
\ln\frac{Q^2}{\mu^2}\,=\int_{\mu^2}^{Q^2} \frac{\rmd k^2}{k^2}\,,
 \eeq
where the logarithm has been generated by integrating over the phase--space for the
bremsstrahlung of a quantum with virtuality $k^2$ between $\mu^2$ and $Q^2$.
(For very large times such that ${\lambda}\ln(Q t) > 1$, the higher order 
corrections become important and will be estimated in Sect.~\ref{sect:jet}.)
This argument also shows that the typical virtuality $\mu$ of the quanta composing the
jets is such that $\lambda\ln ({Q^2}/{\mu^2})\sim 1$, which at weak coupling
implies $\mu\ll Q$. This confirms that the typical quanta are nearly on--shell 
and thus propagate quite fast: $\gamma=xQ/\mu\gg 1$.

Consider now the situation at relatively strong coupling, $\lambda > 1$. Then the
virtual photon splits into a pair of quarks whose virtualities are comparable to their
energies, $\mu\sim xQ$. These quarks are themselves highly virtual and hence
they are slowly moving: $\gamma\sim 1$. They will rapidly decay into
quanta with similar characteristics. We expect this pattern to repeat itself in the 
subsequent steps of the evolution: at each branching, the energy and virtuality of the 
parent parton are quasi--democratically divided among the offspring quanta, which 
therefore emerge at large angles with respect to the direction of propagation 
of their parent. For sufficiently large times $t\gg 1/Q$, this evolution leads to a
parton distribution characterized by a wide dispersion in velocities and angles. For 
the conformal theory ${\mathcal N}=4$ SYM, we expect this distribution to be isotropic 
{\em event--by--event} and to show {\em maximal radial broadening}, that is, to uniformly
cover the whole volume at $r\le t$ which is allowed by causality.

Moving to extremely strong coupling $\lambda\gg 1$, the situation is {\em a priori}
more complicated, since the concept of {\em partons} (elementary quanta 
representing excitations of the quantum fields in the Lagrangian which are point--like 
and nearly on--shell) is probably not useful anymore: the matter distribution produced
by the decaying photon is made with collective excitations whose composition in terms
of elementary quanta can be arbitrarily complicated. Yet, since isotropy and maximal
broadening are already reached for moderate values of the coupling $\lambda\sim\order{1}$, 
it is natural to expect these features to remain valid when $\lambda\to\infty$.

These expectations are indeed supported, at least indirectly, by a series of calculations 
at infinitely strong coupling using AdS/CFT. These include studies of the decay of a 
virtual photon using the ultraviolet/infrared (UV/IR) duality \cite{HIM3}, calculations 
of the associated angular correlations which demonstrate isotropy \cite{Hofman:2008ar},
studies of the jet fragmentation showing the absence of  point--like partons 
\cite{Hatta:2008tn,Hatta:2008qx},
and also studies of deep inelastic scattering 
\cite{Polchinski:2002jw,Brower:2006ea,BallonBayona:2007qr,HIM1,HIM2,Cornalba:2008sp,Avsar:2009xf} leading to a similar conclusion: the partons cannot
survive in the wavefunction of a hadron, or in a plasma, at strong coupling because 
they efficiently decay towards smaller and smaller values of $x$.

Yet, such previous approaches had not address the issue of the radial, or longitudinal,
distribution of the radiation. For instance, in the study of angular correlations
performed in Ref.~\cite{Hofman:2008ar}, the radial distribution was explicitly integrated
over. Also, most of the other studies alluded to above
were performed in momentum space. The calculation of the backreaction for the 
synchrotron radiation in Ref.~\cite{Athanasiou:2010pv} is the first attempt in that
sense and, as already mentioned, it led to the surprising conclusion about the lack of 
radial broadening. As also mentioned, this conclusion applies to other forms of radiation, 
including our prototype problem --- the energy produced by the decay of a virtual photon ---, 
for which the backreaction predicts the same space--time distribution as at zero coupling: 
a thin spherical shell expanding
at the speed of light with a constant width $t-r\sim \sigma$. This looks puzzling as
it suggests that the situation at (infinitely) strong coupling could be closer to that at zero
coupling, rather than to that at weak or intermediate values of the coupling.
However, this is not the case, as we now argue.

A first indication in that sense comes from the following argument, which refers
to the radiation produced by the decay of a virtual photon. 
The SUGRA calculation of the backreaction amounts to computing a specific 
three--point correlation function in the underlying field theory, which is protected by 
symmetries and hence it is independent of the value of the coupling.
Specifically, this correlator reads $\langle {\hat J}_q^{\dagger}
 \,{\hat T}_{00}(x) \,{\hat J}_q\rangle$, where ${\hat J}_q$ is the  
operator which creates the virtual photon 
(a time--like wave packet of the electromagnetic current operator; 
see Sect.~\ref{sec:WP} for details),
while ${\hat T}_{00}(x)$ is the energy density operator at the `measurement' point 
$x^\mu=(t,\bm{x})$. As well known, three--point functions in a
conformal field theory are fixed by conformal symmetry and the
(quantum) dimensions of the relevant operators, up to a constant
(function of the coupling). For the correlator at hand, the operators ${\hat J}_q$
and  ${\hat T}_{00}(x)$ have no anomalous dimensions and the overall
normalization is fixed by the conservation of the energy. Accordingly, 
this three--point function is independent of  the coupling, as anticipated
\cite{Freedman:1998tz}. This property, that we shall explicitly check by comparing
the respective predictions of the zero--order perturbation theory and
of the backreaction, `explains' the lack of broadening shown by the latter, in the sense
of relating this result to the symmetries of the
underlying CFT. But this also demonstrates that the three--point function is 
unable to capture the quantum evolution responsible for the radial broadening,
since it fails to do so already at weak coupling, where this evolution is well understood
in perturbation theory. This makes it clear that this three--point function is not a good
observable for characterizing the space--time distribution of the radiation.
 
To summarize,  %at least for this problem of the decay,
the lack of broadening predicted by the backreaction is  
not a true feature of the radiation at strong coupling, but merely an artifact of computing
an observable which is not appropriate for that purpose.  This observation
rises several questions: \texttt{(i)} what are the reasons for this failure of the
three--point function,  \texttt{(ii)} what is the actual physical content of a three--point function like 
$\langle {\hat J}_q^{\dagger} \,{\hat T}_{00}(x) \,{\hat J}_q\rangle$, and \texttt{(iii)}  what are the 
observables that one should study in order to understand the space--time 
distribution of the radiation. These are clearly very general questions 
and the answers that we shall provide to them are not necessarily new. 
(Some connections with similar problems in QCD will be later pointed out.)
But precisely because they are so general, these answers are independent
of the non--renormalization property of the three--point function alluded to above.
Most of them apply to any interacting field theory, conformal or not, at either weak or strong
coupling.

%Accordingly, we expect similar arguments to apply for other sources of radiation, like an
%accelerated heavy quark. We shall return to this point towards the end of the Introduction.

Specifically, we shall argue that a three--point function like
$\langle {\hat J}_q^{\dagger} \,{\hat T}_{00}(x) \,{\hat J}_q\rangle$ is truly a {\em forward
scattering amplitude}~: the amplitude that the `target' state created by ${\hat J}_q$
(i.e. the decaying system) survive intact after interacting with 
the localized probe operator ${\hat T}_{00}(x)$. In an interacting field theory, this
amplitude cannot provide information about the internal structure of the target at very
large times\footnote{More precisely, we have in mind times which are sufficiently
large to allow for a well developed evolution; at strong coupling, the condition 
$t\gg \sigma$ is enough in that sense, as we shall later check, 
whereas at weak coupling we require ${\lambda}\ln(Q t)\gtrsim 1$.}
$t\gg \sigma$. % (the time argument of  ${\hat T}_{00}(x)$). 
Indeed, the quanta composing the target at such late
times are very soft, as they are the products of many successive branchings, 
so they cannot provide the high momentum transfer that would be
required by a local measurement. 
(The typical momenta of the quanta in the decaying system keep decreasing with $t$, 
as we shall see, so they can become arbitrarily small for sufficiently large times.
By contrast, the typical momenta $\Delta^\mu$ transferred by the target
to the probe are of order $1/\sigma$ --- the maximal value allowed by 
energy--momentum conservation\footnote{Energy--momentum conservation
implies that a forward amplitude like 
$\langle {\hat J}_q^{\dagger} \,{\hat T}_{00}(x) \,{\hat J}_q\rangle$ can
receive contributions only from the Fourier modes ${\hat T}_{00}(\Delta)$
whose momenta $\Delta^\mu$ are smaller than the uncertainty $\sim 1/\sigma$
in the total energy and momentum of the decaying system.}
--- as clear from the fact that the signal has a small width $t-r\sim\sigma$.)
This argument shows that the narrow signal given by the backreaction cannot be
a part of the radiation in the decaying system at the time $t$ of `measurement'
(the time argument of  ${\hat T}_{00}(x)$). 
Rather, this signal must have been generated at some early time $t_{\rm int}\ll t$,  
before the target had a chance to significantly evolve; at that time, the target was 
composed with only few and relatively hard quanta,
with momenta $k\sim Q\gg \Delta$. But, clearly, such an early emission gives 
no information about the state of the target at the late time $t$ 
(except at zero coupling). 

One expects the
disparity between $t_{\rm int}$ and $t$ to be maximal at strong coupling, since in that
case one needs a very small value for $t_{\rm int}$ in order
to minimize the effects of the evolution.
As we shall see in Sect.~\ref{sec:interp} below, this argument is indeed consistent 
with the calculation of the backreaction in AdS/CFT, 
provided one makes the natural identification between $t_{\rm int}$
and the time at which the gravitational wave in AdS$_5$ (the `backreaction') is emitted
by the bulk excitation representing the decaying system (a SUGRA vector field).

%Accordingly, if the interaction between the probe and the target occurs at some late
%time, close to the `measurement time' $t$,
%the internal state of the target is disturbed by this process and there is 
%no contribution to the forward amplitude. Conversely, in order to generate such a 
%contribution, the interaction must 

The above arguments, which explain the failure of the three--point function as a 
local measurement, have some other interesting consequences.
First, they suggest what should be the simplest observable which allows one to 
study the space--time distribution of the decay: this is a 
{\em four--point function} like $\langle {\hat J}_q^{\dagger}
 \,{\hat T}_{00}(x_1){\hat T}_{00}(x_2) \,{\hat J}_q\rangle$,  in which the 
momentum $\Delta$ transferred to the target by the first insertion ${\hat T}_{00}(x_2) $ 
of the probe operator is then taken away by the second insertion  
${\hat T}_{00}(x_1)$. This makes it possible to probe the target with a good resolution
(i.e. a relatively large momentum transfer $\Delta$ ) without affecting its 
properties\footnote{One should notice the difference between the four--point 
function that we propose here and the $n$--point 
functions with $n\ge 3$ used in Ref.~\cite{Hofman:2008ar}.
The probe operators in Ref.~\cite{Hofman:2008ar} are soft, non--local operators,
like the total energy radiated per unit solid angle along a given direction $\bm{n}$~:
$\hat{\mcal{E}} (\bm{n})\equiv \lim_{r\to \infty} \,r^2\int^\infty_0 \rmd t\, n_i
{\hat T}^{0i}(t,r\bm{n})$.
Such operators do not probe the radial distribution of the radiation, but only its
angular correlations.}.
Such a measurement gives us informations about the state of the target around the 
space--time point $x=(x_1+x_2)/2$, with a resolution fixed by the difference $x_1-x_2$. 

The previous discussion also tells us under which circumstances a
three--point function can still act as a measurement: this is possible
provided one gives up any radial (or longitudinal)
resolution, that is, if one integrates over the radial profile of the distribution to get
the total energy (or, more generally, the energy radiated per unit solid angle), as done
e.g. in Ref.~\cite{Hofman:2008ar}. Indeed, an operator like the total energy
$\hat E\equiv \int \rmd^3 \bm{x} {\hat T}_{00}(t,\bm{x})$ involves arbitrarily soft
Fourier modes, hence it can measure the target without disturbing it.
The result of this particular measurement is, of course, {\em a priori} known:
by energy conservation, $\langle {\hat J}_q^{\dagger} \,{\hat E} \,{\hat J}_q\rangle
=Q$, with $Q$ the energy of the original photon. Less trivial situations
occur in the applications of the backreaction method to finite--temperature problems. In such
cases, one is typically interested in the energy deposition in the plasma by a 
`hard probe' (a heavy quark, a gluon, or a virtual photon), as measured over relatively 
large space--time scales $\Delta r,\,\Delta t\gtrsim 1/T$, with $T$ the temperature
\cite{Friess:2006fk,Gubser:2007xz,Chesler:2007an,Chesler:2007sv,Gubser:2008as,
HIM3,Chesler:2008wd,Arnold:2010ir,Arnold:2011hp,Chesler:2011nc}.
Then the method of the backreaction is again reliable, since $1/T$ is the largest scale
for quantum broadening in that case. (Indeed, this is the typical value of the broadening 
by the time when the radiation gets thermalized in the plasma.)

As anticipated, the previous considerations are quite general and in particular
they are reminiscent of some of the strategies used to study the hadron structure and
interactions in QCD. Namely, the three--point function and the 
four--point function above introduced are very similar to the {\em electromagnetic form
factors} and, respectively, the {\em structure functions} for deep inelastic scattering (DIS),
which can both be viewed as measures of the electric charge distribution in a nucleon,
but on very different resolution scales. A `form factor' is a matrix element like
$\langle P'| {\hat J}^\mu(x)| P\rangle$ where $| P\rangle$ denotes the proton
state with 4--momentum $P^\mu$ and ${\hat J}^\mu(x)$ is the electromagnetic current
operator. For relatively low momentum transfers $|\Delta|\lesssim 1/R$, 
where $\Delta\equiv P'-P$ and $R$ is the proton (charge) radius, 
this form factor, which can be studied via low--energy electron--proton scattering, 
provides a good measurement of the proton radius $R$. But if one is interested
on the proton structure on much shorter scales $r\ll R$, as probed by a hard scattering
which typically breaks the proton, one should rather compute a matrix element like
$\Pi^{\mu\nu}(\Delta)\equiv \int \rmd^4 x\,\rme^{-ix\cdot\Delta}
\langle P| {\hat J}^\mu(x){\hat J}^\nu(0)| P\rangle$, where 
 %$P\gg M$ (the proton mass) and 
$\Delta^\mu$ can now be arbitrarily high. 
This is a forward scattering amplitude which via the optical theorem
can be related to the total cross--section (or `structure function') for 
DIS. The experimental measurement of the latter gives us the most direct access to 
parton distributions on short distances.

Inspired by the above, we shall use here a similar strategy to investigate the space--time
distribution of the radiation produced by the decay of the virtual photon: 
we shall compute the four--point function describing the DIS between the decaying 
system and an electromagnetic current with space--like virtuality
in a boosted frame where the virtual photon propagates %along the $x_3$ axis
at nearly the speed of light.  (In the context of a decay, 
this four--point function is also known as the `fragmentation function'.)
The boost is useful (at least, at weak coupling) to render
the parton picture of DIS manifest, but our final conclusions at strong coupling can be 
easily translated to the photon rest frame. These results will confirm and substantiate
the picture of quantum broadening that we previously exposed.

In the boosted frame, the decaying system looks like a jet --- the matter 
is concentrated within a small solid angle $\Delta\Omega\sim 1/\gamma^2$ around the
longitudinal axis ($x_3$) and within a comparatively small longitudinal interval 
$\Delta x_3\ll t$ behind the light--cone ($x_3=t$) --- for any value of the coupling.
However, at strong coupling this `jetty' picture is merely the effect of the boost: the 
respective `jet' is recognized as the boosted version of a distribution
which in the photon rest frame looks like a uniformly filled sphere with radius $r=t$. 
In the boosted frame, this is visible in the fact that the longitudinal width $\Delta x_3$  
of the distribution increases {\em linearly} with $t$~: $\Delta x_3\simeq t/2\gamma^2$ 
(see Fig.~\ref{fig:lorentz}). This should be compared to the situation at zero coupling,
where $\Delta x_3\simeq \sigma/\gamma$ (the Lorentz--contracted version of a radial
width $t-r\simeq\sigma$ in the rest frame), and also at weak coupling $\lambda \ll 1$, 
where $\Delta x_3$ increases very slowly with $t$, 
as shown in the second line of the equation below (the all--order generalization
of \eqn{spread0})
\begin{equation}\label{Deltax3}
    \Delta x_3\,\equiv\,(t-x_3)_{\rm max}\,\simeq\,
    \begin{cases}
        \displaystyle{\sigma/\gamma} &
        \text{ for\,  $\lambda= 0$}
        \\*[0.2cm]
        \displaystyle{\frac{1}{\gamma Q}
 \left(\frac{Q t}{\gamma}\right)^{\lambda/24}
      } &
        \text{ for\,  $0 < \lambda \ll 1$}
        \\*[0.4cm]
        \displaystyle{
        \frac{t}{2\gamma^2}} &
        \text{ for\,  $\lambda\to\infty$}.
    \end{cases}
 \end{equation}
%The behaviour shown in the first two lines of this equation, i.e. a very slow increase
%of $\Delta x_3$ with $t$, is the hallmark of a genuine jet.  
The above result at strong coupling (the third line in \eqn{Deltax3})
can be rephrased  
in a boost--invariant way by referring to the typical virtuality $\mu$
of the modes in the decaying system: at large times, this decreases as 
$\mu\simeq 1/t$ .
 
Moreover, our analysis of the four--point function will also show that, at strong coupling, 
the matter is uniformly distributed {\em event--by--event} within the region of space occupied
by the jet, meaning that there are no localized substructures, like partons. Indeed, if one
tries to scrutinize this matter on longitudinal and transverse scales much smaller than
its overall respective sizes, $\Delta x_3\simeq t/2\gamma^2$ and 
$\Delta x_\perp\simeq t/\gamma$, then one finds that the fragmentation function is 
exponentially suppressed: it is proportional to $\exp\{-\Delta_\perp t/\gamma\}$, with
$\Delta_\perp$ the transverse momentum transferred by the probe current in DIS.
By contrast, at weak coupling the fragmentation function is essentially independent of 
$\Delta_\perp$, meaning that partons exist and they are point--like.

So far, we have not been very explicit about the formalism that we shall use and
the specific calculations that we shall perform. This will be shortly mentioned below,
when presenting the structure of the paper, and then discussed in more detail in
the appropriate sections. As a general strategy, we shall perform all our calculations
in the framework  of the ${\mathcal N}=4$ SYM theory, either by using perturbation
theory at weak coupling, or the SUGRA approximation to the dual string theory at
infinitely strong coupling. In particular, we shall use the technique of Witten diagrams
to evaluate the four--point function describing the fragmentation of the time--like 
photon at strong coupling. A similar calculation has been previously performed
in Ref.~\cite{Hatta:2010kt}, but only for light--like kinematics (for the `probe'
currents), corresponding to the production of on--shell photons. 
Here, we shall rather focus on the space--like kinematics, which is better suited
to measure the internal space--time structure of the decaying system. In this paper,
we shall not address the issue of the stability of the SUGRA approximation against
(longitudinal) string fluctuations. It has been argued in Ref.~\cite{Hatta:2010dz} that
such fluctuations are potentially large and unsuppressed in the infinite 
coupling limit. However, their effects cannot be properly computed by lack of a consistent
quantization scheme for the string fluctuations in a curved space--time.
(The heuristic estimates given in Ref.~\cite{Hatta:2010dz} 
are plagued with severe ultraviolet divergences.)

Let us also make some comments on the related problem of the radiation by
an accelerated heavy quark in the fundamental representation of the colour group.
%, that we shall not further address in this work (as 
%this would require different mathematical manipulations).
There are clearly some differences w.r.t. the problem of the decay
--- notably the fact that the dual object at strong coupling is a Nambu-Goto string,
instead of a SUGRA field ---
but we are confident that our main conclusions should apply to this problem as well. 
Indeed, the conclusions concerning the quantum evolution at strong 
coupling, like the maximal broadening, the absence of jets, and the absence of 
partons or other substructures, are universal properties of the 
radiation at strong coupling, which hold independently of the nature 
of its source. The fact that the radial broadening is not visible in the results of the
backreaction is again to be attributed to the inability of this method 
to faithfully capture the space--time distribution of the radiation. To shed more light
on this point, it is useful to exhibit the CFT correlator which is implicitly computed
(in the strong coupling limit) via the backreaction. The operator describing
the interactions between the massive quark and its comparatively soft radiation in the eikonal
approximation is the Wilson line ${\hat{\mcal{U}}}(C)$, with $C$ the trajectory of the quark.
Hence, the result of the backreaction is proportional to the following correlator in CFT:
 \beq\label{Wilson}
 \frac{1}{N_c} \left\langle {\rm Tr} \big\{{\hat{\mcal{U}}}^\dagger(C)\,{\hat T}_{00}(x)\,
   {\hat{\mcal{U}}}(C)\big\}\right\rangle\,,
   %\qquad {\hat U}_{\mcal{C}}\,=\,\rm{P}
 %\,\exp\left\{ig\int\rmd t\big(\dot x^\mu(t) A_\mu^at^a
 \eeq
which is recognized as a generalization of the three--point function 
$\langle {\hat J}_q^{\dagger} \,{\hat T}_{00}(x) \,{\hat J}_q\rangle$ in which the local
operator ${\hat J}_q$ is replaced by the non--local operator ${\hat{\mcal{U}}}(C)$. 
We implicitly assume
here a large spatial separation between the trajectory $C$ of the quark
and the position $\bm{x}$ of the probe operator. (If $C$ is restricted to some
bounded region in space with the largest size $R$, 
then we assume $r\equiv|\bm{x}|\gg R$.)
%(For instance, in the case of a quark following
%a circular trajectory with radius $R$, we assume that $r\equiv|\bm{x}|\gg R$.) 
Unlike for $\langle {\hat J}_q^{\dagger} \,{\hat T}_{00}(x) \,{\hat J}_q\rangle$, we are not
aware of general non--renormalization properties\footnote{This being said,
there is empirical evidence that such a property must hold: the results of the backreaction
in Refs.~\cite{Athanasiou:2010pv,Hatta:2011gh}, which include the case of an
arbitrary motion for the heavy quark, coincide with the respective results at
{\em zero} coupling up to the replacement  $\lambda \to 4
\sqrt{\lambda}$ in the overall factor and up to an additional piece at (infinitely) strong coupling,
which is however a total time derivative and hence averages out for a periodic motion.
A similar property at the level of the radiated power has been previously observed in 
Ref.~\cite{Mikhailov:2003er}. Such non--renormalization properties for the radiation in
${\mathcal N}=4$ SYM, whose precise origin remains to be understood, may be viewed
as generalizations of similar properties which are known to hold, by
conformal symmetry, in Euclidean space--time and for simple Wilson loops, like
the circular one (see e.g. \cite{Buchbinder:2012vr} and references therein).
}
for the correlator \eqref{Wilson},
but this is not essential for our purpose. All that matters is that \eqn{Wilson} describes
an elastic scattering process in which the radiation generated by the heavy quark 
interacts with the probe operator ${\hat T}_{00}(x)$ without being 
significantly disturbed. Then the arguments previously used for 
$\langle {\hat J}_q^{\dagger} \,{\hat T}_{00}(x) \,{\hat J}_q\rangle$ can be taken over. 
Namely, the interaction with a localized operator is a relatively hard process, which 
requires a high momentum transfer from the target to the probe. The signal carrying
such a high momentum can only be emitted by quanta which are in the early stages 
of their evolution, when they are still hard. Such quanta have been freshly 
emitted by the heavy quark and hence they are located in the vicinity of 
the quark trajectory $C$. Accordingly, the signal carries no information about  
the structure of the radiation at the comparatively remote 
`measurement' point $x^\mu$. This argument is corroborated by the backreaction calculation
\cite{Hatta:2011gh} which shows that the emission time $t_{\rm int}$ (identified, once again, as
the time at which the gravitational wave in AdS$_5$ is emitted by the string) coincides
with the retardation time $t_r\simeq t-r$ in the corresponding classical problem --- that is,
the time at which a signal propagating at the speed of light should be emitted by the source
in order to reach the measurement point $r$ at time $t$.

It is finally interesting to mention that correlation functions similar to \eqn{Wilson}
are commonly used in perturbative QCD to compute the soft radiation produced
by energetic partons (represented by the Wilson lines), notably in studies
of the shape of a jet (see e.g. Ref.~\cite{Belitsky:2001ij}). However, in such cases the
local operator ${\hat T}_{00}(x)$ is replaced with a non--local one, such as the total
energy radiated per unit solid angle, which is a soft, acceptable, probe.

The plan of the paper is as follows: In Sect.~\ref{sec:WP} we shall introduce some general
elements of the formalism, like the wave--packets describing the virtual photon and the
probe operator (in both the rest frame of the decay and in a highly boosted frame), 
and the three--point and four--point functions that we shall later use 
to study the decay. We shall explain in more detail why a three--point function is
not suitable for a local measurement. Also, we shall describe the causality constraints
on the space--time distribution of the matter produced by the decay. In Sect.~\ref{sect:3P}
we shall present the result of the backreaction for the three--point function at
infinitely strong coupling. With this occasion, we shall correct the original calculation in
Ref.~\cite{Hatta:2010dz} by adding one term that has been missed there. We shall emphasize
the lack of radial broadening of the final result and pinpoint the origin of
this property in the process of the calculation. We shall attempt a physical interpretation
for this result in CFT. We shall also perform the Fourier transform of the result to a mixed
Fourier representation, which is tantamount to using a wave--packet for the probe operator.
In Sect.~\ref{sec:zero} we shall present the calculation of the three--point function 
in ${\mathcal N}=4$ SYM at zero coupling (using the mixed Fourier representation, once again)
and thus obtain exactly the same result as that of the backreaction at
infinitely strong coupling. Starting with Sect.~\ref{sect:jet}, we shift our attention towards
the four--point function that describes the DIS of a virtual  $\mcal{R}$--current 
off the decaying system. We first consider the situation at weak coupling but late times,
where we rely on a leading logarithmic approximation to resum perturbative corrections 
to all orders in $\lambda\ln (Qt)$. This will allow us to derive the result for 
longitudinal broadening shown in the second
line in \eqn{Deltax3} and to demonstrate that weakly--coupled partons are point--like.
Finally, Sect.~\ref{sec:Witten} contains our main new results in this paper, namely the
calculation of the four--point function at infinitely strong coupling from Witten diagrams.
For simplicity, that is, in order to avoid a proliferation of diagrams with complicated vertices,
we shall restrict ourselves to a toy--version of SUGRA --- a scalar field theory with trilinear
couplings. This reproduces the relevant topologies for the Witten diagrams and thus correctly
captures the physical information which is important for us here: the support of the
space--time distribution of the radiated matter. We thus find that this matter is uniformly
distributed over the whole region in space and time which is allowed by causality.

  \section{Preliminaries: observables for decaying states}
\label{sec:WP}

As announced in the Introduction, our goal is to study the matter distribution created at 
large times by the decay of an unstable excitation of the ${\mathcal N}=4$ SYM theory.
For convenience, we choose this excitation to be a {\em time--like photon}.
We follow the standard strategy for introducing electromagnetism in ${\mathcal N}=4$ SYM,
which consists in gauging one of the U$(1)$ subgroups of the global SU$(4)$ 
$\mcal{R}$--symmetry.
Then, the electromagnetic vector potentials $A^\mu_{\rm em}(x)$ couple to the
conserved $\mcal{R}$--current, $J_\mu(x)$, associated with the generator of that particular 
U(1) subgroup, via the action $S_{\rm int}=\int\rmd^4 x\,A^\mu_{\rm em}(x) J_\mu(x)$.

A photon state with given 4--momentum $q^\mu$,  
as represented by a plane--wave\footnote{We use a  metric convention
with the minus sign for the temporal components; e.g. $q\cdot x\equiv -q^0 x^0+
{\bm q}\cdot{\bm x}$.}  $\rme^{iq\cdot x}$, will be on--shell and stable
if it has zero virtuality, $q^2=0$,  but it will 
be off--shell and unstable when its virtuality is {\em time--like}, $q^2<0$.
(The virtuality $q^2$ is defined as $q^2\equiv q^\mu q_\mu = -q_0^2 + {\bm q}^2$.)
The unstable photon will decay into the quanta of 
${\mathcal N}=4$ SYM which enter the structure of the $\mcal{R}$--current 
(massless fermion and scalar fields in the adjoint representation of the colour group
SU$(N_c)$). These quanta will be time--like too, as they share the virtuality
of the original photon, so they will themselves decay into other quanta of
${\mathcal N}=4$ SYM  (including gluons), which will then split again and again, thus 
progressively evacuating the original virtuality via successive branchings. 
In a conformal field theory like ${\mathcal N}=4$ SYM this branching process 
will in principle go on for ever. If the coupling is weak, the 
probability for having many splittings is however small and the evolution is slow. 
Then the evolution can be studied in perturbation theory, as we shall discuss in 
Sects.~\ref{sec:zero} and \ref{sect:jet}.
But at strong coupling, we expect this evolution to proceed as fast as permitted by the 
energy--momentum conservation together with the uncertainty principle. 
Its study can then be addressed within the framework of the AdS/CFT 
correspondence, and some results in that sense will
be presented below, in Sects.~\ref{sect:3P} and \ref{sec:Witten}.

To be able to follow the space--time evolution of the decaying system, we need to 
start with a perturbation which is {\em localized} in space and time.  This is conveniently
described by a {\em wave--packet} (WP). Namely, we shall assume that the time--like photon 
is created by the following operator (for more clarity we shall use a hat to denote quantum 
operators in the CFT)
 \beq \label{Jqdef}
 {\hat J}_q\,\equiv\,\int \rmd^4x\,A^\mu_{q}(x)\,{\hat J}_\mu(x)\,,\eeq
where the $\mcal{R}$--current operator  ${\hat J}_\mu(x)$ is convoluted with a
Gaussian WP $A^\mu_{q}(x)$ which encodes the information about the 4--momentum,
the polarization, and the space--time localization of the original perturbation.

It is instructive to construct this WP in the rest frame of the photon, but then study it
in a highly boosted frame. 
This is useful since a boost with a large Lorentz factor $\gamma\gg 1$ renders the
physical interpretation of the quantum evolution more transparent by
enhancing the lifetime of the virtual excitations (by Lorentz time dilation).
In the photon rest frame, the WP is chosen as $A^\mu_Q(x)=\varepsilon^\mu_{(\lambda)}
 \phi_Q(x)$ where $\varepsilon^\mu_{(\lambda)}$ with $\lambda= 0,\pm1$ are the 
 three polarization states allowed to a time--like photon (the polarization
 index will be omitted in what follows) and 
 \beq\label{AWPRF}
  \phi_Q(x)\,\equiv\,\mcal{N}\,
 \rme^{-iQ t}\,\exp\left\{-\frac{t^2+r^2}
 {2\sigma^2}\right\}\,,\qquad\int\rmd^4 x\,
 |\phi_Q(x)|^2\,=\,1\,,
 \eeq
is a normalized WP with central 4--momentum $q^\mu=(Q,0,0,0)$, which is 
localized near the origin of space--time ($t=r=0$) with an uncertainty $\sigma$.
We assume $\sigma Q\gg 1$, in such a way that the Fourier modes 
$k^\mu=(k^0,\bm{k})$  included in the WP have a typical energy 
$k^0\simeq Q$ and a typical virtuality $k^2\simeq q^2 = -Q^2$. 
One has indeed $k^0=Q+ \order{1/\sigma}$, $k^i=\order{1/\sigma}$, $i=1,2,3$.

Consider now the wave--packet in a boosted frame (the `laboratory' frame)
in which the photon propagates along the $x_3$ axis nearly at the speed of light. 
In this frame, the WP has a central 4--momentum  $q^\mu=(q^0,0,0,q^3)$ with
$v\equiv q^3/q^0\simeq 1$. It is then convenient to introduce light--cone components
$q_{\pm}\equiv (q^0\pm q^3)/{\sqrt{2}}$, in terms of which $q^\mu=(q_+,q_-,{\bm 0}_\perp)$
and the virtuality can be expressed as $Q^2=q_0^2-q_3^2 = 2q_+q_-$. We shall also
need the boost factor,
 \beq
 \gamma\,\equiv\,\frac{1}{\sqrt{1-v^2}}\,=\,\frac{q^0}{Q}\,\equiv\,
 \cosh \eta\quad \Longrightarrow\quad
 q_+=\frac{Q}{\sqrt{2}}\,\rme^\eta \,\simeq\,\sqrt{2}\gamma Q\,,
 \eeq
which is very large: $\gamma\gg 1$.  The boosted version of the WP
reads $A^\mu_q(x)=\varepsilon^\mu\,\phi_q(x)$ 
where\footnote{To simplify writing, we shall not distinguish between
lower and upper light--cone components; e.g. $A^+\equiv A_+ \equiv
(A^0+A^3){\sqrt{2}}$. Also, we use the same notations for the polarization
vectors in the rest frame and in the laboratory frame, although the longitudinal
polarization is of course affected by the boost.}
 \beq\label{ATLWP}
 \phi_q(x)\,=\,\mcal{N}\, 
 \rme^{-iq_+x_--iq_-x_+}\,\exp\left\{-\frac{x_+^2}{2\sigma_+^2}
 -\frac{x_-^2}{2\sigma_-^2}-\frac{x_\perp^2}{2\sigma_\perp^2}\right\},
 \eeq
with the various widths related to the width $\sigma$ in the rest
frame via the following relations,
 \beq
 \sigma_+\,\simeq\,2\gamma\,\sigma,\qquad
 \sigma_-\,\simeq\,\frac{\sigma}{2\gamma}\,,
 \qquad \sigma_\perp\,=\,\sigma\,,\eeq
which express the Lorentz dilation
(contraction) of the WP in the $x_+$ ($x_-$) direction.
These relations imply the inequalities
  \beq\label{Asigma}
  \sigma_+ q_-\,\gg\,1,\qquad \sigma_- q_+\,\gg\,1,\qquad
  \sigma_\perp Q\,\gg\,1,\eeq
which in turn guarantee that $k^\mu\simeq q^\mu$ for the typical modes included in the WP. 
The WP \eqref{ATLWP} is normalized to unity in the sense of \eqn{AWPRF} if we choose 
$|\mcal{N}|^2 = 1/(\pi^2 \sigma_+ \sigma_- \sigma_{\perp}^2)$.

In order to study the matter distribution produced by the decaying system at
late times, we shall compute one--point functions like\footnote{The `average
electric charge density' $\mcal{J}_q$ is included here only for illustration: for the problem
at hand, where the decay is initiated by a electrically neutral photon, we have
$\mcal{J}_q=0$ in the conformal ${\mathcal N}=4$ SYM theory.}
 \beq \label{Edensity}
 \mcal{E}_q(x)\,\equiv\,\langle {\hat J}_q^{\dagger}
 \,{\hat T}_{00}(x)
 \,{\hat J}_q\rangle\,,\qquad
 \mcal{P}_q(x)\,\equiv\,\langle {\hat J}_q^{\dagger}
 \,{\hat T}_{++}(x)
 \,{\hat J}_q\rangle\,,\qquad
  \mcal{J}_q(x)\,\equiv\,\langle {\hat J}_q^{\dagger}
 \,{\hat J}_{+}(x)
 \,{\hat J}_q\rangle\,,
 \eeq
and two--point functions of the type
 \beq \label{2point}
 \mcal{P}^{(2)}_q(x_1,x_2)\,\equiv\,\langle {\hat J}_q^{\dagger}
 \,{\hat T}_{++}(x_1){\hat T}_{++}(x_2)
 \,{\hat J}_q\rangle\,,\qquad
  \Pi_q(x_1,x_2)\,\equiv\,\langle {\hat J}_q^{\dagger}
 \,{\hat J}_{+}(x_1){\hat J}_{+}(x_2)
 \,{\hat J}_q\rangle\,,
 \eeq
where it is understood that all the time arguments $x_i^+$ are much larger than $\sigma_+$.
Recalling the definition \eqref{Jqdef} of the operator ${\hat J}_q$ which creates the state,
it should be clear that a `one--point function' like $\mcal{E}_q(x)$ is truly a {\em three}--point
function in the CFT, and similarly $\mcal{E}^{(2)}_q(x_1,x_2)$ and $\Pi_q(x_1,x_2)$
are truly  {\em four}--point functions.

The integrated quantities
\beq \label{Etotal}
 E_q\equiv \int \rmd^3 \bm{x} \,\mcal{E}_q
 (x_+,x_-,x_\perp)\,\qquad
  P_q\equiv \int\rmd x_- \rmd^2  \bm{x}_\perp \,\mcal{P}_q
 (x_+,x_-,x_\perp)\,
 %\,=\,\langle {\hat J}_q^{-\,\dagger}\,{\hat P}^{+}(x^+)
 %\,{\hat J}_q^-\rangle\,,\qquad {\hat P}^{+}(x^+)\equiv
 %\int\rmd x^- \rmd^2 x_\perp \,{\hat T}_{++}(x)
 \eeq
represent the total energy and the total (light--cone) longitudinal momentum
of the state created by the operator ${\hat J}_q$, and are {\em a priori} known:
by energy--momentum conservation, they are the same as the respective quantities, $q_0$ 
and $q_+$, of the original, time--like, photon. In view of this, it might be tempting to 
interpret the integrands in \eqn{Etotal}, i.e. $\mcal{E}_q(x)$ and $\mcal{P}_q(x)$, as the 
corresponding average {\em densities}. 
But this interpretation would be generally incorrect, as we now explain.
The correlation functions introduced in Eqs.~\eqref{Edensity} and \eqref{2point} 
are truly {\em (forward) scattering amplitudes}, which describe the interaction between
a `probe' (operator insertions like ${\hat T}_{++}(x)$ or ${\hat J}_{+}(x_1){\hat J}_{+}(x_2)$)
and a `target' (the decaying system created by ${\hat J}_q$). In the case of the
three--point functions, this interaction will generally modify the internal structure
of the target and thus it cannot represent a fine measurement of this structure
at the time of scattering. The four--point functions, on the other hand, {\em can} 
be used to define a proper measurement, in the following sense: if the space--time
coordinates $x_1$ and $x_2$ of the two operator insertions are sufficiently close to 
each other, then the quantity  $\Pi_q(x_1,x_2)$ is a measure of the density of 
$\mcal{R}$--charge squared at the central point $(x_1+x_2)/2$ as probed with a 
resolution scale fixed by the difference $x_1-x_2$ (and similarly for the other
four--point functions).

The above considerations, to be developed at length in what follows, show that the
notion of {\em resolution} is central to any quantum measurement. This is best appreciated
by working in momentum space. Then the resolution is controlled by the 4--momentum 
$\Delta^\mu$ transferred by the probe to the target, i.e. the momentum carried by
the  Fourier modes ${\hat T}_{++}(\Delta)$ of the probe operator ${\hat T}_{++}(x)$. 
For a three--point function like $\mcal{P}_q(x)$, energy--momentum conservation
requires this transferred momentum $\Delta$ to be smaller 
than the uncertainty $\sim 1/\sigma$ in the target momentum. (For brevity, we use
$\sigma$ to collectively denote any of the widths of the target WP, \eqn{ATLWP}. 
More precisely, the conditions on the 4--momentum $\Delta^\mu$ of the probe 
should read as follows: $\Delta_+\lesssim 1/\sigma_-\ll q_+$, 
$\Delta_-\lesssim 1/\sigma_+\ll q_-$, and $\Delta_\perp\lesssim 1/\sigma$.) 
Yet, in general, it would be wrong to conclude that the quantity $\mcal{P}_q(x)$ can be 
interpreted as the average longitudinal--momentum density at $x$ coarse--grained 
over a distance $\sigma$. Indeed, even a relatively soft momentum $\Delta\sim 1/\sigma$
is still too hard to be absorbed by the target at some large time $x_+\gg\sigma_+$
{\em and let the state
of the latter unchanged} (within the limits of the uncertainty principle). This is so because,
for sufficiently large times, the decaying system contains {\em arbitrarily soft quanta}.

This is most easily seen at weak coupling, 
where one can explicitly follow the evolution of the system via successive branchings. 
One thus finds that the typical longitudinal momenta, $k_+$, of the partons composing
the system keep decreasing with time, as expected for a branching picture 
(see Sect. \ref{sect:jet} for details). In order to `see' such partons,
a probe should transfer to them a longitudinal momentum $\Delta_+$ of the
order of their own respective momentum $k_+$. (If $\Delta_+\gg k_+$, 
there is not enough overlap between the probe and the partons to allow for significant
interactions. If, on the other hand, $\Delta_+\ll k_+$, the probe cannot discriminate 
the individual partons, but only their collective properties averaged over a distance 
$\delta x_-\sim 1/\Delta_+$.)  Clearly, an interaction with $\Delta_+\sim k_+$ will
strongly affect the struck parton and hence it cannot contribute 
to an {\em elastic} scattering {\em unless} the momentum transfer $\Delta_+$  
is taken back away by a subsequent interaction. This can happen in a measurement represented 
by a four--point function, like $\Pi_q(x_1,x_2)$, in which case the momentum $\Delta$ 
transferred to the target by the first insertion ${\hat J}_{+}(\Delta)$ of the probe operator 
is then taken away by the second insertion\footnote{More generally, the 4--momenta
$\Delta_1$ and $\Delta_2$ introduced by the two successive insertions, ${\hat J}_{+}(\Delta_1)$
and ${\hat J}_{+}(\Delta_2)$, can be arbitrary but such that their sum $\Delta_1+\Delta_2$
is at most of order $1/\sigma$. Via Fourier transform, this sum  $\Delta_1+\Delta_2$
is conjugated to the 
central coordinate $(x_1+x_2)/2$ of  the measurement process, whereas the difference
$(\Delta_1-\Delta_2)/2$ is conjugated to the coordinate separation $x_1-x_2$ and
fixes the resolution.} ${\hat J}_{+}(-\Delta)$.
But this cannot be the case for three--point functions like those shown in \eqn{Edensity}.

We thus conclude that, in order to measure a {\em local} quantity, like a density, 
one can use four--point functions,  but not also three--point functions. 
Yet, the latter can be used to measure {\em global} properties, like the total energy 
\eqref{Etotal}~: the respective measurement involves no momentum transfer, so it cannot 
affect the decaying system. In general, such a global measurement contains no
information about the fine spatial distribution of the energy. In some cases, one can recover
part of this information by exploiting the symmetries of the problem. For instance,  the average
matter distribution produced by the decaying photon has spherical symmetry in the photon
rest frame. Accordingly, the energy density per unit solid angle is simply obtained as
${\rmd E}/{\rmd\Omega}= E_q/4\pi$ with $E_q$ the total energy in \eqn{Etotal}. But 
the {\em radial} distribution of the energy depends upon the detailed  dynamics and cannot 
be inferred in such a simple way. Similarly, the {\em longitudinal} distribution of the energy
in the laboratory frame, i.e. its dependence upon $x_-=(t-x_3)/\sqrt{2}$, cannot be deduced
without an explicit calculation. In what follows, we shall present such calculations for both 
three--point and four--point functions, at both weak and strong coupling.

The above discussion shows the importance of simultaneously controlling the
{\em localization} of the probe and its {\em resolution}. This can be done by 
introducing a corresponding wave--packet, i.e. by using smeared versions of 
the probe operators, defined by analogy with \eqn{Jqdef}; e.g.,
  \beq \label{TDelta}
  \hat T_\Delta(\tau)\,=\,\int \rmd^4 y\, \psi_\Delta(y; \tau)\,
 \hat T_{++}(y)\,.\eeq
The probe wave--packet $\psi_\Delta(y;\tau)$ must explore, with the desired resolution, 
the whole region of space where the decaying system can be located at the time of measurement
$x^+\equiv \tau$, with $\tau\gg\sigma_+$. A convenient form for the WP is
the following Gaussian
\beq\label{PROBEWP}
 \psi_\Delta(y;\tau)\,=\,\mcal{C}\,
 \rme^{i\Delta\cdot y}\,\exp\left\{-\frac{(y_+-\tau)^2}{2\tilde\sigma_+^2}
 -\frac{y_-^2}{2\tilde\sigma_-^2}
 -\frac{y_\perp^2}{2\tilde\sigma_\perp^2}\right\}.
 \eeq
As usual, the central four--momentum $\Delta^\mu=(\Delta_+, \Delta_-, \bm{\Delta}_\perp)$ 
specifies the space--time resolution of the probe, whereas the Gaussian controls its localization.
The latter is centered at $y_+=\tau$, with a  temporal
width which obeys $\tilde\sigma_+\ll\tau$ (for the time of measurement to be well defined).
It is furthermore centered at $y_-=0$ and $y_\perp=0$, with spatial widths
$\tilde\sigma_-$ and $\tilde\sigma_\perp$ which are large enough for the
spatial momenta of the typical Fourier components to have only little spread
around the respective central values: 
$\tilde\sigma_- \Delta_+\gg 1$ and $\tilde\sigma_\perp \Delta_\perp \gg 1$
(compare to \eqn{Asigma}). It might be tempting to try and enforce the similar condition 
$\tilde\sigma_+ \Delta_-\gg 1$ on the minus component (the light--cone energy), but
it turns out that this is not always possible. Indeed, the time variable in \eqn{PROBEWP}
takes a typical value $y_+=\tau$, which is large. In order to avoid the rapid oscillations 
of the complex exponential $\rme^{-i\Delta_-y_+}$ we shall sometimes need to 
require $\Delta_-$ to be small,
$\Delta_-\tau\lesssim 1$. Then the condition $\tilde\sigma_+ \Delta_-\gg 1$ cannot be satisfied
simultaneously with  $\tilde\sigma_+\ll\tau$. But this is not
a serious limitation, since we do not need any other temporal resolution scale besides 
the width  $\tilde\sigma_+$. To summarize, the WP \eqref{PROBEWP} 
with the constraints alluded to above  provides a measurement at time $\tau$ 
with spatial resolutions $\delta x_\perp \sim 1/ \Delta_\perp$ and $\delta x_-\sim 1/ \Delta_ +$.

\begin{figure}
\begin{center}
\includegraphics[scale=0.65]{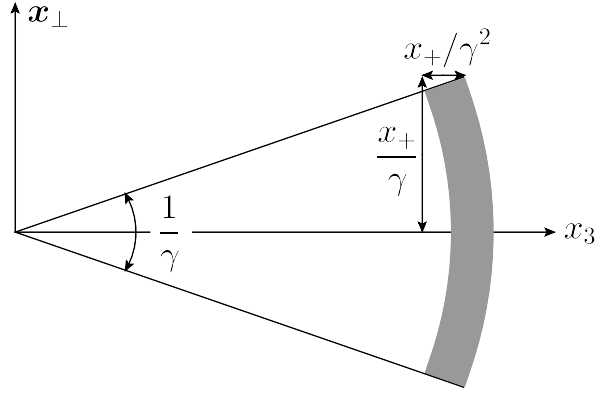}
\end{center}
\caption{The region in space--time allowed by causality and special
relativity for the matter distribution  produced by  the decaying photon 
in a highly boosted frame. More precisely, the gray band represents
the boosted version of the half sphere $\{r\le t; x_3>0\}$ (the region which,
in the rest frame, includes the quanta whose velocities have a
positive third component ($v_3>0$).}
\label{fig:lorentz}
\end{figure}

It is finally convenient, before concluding this section, to anticipate the typical resolution
scales that we shall need in order to probe the structure of the decaying system. This can
be fixed by comparison with the maximal (transverse and longitudinal) sizes occupied by the
system, that we shall now estimate.  For simplicity, we start in the rest
frame of the time--like photon, where the matter produced by its decay is restricted to the
sphere $r\le t$, simply by causality. When boosting this spherical distribution with a large
$\gamma$ factor, its transverse size remains unchanged, that is, $\Delta x_\perp \sim t_{RF}
\sim t/\gamma$. (We used the fact that the time $t$ in the laboratory frame is larger by
a factor $\gamma$ than the time $t_{RF}$ in the rest frame.) As for the longitudinal extent
$\Delta x_3$, this is subjected to Lorentz contraction, yielding $\Delta x_3\sim  t_{RF}/\gamma
\sim t/\gamma^2$. The fastest partons propagate at the speed of light, so they will be located
on the light--cone $x_3=t$ (or $x_-=0$). Most of the other partons, which are expected to be
time--like and thus have velocities smaller than one, will be distributed within a region
$\Delta x_3\sim  t/\gamma^2$ behind the light--cone. Hence, the matter produced by the 
decay at light--cone time $\tau$ will be located within a small solid angle $\Delta\Omega\sim 
(\Delta x_\perp/\tau)^2 \sim 1/\gamma^2$ around the $x_3$ axis and within a (relatively)
thin longitudinal shell $\Delta x_-\sim \tau/\gamma^2$ around $x_-=0$. This region
is represented as a grey band in Fig.~\ref{fig:lorentz}.
To be able to explore its internal structure, we need a probe with sufficiently large spatial 
momenta $\Delta_+\gtrsim \gamma^2/\tau$ and $\Delta_\perp \gtrsim \gamma/\tau$.
But the opposite case, with $\Delta_+\ll \gamma^2/\tau$, is also interesting, since then
the probe measures the matter distribution {\em integrated} over the longitudinal (or radial)
axis. From the previous discussion, we expect a three--point function to be a good measurement
(say, of the energy) when $\Delta_+\ll \gamma^2/\tau$ --- in which case it correctly provides
the energy density {\em per unit transverse area} (or per unit solid angle in the photon rest frame) 
---, but not also in the opposite case  ($\Delta_+\gtrsim \gamma^2/\tau$), where the longitudinal resolution is relatively high. These expectations will be confirmed by the subsequent
calculations, at both strong and weak coupling.

\section{The three--point function at infinitely strong coupling}
\label{sect:3P}

In this section we shall briefly review a recent calculation \cite{Hatta:2010dz}
of the three--point function introduced 
in \eqn{Edensity} in the $\mcal{N}=4$ SYM theory at (infinitely) strong coupling,
which uses the method of the `backreaction' within the dual supergravity theory.
An alternative method, which relies on Witten diagrams for supergravity and is perhaps
more straightforward to use for the calculation of the four--point functions, will be presented 
in Sect.~\ref{sec:Witten}.

\subsection{Backreaction in supergravity}
\label{sec:AdS}

Within the AdS/CFT correspondence, a time--like photon decaying in the
vacuum of the ${\mathcal N}=4$ SYM theory with infinitely strong 't Hooft
coupling ($\lambda\equiv g^2 N_c\to\infty$) is dual to a supergravity
(SUGRA) vector field $A^\mu(x,z)$ which propagates into the bulk of
AdS$_5$ and whose boundary value $A^\mu(x,0)$ at the Minkowski boundary
($z=0$) is identified with the classical field  $A^\mu_{q}(x)$ 
representing the perturbation on the gauge theory side:
$A^\mu_q(x)=\varepsilon^\mu\,\phi_q(x)$ with  the wave--packet
$\phi_q(x)$ given in \eqn{ATLWP}.

Within this franework, the first three--point function in \eqref{Edensity} 
(the `energy density' $\mcal{E}_q(x)$) can be determined
via a {\em backreaction calculation}. This refers to the linear response
of the metric of AdS$_5$ to the small perturbation represented by the
bulk excitation induced by the boundary WP \eqref{ATLWP}. In turn, this
bulk excitation can be obtained by propagating the boundary field with
the help of the relevant bulk--to--boundary propagator (the Green's
function for the Maxwell equation in AdS$_5$)~:
  \beq\label{Abulk}
 A^\mu_q(x,z)\,=\,\int\,\rmd^4y\,D^{\mu\nu}(x-y,z)\,
  \varepsilon_\nu\,\phi_q(y)\,=\,
  \int\,\frac{\rmd^4p}{(2\pi)^4}\,\rme^{ip\cdot x} \,\varepsilon_\nu\,
  D^{\mu\nu}(p,z)\,\phi_q(p)\,,
  \eeq
where $D^{\mu\nu}(x-y,z)$ is the Maxwell propagator in AdS$_5$ 
and in the `radial' gauge $A_z=0$. Here, $z$ denotes the radial (or `fifth') dimension
in AdS$_5$ and we are using the metric (with $L$ the curvature radius of
AdS$_5$)
  \beq\label{metric}
\rmd s^2\,\equiv\,G_{MN}\,\rmd x^M \rmd x^N
 \,= \,
\frac{L^2}{z^2} \,\big[ -\rmd t^2 + \rmd {\bm r}
^2 + \rmd z^2\big],
 \eeq
(with $M=\mu$ or $z$) in terms of which the Minkowski boundary lies at $z=0$, as anticipated.

The SUGRA field \eqref{Abulk} will be explicitly constructed in
Sect.~\ref{sect:TL} below, from which we anticipate here the salient
features (see also Ref.~\cite{Hatta:2010dz}). Namely, the bulk excitation
is a Gaussian WP which propagates in AdS$_5$ at the 5--dimensional speed
of light, with longitudinal velocity equal to $v$ and radial velocity
$v_z=\sqrt{1-v^2}=1/\gamma$. More precisely, at  time\footnote{Note that
$x_+\simeq \sqrt{2}t$ for space--time points located near the center of
the WP.} $t\gtrsim \sigma_+$, the center of the Gaussian is located at
 \beq\label{centerBWP}
 z=\,\frac{t}{\gamma}\,,\qquad %=\,\frac{x^+}{\sqrt{2}\gamma}\,,\quad
  \bm{x}_\perp=0,\,\qquad x_3=vt\,,
 %\quad\bigg(\mbox{and hence}\quad
 %x_-^*=\,\frac{x_+}{4\gamma^2}\bigg)
 \eeq
with (roughly) time--independent widths fixed by the original Gaussian
\eqref{ATLWP} (see Sect.~\ref{sect:TL} for details). The physical meaning
of the bulk trajectory \eqref{centerBWP} can be understood with the help
of the {\em UV/IR correspondence} \cite{Susskind:1998dq,Peet:1998wn}~: 
the penetration $z$ of the WP in the
bulk is related to the virtuality $K=\sqrt{|k^2|}$ of the typical quanta composing the
decaying WP in the boundary gauge theory: $z\sim 1/K$. (For the situation at
hand, these quanta are typically time--like: $k^2 <0$.) Hence, the fact
that $z$ is localized near ${t}/{\gamma}$ means that the decaying system
at time $t\gg\sigma_+$ involves quanta with a typical virtuality
$K(t)\sim\gamma/t$ and hence a typical longitudinal momentum $k_+(t)
=\gamma K(t)\sim\gamma^2/t$. By the uncertainty principle, such quanta 
occupy a region with transverse area
$(\Delta x_\perp)^2 \sim (t/\gamma)^2$ and longitudinal extent 
$\Delta x_3\sim t/\gamma^2$ behind the light--cone ($x_3=t$). 
Note that this is the maximal region allowed
by causality and special relativity (cf. the discussion towards the end
of Sect.~\ref{sec:WP}).
This qualitative picture for the decaying system at strong coupling will be later 
substantiated, in Sect.~\ref{sec:Witten},  by a proper `measurement' which 
involves the calculation of a four--point function.
On the other hand, this picture is not manifest at the level of the 
three--point function $\mcal{E}_q(x)$, to which we now turn.

The calculation of the backreaction amounts to solving
the linearized Einstein equations for the (small) change $\delta G_{MN}$
in the metric of AdS$_5$ which is generated by the energy--momentum
tensor $t^{MN}$ associated with the bulk excitation. Finally, the
three--point function \eqref{Edensity} is inferred from the near boundary
behaviour of $\delta G_{MN}$. Mathematically, this is obtained by
propagating the metric perturbation from the location $(\acute x^\mu, z)$
of its source in the bulk to the measurement point $x^\mu$ on the
boundary ($z=0$), with the help of the retarded bulk--to--boundary
propagator. Strictly speaking, this calculation will yield a {\em
retarded} three--point function --- the retarded version of the Wightman
function introduced in \eqn{Edensity}. But this retarded three--point
function is precisely the physical response function whose space--time
localization we would like to study.

For simplicity, we shall replace the bulk WP by a 4--dimensional
$\delta$--function with support at the central coordinates shown in
\eqn{centerBWP}~: $t^{MN}(\acute x, z)\propto\delta(\acute x_3-v
\tp)\delta^{(2)}(\acute{\bm{x}}_{\perp}) \delta(z-\tp/\gamma)$. This
means that we probe physics on space--time resolution scales which are
soft compared to the respective widths of the Gaussian WP, which is
indeed sufficient for our purposes here.  This
facilitates the calculation of the backreaction, which in general
involves an integral over the support of the bulk excitation. The result
of this calculation reads (see Ref.~\cite{Hatta:2010dz} and also the Appendix
\ref{sec:EB} to the present paper for details)
\begin{align}\label{EApart}
 \mcal{E}_q(t,\bmx) \,=\,
 &\frac{2q_0}{\pi}\,\frac{t+vx_3}{\gamma^2}\
 \del^2_{r^2} \int_0^\infty \dif \tp\, \tp \
 \delta\big(t^2 -r^2 - 2(t-v x_3)\tp\big)
 \nn 
 & + \,\frac{2q_0}{\pi}\,\frac{v^2 {x}_{\perp}^2}{\gamma^2}\
 \del^3_{r^2} \int_0^\infty \dif \tp\, \tp^{\,2} \
 \delta\big(t^2 -r^2 - 2(t-v x_3)\tp\big),
 \end{align}
where $r=|\bm{x}|$ and $q_0$ is the total energy carried by the original
WP \eqref{ATLWP} (and therefore also the total energy of the evolving
partonic system produced by its decay).  Below we shall denote
the two terms in \eqn{EApart} as $\mcal{E}_A$ and $\mcal{E}_B$,
respectively, with $\mcal{E}_q=\mcal{E}_A+\mcal{E}_B$. In the original
calculation in Ref.~\cite{Hatta:2010dz}, the second term $\mcal{E}_B$
has actually been missed, so for completeness we shall explicitly derive
this term in Appendix \ref{sec:EB}.

\eqn{EApart}  can be
understood as follows: at time $\tp$, the bulk excitation localized at
$z=\tp/\gamma$, $\acute x_3=v\tp$, and $\acute x_\perp=0$ emits a
gravitational wave $\delta G_{MN}$ which propagates through AdS$_5$ at
the respective speed of light up to the measurement point
$x^\mu=(t,\bm{x})$ on the boundary. The $\delta$--function in the
integrand represents the support of the retarded bulk--to--boundary
propagator for the Einstein equations in AdS$_5$. Its argument follows
from causality together with the condition of propagation at the 5D speed
of light, for both the bulk excitation and the gravitational wave:
 \beq\label{tp}
 (t-\tp\,)^2={z}^{\,2}+(x_3-\acute x_3)^2+x_\perp^2=
{{\tp}^{\,2}+ r^2 -2 x_3v\tp}\ \Longrightarrow\
 t^2 -r^2 = 2(t-v x_3)\tp\,.
 \eeq
A physical interpretation for this  condition back in the original gauge theory 
will be proposed in Sect.~\ref{sec:interp}.

A priori, \eqn{EApart} involves an integral over all the positive values
of $\tp$, meaning over all the values $z=\tp/\gamma$ of the radial
coordinate of the bulk excitation. However, the presence of the external
derivatives, $\del^2_{r^2}$ in the first term and respectively $\del^3_{r^2}$ 
in the second one, introduces an important simplification: it implies
that the net result for $\mcal{E}_q$ comes exclusively from $\tp=0$, that
is, from the early time when the bulk excitation had been just emitted
and was still localized near the boundary ($z\simeq 0$). Indeed, after
using the $\delta$--function to integrate over $\tp$, one finds
  \begin{align}
 \mcal{E}_q(x) \,=\, &\frac{q_0}{2 \pi\gamma^2}\,\frac{t+vx_3}{(t-v x_3)^2}\
 \del^2_{r^2} [(t^2 - r ^2) \Theta(t^2-r^2)]
 \nn
 & +\, \frac{q_0}{4 \pi\gamma^2}\,\frac{v^2 {x}_{\perp}^2}{(t-v x_3)^3}\
 \del^3_{r^2} [(t^2 - r ^2)^2 \Theta(t^2-r^2)],
 \end{align}
where the $\Theta$--function enforcing $r\le t$ (generated via the
condition that $\tp\ge 0$) is the expression of causality. In the first term,
this $\Theta$--function is multiplied by the factor $(t^2 - r ^2)$ which is
linear in $r^2$; so the only way to obtain a non--zero result after
acting with $\del^2_{r^2}$ is that one of the two derivatives act on the
$\Theta$--function and thus generate a $\delta$--function at $t=r$.
A similar discussion applies to the second term, which involves an additional
factor of $(t^2 - r ^2)$ inside the square brackets and also an additional external
derivative. Combining the two terms, one finds
\beq\label{Eparticle}
 \mcal{E}_q(x) \,=\,
 \frac{q_0}{2 \pi\gamma^4}\,\frac{t^2}{(t-v x_3)^3}\,\delta(t^2-r^2)\,.
 \eeq
This describes a spherical shell of zero width which propagates at the
{\em 4--dimensional} speed of light. Returning to the constraint
\eqref{tp} on the emission time $\tp$, one sees that a signal which at
time $t$ is located at $r=t$ has been necessarily generated at $\tp=0$
and hence $z=0$, as anticipated. 

Now, as it should be clear from the previous discussion, these extremely sharp 
localization properties --- the fact that the signal is strictly light--like ($r=t$) and the 
(related) fact that the whole contribution to the backreaction comes from $z=0$ --- 
are to be understood up to a smearing  on the scale set by the width 
$\sigma$ of the original WP : in reality, the spherical shell has a non--zero
radial width $t-r\sim \sigma$ and the values of $z$ contributing to this
result are not exactly zero, but of order $\sigma$.
Yet, these results --- in particular, the fact that the signal appears to propagate 
{\em without broadening} (i.e. by preserving a constant radial width up to 
arbitrarily large times) --- would be extremely curious if they were to represent 
the distribution of matter produced by a decaying system {\em at strong coupling},
as we now explain.

A thin shell of energy propagating at the speed of light is the result
that would be naturally expected in a {\em non--interacting} quantum field 
theory, or, more generally, to {\em zeroth order} in perturbation theory for a
field theory at weak coupling. In that limit, the time--like photon would
decay into a pair of (massless) {\em on--shell} partons which would then
propagate at the speed of light. In a given event and in the rest frame
of the virtual photon, such a decay yields two particles propagating back
to back. After averaging over many events, the signal looks like a thin
spherical shell expanding at the speed of light. In fact, it is straightforward 
to check (and we shall explicitly do that in the next sections)
that the result \eqref{Eparticle} of the AdS/CFT calculation at
{\em infinitely strong coupling} is exactly the same as the corresponding
prediction of  the $\mcal{N}=4$ SYM theory at {\em zero coupling}. 
By itself, this `coincidence' should not be a surprise: as explained in the
Introduction, the three--point function under consideration cannot
receive quantum corrections, as it is protected by conformal symmetry and
energy conservation. So, the corresponding result, as shown in \eqn{Eparticle},
is {\em a priori} known to be independent of the coupling. 
But whereas this situation looks natural in view of the underlying conformal 
symmetry, it might still look puzzling from a physical perspective: at non--zero 
(gauge) coupling, the decay of the time--like photon should also involve 
virtual quanta which propagate slower than light. Then,
the emerging matter distribution should also have support at points inside
the sphere $r\le t$, and not only on its (light--like) surface.

The solution to this puzzle is that, as already argued in Sect.~\ref{sec:WP} and
will be demonstrated via explicit calculations in what follows, this
three--point function is {\em not} a good measurement of the
energy density produced at late times by the decaying photon. It is clearly a good
measurement of its {\em total} energy, and also of its angular distribution in 
the photon's rest frame ($v=0$), where \eqn{Eparticle} yields the expected result
for ${\rmd E}/{\rmd\Omega}$ (recall that $q_0\to Q$ in the rest frame)~:
 \beq\label{shell} \mcal{E}_q(t,r)\, = \,\frac{Q}{4 \pi r^2}\,
 \delta(t-r)\qquad \Longrightarrow\qquad
 \frac{\rmd E}{\rmd\Omega}\,\equiv\int \rmd r \,r^2
 \mcal{E}_q\,=\,\frac{Q}{4 \pi}\,.
 \eeq
But the radial distribution of the energy density is {\em not} correctly represented by
\eqn{Eparticle}, in any frame.  The correct respective distribution will be later computed,
at both weak and strong coupling, 
from four--point functions like those introduced in \eqn{2point}. 
The results to be thus obtained will be very different in the two cases and in particular
they will exhibit strong radial broadening at strong coupling,
in agreement with our general expectations.
This being said, it would be interesting to understand `how the conformal symmetry
works in practice', meaning how is that possible that such a sharply localized result,
\eqn{Eparticle}, can emerge from a calculation {\em at strong coupling}. 
A possible interpretation for that will be provided in the next subsection.

\begin{figure}
\begin{center}
\includegraphics[scale=0.65]{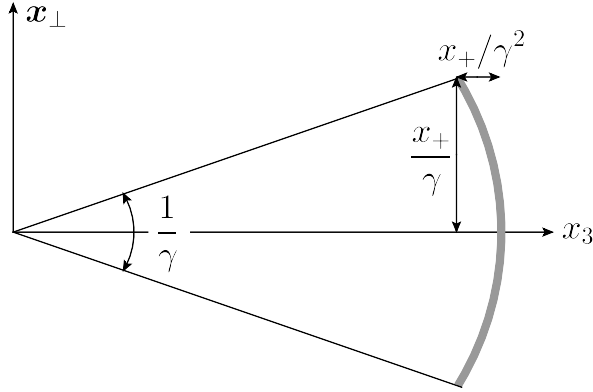}
\end{center}
\caption{Signal generated from the decaying photon in a highly boosted frame. The width of the grey band is $\sigma_-\sim \sigma/\gamma$ (the Lorentz contracted version of the radial width
in the rest frame).}
\label{fig:signal}
\end{figure}

For what follows, it will be useful to have a version of the three-point function \eqref{Eparticle} adapted 
to a highly boosted frame ($\gamma\gg 1$). In that case, it is preferable to work with the probe 
operator ${\hat T}_{++}(x)$ and the associated three--point function $\mcal{P}_q(x)$, as introduced
in \eqn{Edensity}. At high energy, the latter can be estimated as  
$\mcal{P}_q(x)\simeq 2 \mcal{E}_q(x)$
with $\mcal{E}_q(x)$ conveniently rewritten in light--cone coordinates. 
Using $1-v\simeq 1/2\gamma^2$, $1+v\simeq 2$, and hence
 \beq\label{gammaL} 
 t-vx_3\,=\,\frac{x_+(1-v)+x_-(1+v)}{\sqrt{2}}\,\simeq\,
 \frac{x_+}{2\sqrt{2}\gamma^2}\,+\sqrt{2}x_-\,,\eeq
one finds (with $q_+\simeq\sqrt{2}q_0$)
 \beq\label{EIMF}
 \mcal{P}_q(x) \,\simeq\,
  \frac{q_+}{8 \pi \gamma^4}\,
 \frac{x_+^2}{\Big(x_- + \frac{x_+}{4 \gamma^2}\Big)^3}\,
 \delta\left(2 x_+ x_- - {x}_{\perp}^2\right).
 \eeq
The denominator in this equation is the reflection of Lorentz
contraction, as discussed at the end of Sect.~\ref{sec:WP}~: it
restricts the longitudinal coordinate $x^-$ to (positive) 
values satisfying $x_- \lesssim {x_+}/4\gamma^2$. 
But the presence of the $\delta$--function in \eqn{EIMF} entails a much
stronger constraint: it implies  that the signal is localized within an arc of a 
spherical shell of zero width, or more precisely of width $\sigma_-\sim\sigma/\gamma$ 
(the Lorentz--contracted version of the respective width in the rest frame). 
This distribution is illustrated in Fig.~\ref{fig:signal} which should be compared 
with Fig.~\ref{fig:lorentz}. One sees that, in the boosted frame, the lack
of radial broadening mostly manifests itself as a {\em lack of longitudinal broadening}: 
the signal \eqref{EIMF} deviates from the light--cone ($x_-=0$) by a 
distance $x_-= x_\perp^2/2x_+ $ (modulo the width $\sigma_-$ of the shell) 
which for sufficiently small $x_\perp\ll x_+/\gamma$ is much smaller than the maximal 
value $\sim{x_+}/\gamma^2$ permitted by Lorentz contraction. Conversely, this
argument implies that $x_\perp$ is restricted to values $x_\perp\lesssim x_+/\gamma$,
which in turn implies that the solid angle subtended by the shell
is $\delta\Omega \sim 1/\gamma^2$.
Note finally that \eqn{EIMF} yields the correct result for the {\em total} longitudinal 
momentum (cf. \eqn{Etotal}),  as expected: 
$ P_q\equiv \int\rmd x^- \rmd^2 x_\perp \,\mcal{P}_q= q_+$.

\subsection{A physical interpretation for the `backreaction'}
\label{sec:interp}

As already noticed, the SUGRA results for the three--point function, \eqref{Eparticle} or
\eqref{EIMF}, are characterized by two remarkable and perhaps surprising facts: \texttt{(i)}
the signal propagates at the speed of light without (radial or longitudinal) broadening, and
\texttt{(ii)} the whole contribution to the backreaction comes from small values of $z\lesssim
\sigma$. Within the AdS/CFT calculation, these two features are related to each other, as
we have seen. Namely, the `backreaction' has support only at points satisfying  \eqn{tp},
which for small $z=\tp/\gamma \lesssim\sigma$ implies that $t-r$ (or $t-x_3$ in a
boosted frame) is small as well: $t-r\lesssim\sigma$ in the 
center--of--mass frame and respectively $x_-\lesssim \sigma/\gamma$ in the
frame where $\gamma\gg 1$. That is, the smallness of $z$ (or of $\tp$) 
implies the propagation of the signal at the speed of light. In what follows,
we would like to propose a physical interpretation for these facts in the CFT.

Namely, we shall argue that the interactions responsible for the three--point function are
{\em highly delocalized in time}. The high--momentum transfer between
the target and the probe is carried by a signal which is emitted by the decaying system 
at an early time, well before the measurement time $t$ at which the signal is 
absorbed by the probe. This physical emission time (denoted as $t_{\rm int}$
in the Introduction) can be identified with the time  
$\tp$  at which the gravitational wave is emitted by the bulk 
excitation in the calculation of the backreaction. 
With this interpretation, \eqn{tp} represents the matching 
condition between the resolution of the probe and the kinematics 
of the target `partons' which emit the signal. 
%In particular, for a probe with very high resolution 
%(like the localized operator ${\hat T}_{++}(x)$), the interaction time 
%$\tp$ is extremely small, since it is only at such early times that 
%the target still contains `partons' which are hard enough to match the resolution of the probe.
Furthermore, the gravitational wave in the `backreaction' is the AdS dual of the 
physical signal --- a nearly light--like mode with the quantum numbers of the probe 
operator, which propagates at the speed of light from $\tp$ up to $t$.

In order to establish this interpretation, we shall have a new look at \eqn{tp} which we recall
is the condition that the gravitational wave propagate at the speed of light in AdS$_5$.
For a given observation point $x^\mu$ on the boundary, this condition determines the time
$\tp$ at which the gravitational wave is emitted, hence the radial penetration $z=\tp/\gamma$
of the bulk excitation at that time and, ultimately, the virtuality $K=\sqrt{|k^2|}$ of the typical
quanta composing the decaying system at time $\tp$\,: the UV/IR correspondence implies 
$K\simeq 1/z=\gamma/\tp$. For what 
follows it is convenient to fix the transverse coordinate of the observation point --- namely,
we choose $x_\perp\simeq 0$ (with uncertainty $\sigma$) --- and explore the longitudinal 
region behind the light--cone ($x_-=0$) on a resolution scale $\delta x_-$ which is allowed 
to vary.  As usual, this resolution is controlled by the longitudinal momentum of the probe,
$\delta x_-\simeq 1/\Delta_+$, and is limited by the longitudinal width 
$\sigma_-\sim \sigma/\gamma$ of the original wave--packet. The best possible 
resolution $\delta x_-\simeq \sigma_-$ (corresponding to a maximal momentum transfer 
$\Delta_+\sim \gamma/\sigma$)
has been implicitly used in the calculation of the three--point function 
`at a given space--time point', cf. \eqn{Eparticle} and \eqref{EIMF}. 
But for the present purposes we shall also allow for less precise measurements, 
with $\Delta_+\ll 1/\sigma_-$.

Starting with \eqn{tp}, inserting $x_\perp=0$ and $t- x_3=\sqrt{2}/\Delta_+$, and switching
to light--cone coordinates, one easily finds (we use the notation $\tau\simeq\sqrt{2}t$ for 
the light--cone time of measurement and similarly 
$\acute\tau\simeq\sqrt{2}\tp$ for the emission time)
 \beq\label{acutetau}
 \acute\tau\,\simeq\,\frac{\tau}{1+\frac{\tau\Delta_+}{4\gamma^2}}\,.\eeq
There are two interesting limiting cases:

 \texttt{(i)} {\em High longitudinal resolution:~} $\Delta_+\gg \gamma^2/\tau$. In this case,
 the probe can discriminate longitudinal distances $\delta x_-\simeq 1/\Delta_+$ 
 which are much smaller
 than the upper limit  $\Delta x_-\sim \tau/\gamma^2$ enforced by causality and Lorentz 
 contraction. Then \eqn{acutetau} implies
  \beq\label{acute1}
 \acute\tau\,\simeq\,\frac{4\gamma^2}{\Delta_+}\,\ll\,\tau.\eeq
It is also interesting to estimate (using the UV/IR correspondence) 
the typical virtuality and longitudinal momentum of a 
quantum from the target at time $\acute\tau$~:
 \beq\label{K1}
 K\,\simeq\,\frac{\sqrt{2}\gamma}{ \acute\tau}\,\simeq\,
 \frac{\Delta_+}{2\sqrt{2}\gamma}\,,\qquad k_+\,\simeq\, \sqrt{2}\gamma\,K
 \simeq\, \frac{\Delta_+}{2}\,.\eeq
The last condition ($k_+\sim \Delta_+$) is very interesting: this is the expected
matching condition between the quanta from the target which emit the 
relatively hard signal and the resolution of the probe. Remarkably, this condition 
holds here at the  `emission time'  $\acute\tau$ and not at the measurement time
$\tau$. This is in agreement with our expectation that such hard quanta can only
exist at very early times in the history of the decay. In fact, the above results can be
combined to yield $\acute\tau\simeq k_+/K^2$, which is the time interval required
via the uncertainty principle for the emission of a quantum with 
longitudinal momentum $k_+$ and virtuality $K^2$.
The above discussion makes it natural to identify the `emission time' $\tp$ 
(or $\acute\tau$) in the SUGRA calculation with the time at which the signal measured
by the probe operator at time $t$ (or $\tau$) has been {\em actually emitted} by
the decaying system in the underlying quantum field theory.

%The fact that the signal appears to propagate at the speed of light without
%broadening is also natural from this viewpoint: the matching condition \eqref{acutetau}
%selects a mode of the probe having the same virtuality as the struck parton, that is, 
%$K\sim 1/\sigma$. This value is small compared to $\Delta_+\sim \gamma/\sigma$,
%so this mode is ultrarelativistic. This light--like mode of the probe is dual to the gravitational
%wave which in the corresponding SUGRA calculation remains very close to the
%boundary (i.e., it propagates at the {\em four--dimensional} speed of light).

Returning to \eqn{acutetau}, let us also consider the other interesting limiting case, namely :

\texttt{(ii)} {\em Low longitudinal resolution:~} $\Delta_+\ll \gamma^2/\tau$. In this case, 
the probe cannot discriminate any longitudinal substructure, but, interestingly,
it  can explore the state of the system at relatively late times,
close to the time $\tau$ of measurement.  Indeed,
 \eqn{acutetau} implies
 \beq\label{acute2}
\acute\tau\,\simeq\,\tau\,,\qquad K\,\sim\,\frac{\gamma}{\tau}\,,
\qquad k_+\,\sim\,\frac{\gamma^2}{\tau}\,.
\eeq
This can be understood as follows:
the probe is now much softer ($\Delta_+\ll k_+$) than the typical quanta in the 
decaying system at time $\tau$, so it can interact with the latter without 
significantly disturbing them. 

\comment{
Note also that the longitudinal momentum $\Delta_+$ of the probe is not necessarily larger 
than its virtuality $\Delta\sim K\sim {\gamma}/{\tau}$ (and in any case is much smaller
than $\gamma\Delta$), so the kinematics of the probe is typically space--like.

Note finally that by progressively decreasing $\Delta_+$ from the maximal value 
$\gamma/\sigma$ towards $\gamma^2/\tau$, one can increase the value of the interaction 
time \eqref{acutetau} and thus explore the state of the decaying system at different stages 
of its evolution. For a fixed and relatively high resolution $\Delta_+\gg \gamma^2/\tau$,
the signal will look very much like in Fig.~\ref{fig:signal}, that is, an arc of a spherical
shell which propagates at the speed of light with constant width $\delta x_-\simeq 1/\Delta_+$.
Due to the conformal nature of the  ${\mathcal N}=4$ SYM theory, the very same 
result would be obtained for any value of the coupling, but the underlying physics
would be different at weak coupling. In particular, at zero coupling, there is 
no quantum broadening; so, if the signal corresponding to $\Delta_+\ll \gamma/\sigma$ looks
quite wide ($\delta x_-\gg \sigma_-$), it is just because of the comparatively poor
resolution of the probe. At strong coupling, on the other hand, we expect strong broadening,
yet the signal keeps a constant width $\delta x_-\simeq 1/\Delta_+$ because 
the probe `selects' partons with $k_+\simeq\Delta_+$ in the early history of the target.
}

So far, we have considered a probe with a fixed longitudinal resolution
$\delta x_-\simeq 1/\Delta_+$, that is, we have focused on a single Fourier mode,
with longitudinal momentum $\Delta_+$, of the probe operator. But a similar discussion
applies to a three--point function in coordinate space, like $\mcal{E}_q(x)$.
The associated Fourier 
decomposition involves an integral over all values of $\Delta_+$, but in practice this
integral is dominated by its upper limit $\Delta_+\sim \gamma/\sigma$, which is the maximal
value allowed by energy--momentum conservation. Then, \eqn{acute1} implies 
$\acute\tau\sim\gamma\sigma$ and therefore $z\sim\sigma$. This explains
why the whole contribution to the backreaction `at a given space--time point' comes from
very small $z\lesssim\sigma$. 
The fact that the signal propagates at the speed of light and without broadening
can be qualitatively understood as a consequence of kinematics. Given that this signal is
carried along by essentially a single mode of the probe --- the one with the maximal value of 
$\Delta_+$ ---, it naturally preserves a constant width $\delta x_-\sim \sigma/\gamma$.
And a signal which propagates with $t-x_3=$~const. over a large period of time is 
necessarily luminal.  

\comment{A more formal argument can be constructed as follows: the relevant
Fourier mode of the signal involves the phase
 \beq
 \rme^{i\Delta\cdot x}\,=\,\rme^{-i\Delta_-x_+  -i\Delta_+x_- +i\bm{\Delta}_\perp
 \cdot \bm{x}_\perp}\,\eeq
where $x_+\equiv\tau$ and $\Delta_+\sim \gamma/\sigma$ (since the integral over
$\Delta_+$ is dominated by its upper limit, as alluded to above). This immediately
implies that the longitudinal width remains constant: $x_-\sim 1/\Delta_+ \sim \sigma_-$.
(There is no broadening since larger values $x_-\gg  \sigma_-$ are suppressed by
rapid oscillations.) Furthermore, for large times $\tau\gg\sigma_+\sim\gamma\sigma$,
the integrals over $\Delta_-$ and ${\Delta}_\perp$ are dominated by $\Delta_-\sim 1/\tau$
and ${\Delta}_\perp\sim \gamma/\tau$, respectively (recall that $x_\perp\lesssim \tau/\gamma$).
Accordingly,  the virtuality of the propagating mode is $\Delta^2= -2\Delta_+\Delta_-
+{\Delta}_\perp^2\sim  - \gamma/\sigma\tau$, which is negative (time--like), but small, 
meaning that the mode is {\em nearly light--like}. Specifically, the ratio $|\Delta^2|/\Delta_+^2
\sim \sigma/\gamma\tau$ can be made arbitrarily small by increasing $\tau$, which shows
that the mode propagates indeed at the speed of light.
The lack of broadening in the three--point function at strong (more generally,
finite) coupling follows from the fact 
the width of the signal is controlled by the resolution of the probe and not by the dynamics
of the target. }

To summarize, a three--point function with high longitudinal resolution explores the state 
of the target at very early time, much smaller than  the time of measurement $t$. 
Conversely, the only way how a three--point function can measure the state of 
the target at $t$ is by giving up any precision in the longitudinal (or radial) direction.
These conclusions will be corroborated by the Fourier decomposition of
the three--point function to be presented in the next subsection.

\subsection{Momentum--space analysis of the backreaction}
\label{sec:Delta}

In this subsection,
we shall compute the Fourier transform of the result in \eqn{EIMF} for the
three--point function  in a highly boosted frame. We shall use a mixed Fourier representation
which involves the component ${\hat T}_{++}(\tau,\Delta_+,{\bm \Delta}_\perp)$ of the
probe operator.  As explained towards the end of Sect.~\ref{sec:WP}, this mixed 
representation contains the essential information that we need about the probe, 
namely the time of measurement $\tau$, assumed to be large ($\tau\gg\sigma_+$), 
and  the associated, longitudinal and transverse, resolutions: $\delta x_-\sim 1/{\Delta_+}$ 
and $\delta x_\perp\sim {1}/{\Delta_\perp}$.

This change of representation is useful for several purposes.
First, it will facilitate the comparison
with the zeroth order calculation at weak coupling, to be presented in the
next section. Second, it will substantiate the argument developed
in the previous subsection, concerning the correlation between the resolution of the 
probe and the time of interaction (cf.  \eqn{acutetau}) .  Third,  it will allow us to
explicitly check that the narrow signal seen in coordinate space corresponds 
to a light--like mode of the probe operator. 
For the latter purposes, it is preferable to perform the Fourier
transform {\em before} computing the integral over $\tp$ in \eqn{EApart}.

Consider for illustration the first term, ${\mcal{E}}_{A}$, in \eqn{EApart}. By using simplifications 
appropriate at high energy, cf. \eqn{gammaL}, and changing the integration variable
from $\tp$ to $z=\tp/\gamma$, we obtain
  \begin{align}\label{tildeEdef}
 \tilde{\mcal{E}}_{A}(\tau,\Delta_+,{\bm \Delta}_\perp)
 &\,\equiv\,\int \rmd x_-\rmd^2 \bmx_\perp
\rme^{i\Delta_+ x_--i {\bm \Delta}_\perp\cdot\bmx_\perp}\,
\mcal{P}_A(\tau,x_-,\bmx_\perp)\\
 \,=\,\frac{2q_+ \tau}{\pi}&
 \int\rmd x_-\rmd^2 \bmx_\perp
  \rmd z\,z\,
  \delta''\bigg(2x_-\big(\tau-\sqrt{2}\gamma z\big)-x_\perp^2 -
  \frac{z\tau}{\sqrt{2}\gamma}\bigg)\,
  \rme^{i\Delta_+ x_--i {\bm \Delta}_\perp\cdot\bmx_\perp}\,
 \,.\nonumber
 \end{align}
The double prime on the $\delta$--function within the integrand denotes
two derivatives w.r.t. its argument. It is convenient to rewrite one of
them as a derivative w.r.t. $z$ and perform an integration by parts to
deduce %(recall that the integral over $z$ runs from 0 to $\infty$)
 \beq\label{tildeE0}
 \tilde{\mcal{E}}_A
  \,=\,{4q_+ \tau}
 \int \frac{\rmd x_-\rmd x_\perp
  \rmd z}{2\sqrt{2}\gamma x_- + \frac{\tau}{\sqrt{2}\gamma}}
  \,
  \delta'\bigg(2x_-\big(\tau-\sqrt{2}\gamma z\big)-x_\perp^2 -
  \frac{z\tau}{\sqrt{2}\gamma}\bigg)x_\perp
  \rmJ_0(\Delta_\perp x_\perp)\,
  \rme^{i\Delta_+ x_-}\,
 \,.
 \eeq
The Bessel function $\rmJ_0(\Delta_\perp x_\perp)$ has been generated by
the angular integration over the azimuthal angle of $\bmx_\perp$. We
shall now express the remaining derivative of the $\delta$--function as a
derivative w.r.t. $x_\perp$ and again perform an integration by parts, to
obtain (recall that $\rmJ_1(x) =- \rmd\rmJ_0/\rmd x$)
 \beq\label{tildeE1}
 \tilde{\mcal{E}}_A
  \,=\,{2q_+ \tau}
 \int \frac{\rmd x_- \rme^{i\Delta_+ x_-}}
 {2\sqrt{2}\gamma x_- + \frac{\tau}{\sqrt{2}\gamma}}
 \left\{\frac{1}{2\sqrt{2}\gamma x_- + \frac{\tau}{\sqrt{2}\gamma}}
 -\,\frac{\Delta_\perp}{2}\int_0^{z_{\rm max}}\rmd z\,
 \frac{\rmJ_1\big(\Delta_\perp X_\perp(z)\big)}{X_\perp(z)}
 \right\}
  \,.\eeq
In writing the above, we have also used the $\delta$--function to perform
the integral over $z$ in the first term within the accolades (the
boundary term) and respectively the integral over $x_\perp$ in the second
term, and we have denoted
 \beq\label{Xz} X_\perp(z)\,\equiv\,\sqrt{2x_-\tau- z\Big(
 2\sqrt{2}\gamma x_- + \frac{\tau}{\sqrt{2}\gamma}\Big)}\,.\eeq
The upper limit $z_{\rm max}$ in the integral over $z$ is determined by
the condition $X_\perp(z_{\rm max})=0$, which yields
 \beq\label{zmax}
 z_{\rm max}\,=\,\frac{1}{\sqrt{2}\gamma}\,
 \frac{\tau}{1+\frac{\tau}{4\gamma^2x_-}}\,.\eeq
Recalling that $z=\tp/\gamma\simeq\acute\tau/\sqrt{2}\gamma$ and
using $x_-\lesssim 1/\Delta_+$, this upper limit is clearly consistent with our previous
estimate for the (maximal) emission time $\acute\tau$ in \eqn{acutetau}.
 
We now change variables in
the integral over $z$ according to $z\to \xi\equiv \Delta_\perp
X_\perp(z)$, which gives
  \begin{align}\label{tildeE2}
 \tilde{\mcal{E}}_A
  \,=\,&{2q_+ \tau}
 \int \frac{\rmd x_- \rme^{i\Delta_+ x_-}}
 {\Big[2\sqrt{2}\gamma x_- + \frac{\tau}{\sqrt{2}\gamma}\Big]^2}
 \left\{1-\int_0^{\Delta_\perp\sqrt{2x_-\tau}}\rmd \xi\,
 \rmJ_1(\xi)\right\}\nn
 \,=\,&{4q_+ } \,\frac{\gamma^2}{\tau}\,
 \int \frac{\rmd x_- \rme^{i\Delta_+ x_-}}
 {\Big[1 + \frac{4\gamma^2 x_-}{\tau}\Big]^2}
 \,\rmJ_0\big(\Delta_\perp\sqrt{2x_-\tau}\,\big)\,.
 \end{align}

The Fourier transform of the second term ${\mcal{E}}_{B}$ in \eqn{EApart} 
can be similarly computed (in particular, this introduces the same upper limit 
$z_{\rm max}$ on $z$ as shown in \eqn{zmax})
and the final result reads
 \begin{align}\label{tildeP}
 \tilde{\mcal{P}}_q\,\simeq\, 2 \big(\tilde{\mcal{E}}_A+\tilde{\mcal{E}}_B\big)
  \,=\,{8q_+ } \,\frac{\gamma^2}{\tau}\,
 \int \frac{\rmd x_- \rme^{i\Delta_+ x_-}}
 {\Big[1 + \frac{4\gamma^2 x_-}{\tau}\Big]^3}
 \,\rmJ_0\big(\Delta_\perp\sqrt{2x_-\tau}\,\big)\,,
 \end{align}
where we have used the relation ${\mcal{P}}_q\simeq\ 2 {\mcal{E}}_q$ valid 
at high energy.

In order to evaluate the remaining integral over $x_-$, we  shall
perform approximations appropriate to the two interesting limiting regimes:
$\Delta_+\gg \gamma^2/\tau$ and respectively $\Delta_+\ll \gamma^2/\tau$.

 \texttt{(i)} {\em High longitudinal resolution:~} $\Delta_+\gg \gamma^2/\tau$.
In this case, the typical values of $x_-$ contributing to the integral in
\eqn{tildeP} obey $x_-\lesssim 1/{\Delta_+}\ll \tau/\gamma^2$, so one can
neglect the second term in the denominator of the integrand. This
yields
 \begin{align}\label{tildeELL}
 \tilde{\mcal{P}}_q
  \,\simeq\ &{8q_+ } \,\frac{\gamma^2}{\tau}\,
 \int_0^\infty {\rmd x_- \rme^{i\Delta_+ x_-}}
 \ \rmJ_0\big(\Delta_\perp\sqrt{2x_-\tau}\,\big)\,=\,
 i \,{8q_+ } \,\frac{\gamma^2}{\tau \Delta_+}\,
 \rme^{-i\frac{\Delta_\perp^2}{2\Delta_+}\tau}\,.
 \end{align}
The complex exponential can be rewritten as  $\rme^{-i\Delta_-\tau}$ with
$\Delta_-={\Delta_\perp^2}/{2\Delta_+}$. This relation
${2\Delta_+\Delta_-}={\Delta_\perp^2}$ is recognized as 
the mass--shell condition for a light--like
mode. (In fact, if one performs the remaining Fourier transform $\tau\to\Delta_-$
in \eqn{tildeELL}, one finds a result proportional to $\delta(2\Delta_+\Delta_--
\Delta_\perp^2)$.) This light--like mode with high longitudinal resolution is emitted at
the early time $\acute\tau\sim \gamma^2/\Delta_+\ll \tau$ and then propagates 
at the speed of light up to the measurement time $\tau$. The Fourier transform
of  \eqn{tildeELL}  back to coordinate space is dominated by the highest possible
values for $\Delta_+$, namely $\Delta_+^{\rm max}\simeq \gamma/\sigma$, which 
explains why the  support of the signal in coordinate space lies on the 
light--cone\footnote{For a generic upper limit $\Delta_+^{\rm max}$, the signal, 
while propagating
at the speed of light, would be shifted from the light--cone by a distance $\delta x_-\sim
1/\Delta_+^{\rm max}$.}
($x_-\simeq x_\perp^2/2x_+$), with an uncertainty $\delta x_-\sim\sigma_-$ 
introduced by the width of the original wave--packet.

\texttt{(ii)} {\em Low longitudinal resolution:~} $\Delta_+\ll \gamma^2/\tau$. In
this case, the typical values of $x_-$ contributing to the integral in \eqn{tildeP}
are determined either by the Bessel function, which implies
$x_-\lesssim 1/(\tau\Delta_\perp^2)$, or by the denominator of the
integrand, which requires $x_-\lesssim \tau/\gamma^2$. The last constraint
implies that $\Delta_+ x_-\ll 1$ irrespective of the value of
$\Delta_\perp$, so we can replace $\rme^{i\Delta_+ x_-}\simeq 1$. The
ensuing integral over $x_-$ can be exactly computed by changing
variables according to $x_-\equiv (\tau/4\gamma^2) u^2$~:
 \begin{align}\label{tildeESL}
 \tilde{\mcal{P}}_q
  \,\simeq\,& 4q_+
 \int_0^\infty \frac{\rmd u \,u}{(1+u^2)^3}
  \,\rmJ_0\bigg(\frac{\Delta_\perp \tau}{\sqrt{2}\gamma}\,u\bigg)
  \,=\,\frac{q_+}{2}\bigg(\frac{\Delta_\perp \tau}{\sqrt{2}\gamma}\bigg)^2
  \rmK_2\bigg(\frac{\Delta_\perp \tau}{\sqrt{2}\gamma}\bigg)
   \,\,,
 \end{align}
with $\rmK_2$ the modified Bessel function of rank 2.  Using $(x^2/2)\mathrm{K}_2(x)
\simeq 1$ for $x\ll 1$, we deduce that $\tilde{\mcal{P}}_q\simeq q_+$ when
$\Delta_\perp \ll \gamma/\tau$.  This is as expected: by causality, the decaying sytem
has a transverse size $\Delta x_\perp\sim\tau/\gamma$ and a longitudinal
size $\Delta x_-\lesssim\tau/\gamma^2$ , so when this is probed with much poorer,
transverse and longitudinal, resolutions, one sees the total energy $q_+$. In the opposite
limit $\Delta_\perp \gg \gamma/\tau$, the signal is exponentially suppressed (we recall
that $\mathrm{K}_2(x)\simeq\sqrt{{\pi}/{2x}}\,\rme^{-x}$ for $x\gg 1$), meaning that the
three--point function does not exhibit any substructure with transverse size much 
smaller than the overall size $\Delta x_\perp\sim\tau/\gamma$. This is again
as expected: when integrated over $x_-$, the three--point function looks 
uniform in the transverse plane (at least, at points $x_\perp\ll\tau/\gamma$) simply
by symmetry, that is, as a consequence of the spherical symmetry of the signal 
in the target rest frame. This can be also verified directly in coordinate space: by
integrating \eqn{EIMF} over $x_-$ or, equivalently, by performing the transverse
Fourier transform in \eqn{tildeESL}, one finds
 \beq\label{EPERP}
  \mcal{P}_q(\tau,x_\perp)\,\equiv \int \rmd x_-\,
  \mcal{P}_q(x) \,\simeq\,
 \frac{q_+}{2\pi \gamma^4}\,\frac{\tau^4}{
 \Big(x_\perp^2 +\frac{\tau^2}{2\gamma^2}\Big)^3}
  \,.
 \eeq

Notice that the low resolution modes are typically {\em space--like}~: one has indeed
$\Delta_\perp \sim \gamma/\tau$ and $\Delta_-\sim 1/\tau$, hence $\Delta_\perp^2
\gg 2\Delta_+\Delta_-$.
Consider also the typical values of $z$ and $\tp$ contributing to the
signal in \eqn{tildeESL}. By using \eqn{zmax} together with $x_-\sim
\tau/\gamma^2$, one finds $z_{\rm max}\sim \tau/\gamma$, which implies that
$\tp=\gamma z$ is commensurable with $\tau$. Thus, as already argued
in the previous subsection, a three--point function with small $\Delta_+$
interacts with the target at times which are close to the time of measurement.
Yet, because of the low longitudinal resolution, this does not bring us any additional
information about the state of the system at $t$. The only physically relevant
information that we can extract from the three--point function is the energy density
per unit transverse area, \eqn{EPERP}, and this is independent of the actual 
interaction time (as it involves an integration over all longitudinal coordinates).

\section{The three--point function at zero coupling}
\label{sec:zero}

In this section, we shall calculate the three--point function
\eqref{Edensity} in $\mcal{N}=4$ SYM in the other extreme limit: that of
a zero coupling. Our main purpose is to verify that the final result is
exactly the same as at infinitely strong coupling, as expected from the
following facts: \texttt{(i)} in a conformal theory like $\mcal{N}=4$ SYM
the general structure of a three--point function is fixed by conformal
symmetry together with the (quantum) dimensions of the involved
operators, and \texttt{(ii)} the ${\mathcal R}$--current and the
energy--momentum tensor are conserved quantities which are not
renormalized, that is, they have no anomalous dimensions. Accordingly,
the matrix element given in \eqn{Edensity} must be independent of the
coupling, and this is what shall explicitly check in what follows.

%Another virtue of the calculations at weak coupling is that they reveal
%the limitations of the three--point function in reflecting the partonic
%structure of the decaying system, in a context where this structure is
%manifest and can be studied via various other calculations (like that of
%the four--point function).

The result of the zeroth order calculation can be easily anticipated. In
this limit the time--like ${\mathcal R}$--current decays into a
fermion--antifermion (or scalar--antiscalar) pair, which then propagates
without further evolution. In the center of mass frame of the decay, two
back--to--back particles moving at the speed of light emerge. The
three--point function is not sensitive to correlations between the directions
of the two decay products, so the answer, in coordinate space, should
look the same as that of a thin spherical shell of energy whose radius
increases with the velocity of light. In the boosted frame in which we
shall actually do the calculation, the energy distribution should be
contracted to the part of the spherical shell having solid angle of size
$1/\gamma^2$ around the longitudinal axis (the $x_3$ axis along which the
decaying current is moving). As we shall see, this simple picture is
indeed faithfully reflected by the zeroth order result for the three--point
function. But as we shall later argue, this ability of the three--point
function to properly reflect the partonic structure of the decaying
system is in fact limited to zeroth order: it does not hold anymore after
including perturbative corrections at weak but non--zero coupling.

%As before, the time--like ${\mathcal R}$--current is created by the
%operator \eqref{Jqdef} which involves a convolution with the wave--packet
%shown in \eqn{ATLWP}. In practice, the only role of this WP is to fix the
%origin of the decay in space and time. Indeed,

As before, we shall assume that the momentum components, $\Delta^\mu$, of
the energy--momentum tensor ${\hat T}_{++}$, are much less than the
momentum of the ${\mathcal R}$--current initiating the decay. Thus,
although we are evaluating a transition matrix element, the insertion of
${\hat T}_{++}$ affects the decay products in such a tiny way that the
matrix element corresponds to a faithful determination of the average
energy flow in the decay. This is of course limited to the present,
zeroth order, calculation, in which the two partons produced by the
original decay do not have the possibility to evolve anymore.

The fact that the three--point function in a conformal field theory is
independent of the coupling means that, in perturbation theory at least,
this quantity cannot correctly describe the flow of energy at non--zero
coupling, where branchings of the decay products occur. The quantum
evolution of the partons is on the other hand manifest in the
perturbative evaluation of the four--point function, to be presented in the
next two sections. As we shall see there, this evolution leads, at both weak
and strong coupling, to the longitudinal broadening of the energy flow in
the decay.

\subsection{The decay rate}
\label{sec:decay}

Our focus in the subsequent calculations at weak coupling will be on the
description of the average properties of the matter distribution produced
by the decay of a time--like ${\mathcal R}$--current in the $\mcal{N}=4$
SYM theory. To that end it will be useful to have an evaluation of the
decay rate $\Gamma$ of the ${\mathcal R}$--current, an operation which
will also allow to introduce our notations. Indeed, in this perturbative
context, the three--point and four--point functions to be later computed need to
be divided by $\Gamma$ to ensure that they describe properties of a {\em
single} decay.

\begin{figure}
\begin{center}
\includegraphics[scale=0.65]{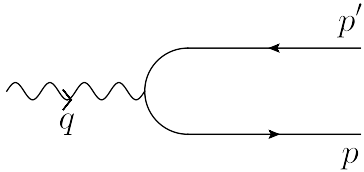}
\end{center}
\caption{Decay of the ${\mathcal R}$--current into a 
fermion--antifermion pair.}
\label{fig:decay}
\end{figure}

To lowest in perturbation theory, meaning to zeroth order in the gauge
coupling $g$ of $\mcal{N}=4$ SYM and to order $e^2$ in the
`electromagnetic coupling' associated with the ${\mathcal R}$--charge,
the ${\mathcal R}$--current can decay into either a fermion--antifermion
pair, or into a pair of scalars. To keep the presentation as simple as
possible, we shall only explicitly evaluate the decay into fermions and
then simply indicate the changes which occur when adding the scalars. As
before, we shall work with an ${\mathcal R}$--current boosted along the
positive $x_3$ axis, with $q_+={Q^2}/{2q_-}\simeq \sqrt{2}\gamma Q$
and we shall evaluate the rate of decay $\Gamma_+$ per unit of
light--cone time $x_+$. To the order of interest and for the decay into a
pair of fermions, this reads
 \beq\label{Gamma0}
 \Gamma_+\,=\int\frac{\rmd^3 p}{(2\pi)^32p_+}\,\frac{\rmd^3 p'}
 {(2\pi)^32p_+^\prime}\,\frac{1}{2q_+}\,\frac{1}{2}
 \sum_{\lambda,\sigma,\sigma'}\Big|e\bar u_\sigma(p)\,
  \gamma\cdot\varepsilon^{(\lambda)}\,v_{\sigma'}(p')\Big|^2
  (2\pi)^4\delta^{(4)}(q-p-p')\,,\eeq
as illustrated in Fig.~\ref{fig:decay}. The indices
$\lambda,\,\sigma,\,\sigma'$ refer to the helicities of the decaying
${\mathcal R}$--current, the fermion, and the antifermion, respectively.
\eqn{Gamma0} includes a sum over final helicities of the fermions and an
average (the factor $1/2$ in front of the sum symbol) over the initial
helicities of the current. (The decay rate being the same for any
helicity state, we consider here only the two transverse helicities:
$\lambda=\pm 1$.) The phase--space reads $\rmd^3 p=\rmd^2\bmp_\perp \rmd
p_+$. To evaluate \eqn{Gamma0} it is convenient to use
 \beq\label{ident}
  \rmd^3 p \,\rmd^3 p'\,\delta^{(4)}(q-p-p')\,=\,q_+^2\,z(1-z)\rmd
 z\rmd \phi\,,\eeq
where $p_+=zq_+$, $p'_+=(1-z)q_+$, $\phi$ is the azimuthal angle of the
fermion, $\bmp_\perp = -\bmp^\prime_\perp$, and 
 \beq\label{perp}
 p_\perp^2\,=\,z(1-z)Q^2\,.\eeq
One furthermore has
 \beq\label{helicit}
 e\bar u_\sigma(p)\,
  \gamma\cdot\varepsilon^{(\lambda)}\,v_{\sigma'}(p')\,=\,
  e\frac{\bmeps_\perp\cdot\bmp_\perp}{\sqrt{z(1-z)}}\,
  \delta_{\sigma\sigma'}\big[\sigma(1-2z)-\lambda\big]\,.\eeq
Using the equations above, one finds
 \beq\frac{1}{2}
 \sum_{\lambda,\sigma,\sigma'}\Big|e\bar u_\sigma(p)\,
  \gamma\cdot\varepsilon^{(\lambda)}\,v_{\sigma'}(p')\Big|^2\,=\,
  2Q^2\big[z^2+(1-z)^2\big]\,,\eeq
and therefore
 \beq\label{Gamma}
 \Gamma_+\,=\,\frac{e^2}{8\pi^2}\frac{Q^2}{2q_+}\int\rmd\phi\int_0^1\rmd
 z \big[z^2+(1-z)^2\big]\,=\,\frac{e^2}{6\pi}\,q_-\,.\eeq
The decay rate is usually written with respect to the ordinary time
variable $t$ in the {\em rest frame} of the decaying system. Using
$\Gamma_+ x_+ = \Gamma t$ and $q_- x_+ \simeq Qt/2$, one finally obtains
 \beq\label{GammaRF}
 \Gamma\,=\,\frac{e^2}{12\pi}\,Q\,,
 \eeq
which is indeed the expected result for the decay of a vector meson with
mass $Q$ and purely vector coupling of strength $e$ into a pair of
massless fermions.

In $\mcal{N}=4$ SYM, we also need to include the respective scalar
contribution. This is done by replacing $z^2+(1-z)^2\to 1$ in the
integrand of \eqn{Gamma}, so we are finally led to
 \beq\label{GammaSUSY}
 \Gamma_+^{\rm SUSY}\,=\,\frac{e^2}{4\pi}\,q_-
 \,=\,\frac{e^2 Q^2}{8\pi q_+}\,.\eeq
This is the factor which will be used to divide the 3 and four--point
functions to get properties of the final state normalized to a single
decay.

\subsection{The three--point function}
\label{sec:3pzero}

We now turn to evaluating the expectation value for the large component
of the energy--momentum tensor, $\hat T_{++}$, at late times in the decay
of the time--like ${\mathcal R}$--current, to zeroth order in the
coupling. As in the corresponding calculation at strong coupling, in
Sect.~\ref{sec:AdS}, we shall assume that the decay is initiated around
the space--time point $x^\mu=0$. In Sect.~\ref{sec:AdS}, this has been
enforced by using the wave--packet \eqref{ATLWP}. However, as we have
seen there, the widths of the WP did not play any role in the
calculations and in particular they dropped out from the final results,
like \eqref{Eparticle},
because the resolution of the probe was comparatively low
($\Delta_+\lesssim 1/\sigma_-\ll q_+$, etc). In that
respect, the situation will be the same at weak coupling. So, to simplify
the discussion, we shall omit the explicit use of a wave--packet for the
incoming ${\mathcal R}$--current, but rather use its (would--be central)
4--momentum $q^\mu=(q^0,0,0,q^3)$ in order to characterize its
localization in space and time. 

A similar discussion applies to the probe:  strictly speaking, one should use
the probe wave--packet introduced in \eqn{PROBEWP}. But as explained
there, the relevant information about the resolution
and the localization of the probe can be economically taken into account
by working in the mixed Fourier representation
$\hat T_{++}(\tau, \Delta_+, \bm{\Delta}_\perp)$.
This is precisely the Fourier component of the `backreaction'
at strong coupling that we have computed in Sect.~\ref{sec:Delta}, 
which will facilitate the comparison between the respective results.

To summarize, in this section we shall compute 
(with ${\bm \Delta}=(\Delta_+,{\bm \Delta}_\perp)$)  \beq\label{Tq3p}
 T_q(\tau, {\bm \Delta})\,\equiv\,
 \frac{e^2}{2q_+}\,\frac{1}{2}\sum_\lambda\int \rmd^4x
 \,\rme^{-iq\cdot x}\,\left\langle \hat J_\mu(x)\,
 \hat T_{++}(\tau, {\bm \Delta})\,\hat J_\nu(0)\right
 \rangle \,\varepsilon_\mu^{(\lambda)\,*}\,
  \varepsilon_\nu^{(\lambda)}\,,
  \eeq
in $\mathcal{N}=4$ SYM at zeroth order in the gauge coupling. The final result of this
calculation, after being divided by the decay rate $\Gamma_+^{\rm SUSY}$,
\eqn{GammaSUSY}, will be shown to be identical with the results
previously obtained at infinitely strong coupling for the quantity
$\tilde{\mcal{P}}_q(\tau, {\bm \Delta})$. 

\begin{figure}
\begin{center}
\includegraphics[scale=0.65]{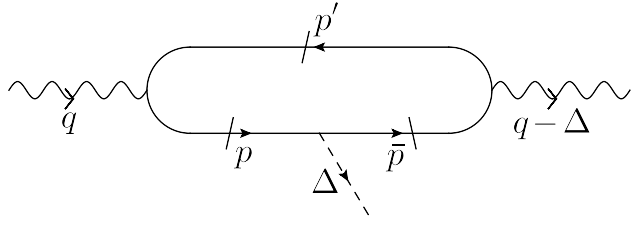}
\end{center}
\caption{The energy--momentum tensor interacting with the fermion line.}
\label{fig:3p0}
\end{figure}

The evaluation of \eqn{Tq3p} proceeds much as for the decay rate
discussed in Sect.~\ref{sec:decay}. The graph in Fig.~\ref{fig:3p0} shows
the energy--momentum tensor interacting with the fermion line, and there
is a corresponding graph where the momentum $\Delta$ comes off the
antifermion line. And there are of course also one--loop graphs involving
scalar fields to be added at the very end. The lines $p$, $p'$ and $\bar
p$ are on--shell, as required by the operator product in \eqn{Tq3p}; this
means e.g. $p_-=p_\perp^2/2p_+$. This also implies that the 4--momentum
$\Delta^\mu$ exchanged with the probe is space--like ($\Delta^2>0$), hence
the sense of the arrow of time on the corresponding leg is purely conventional.
(For definiteness, in Fig.~\ref{fig:3p0} we have chosen this line to be outgoing.)

For the diagram in Fig.~\ref{fig:3p0} (emission from the fermion line),
the only differences with respect to the
calculation given in Sect.~\ref{sec:decay} are a factor of $p_+=zq_+$ 
(the coupling between the fermion and the operator $\hat T_{++}$ is proportional
to the longitudinal momentum $p_+$ of the former) 
and the replacement of the phase space according to
 \beq\label{phase}
  \rmd^3 p \,\rmd^3 p'\,(2\pi)^4\delta^{(4)}(q-p-p')\,\longrightarrow\,
\rmd^3 p \,\rmd^3 p'\,\rmd^3\bar p\,(2\pi)^4\delta^{(4)}(q-p-p')
\,\delta^{(3)}(\bmp-\bar\bmp-\bm{\Delta})\,\rme^{-i(p_--\bar p_-)
 \tau}\,,\eeq
where within the 3--dimensional $\delta$--function, we have denoted
$\bmp=(p_+,\bmp_\perp)$ and similarly for $\bar\bmp$ and $\bm{\Delta}$.
After performing the trivial phase--space integrations using the
$\delta$--functions in \eqn{phase}, adding the $\Delta$--emission from the
anti--fermion line (this introduces an overall factor of 2) 
and including the corresponding scalar
contributions (as before, this amounts to replacing $z^2+(1-z)^2\to 1$
within the integrand), we are left with
 \beq\label{Tqzphi}
 T_q(\tau,{\bm \Delta})\,=\,\frac{e^2Q^2}{8\pi^2} \int \rmd z \rmd\phi\,z
 \,\rme^{-i(p_--\bar p_-)
 \tau}\,,\eeq
where $\phi$ is the azimuthal angle between the transverse vectors
$\bm{\Delta}_\perp$ and $\bmp_\perp$, and
 \beq\label{pminus}
 p_--\bar p_-\,=\,\frac{\bmp_\perp^2}{2p_+}\,-\,
 \frac{(\bmp_\perp-\bm{\Delta}_\perp)^2}{2(p_+-\Delta_+)}\,,\eeq
with $p_+=zq_+$ and $p_\perp$ as given in \eqref{perp}. 
Note that $\Delta_-\equiv p_--\bar p_-$
is the transfer of light--cone energy from the target to the probe.

So far, we have performed no
approximations. At this point we recall that $\Delta_+\ll q_+$, so unless
$z$ is extremely small (which, as we shall see, is generally not the case) we also
have $\Delta_+\ll p_+$. Then we can simplify \eqn{pminus} as
  \beq\label{pminus1}
 p_--\bar p_-\,\simeq\,\frac{\bmp_\perp\cdot\bm{\Delta}_\perp}{p_+}
 \,-\,\frac{\Delta_\perp^2}{2p_+}\,-\,
 \frac{p_\perp^2}{2p_+^2}\Delta_+\,,\eeq
or, after using $\bmp_\perp\cdot\bm{\Delta}_\perp=p_\perp
{\Delta}_\perp\cos\phi$, $q_+\simeq\sqrt{2} \gamma Q$, and the expression
\eqref{perp} for $p_\perp$,
 \beq\label{pminus2}
 p_--\bar
 p_-\,\simeq\,\sqrt{\frac{1-z}{2z}}\,\frac{\Delta_\perp\cos\phi}{\gamma}
 \,-\,\frac{\Delta_\perp^2}{2zq_+}\,-\,\frac{1-z}{4z}\,
 \frac{\Delta_+}{\gamma^2}\,.
 \eeq
Inserting this into \eqn{Tqzphi}, one can perform the integral over $\phi$
and thus find
 \beq\label{Tq3p1}
 T_q(\tau,{\bm \Delta})\,\simeq\,\frac{e^2Q^2}{4\pi} \int_0^1\rmd z\,z
\,\rmJ_0\bigg(\sqrt{\frac{1-z}{2z}}\,\frac{\Delta_\perp\tau}{\gamma}
\bigg)\,\exp\bigg\{i\frac{\Delta_\perp^2}{2zq_+}\tau +i
 \frac{1-z}{4z}\,
 \frac{\Delta_+}{\gamma^2}\tau\bigg\}\,.\eeq

From now on, we shall distinguish, for convenience, between the two
kinematical regimes already introduced in the discussion of the backreaction:
$\Delta_+\ll \gamma^2/\tau$ (low longitudinal resolution) and  respectively $\Delta_+\gg
\gamma^2/\tau$ (high longitudinal resolution).

\subsubsection{Low longitudinal resolution: $\Delta_+\ll \gamma^2/\tau$}
\label{sec:spacelike}

When $\Delta_+\ll \gamma^2/\tau$ and $z$ is not extremely small, both
terms in the exponential are much smaller than one and hence can be
neglected. This is true by assumption for the second term, and it is also
true for the first term since, as we shall shortly see, at large times
one has $\Delta_\perp\lesssim\gamma/\tau$. (Recall that we consider large
times $\tau\gg\gamma\sigma\gtrsim\gamma/Q$.) Then
  \beq\label{Tq3p2}
 T_q(\tau,{\bm \Delta})\,\simeq\,\frac{e^2Q^2}{4\pi} \int\rmd z\,z
\,\rmJ_0\bigg(\sqrt{\frac{1-z}{2z}}\,\frac{\Delta_\perp\tau}{\gamma}
 \bigg)\,,\eeq
which can be exactly integrated (the change of variables $z=1/(1+u^2)$ is
useful in that respect), to finally yield
 \beq\label{Tq3pA}
 T_q(\tau,{\bm \Delta})\,\simeq\,\frac{e^2Q^2}
 {16\pi}\,\left(\frac{\Delta_\perp\tau}{\sqrt{2}\gamma}\right)^2\,
 \rmK_2\bigg(\frac{\Delta_\perp \tau}{\sqrt{2}\gamma}\bigg)\,.\eeq
As anticipated, the integral over $z$ in \eqn{Tq3p2} is not particularly
sensitive to very small values $z\to 0$ and the final result in
\eqn{Tq3pA} has support at $\Delta_\perp\lesssim\gamma/\tau$. Also, one
can check that the typical probe kinematics is {\em deeply space--like}~:
the light--cone energy of the probe $\Delta_-\equiv p_--\bar p_-$ is dominated
by the first term in the r.h.s. of \eqn{pminus2}, which yields $\Delta_-\sim
\Delta_\perp/\gamma$~; hence, for $\Delta_\perp\lesssim\gamma/\tau$ and
$\Delta_+\ll \gamma^2/\tau$, one has indeed $\Delta_\perp^2\gg
2\Delta_+\Delta_-$.

After normalizing by the decay rate \eqref{GammaSUSY}, we obtain the
respective quantity for a single decay:
 \beq\label{Tq3pGamma}
 \frac{T_q(\tau,{\bm \Delta})}{\Gamma_+^{\rm SUSY}}\,\simeq\,
  \frac{q_+}{2}\bigg(\frac{\Delta_\perp \tau}{\sqrt{2}\gamma}\bigg)^2
  \rmK_2\bigg(\frac{\Delta_\perp \tau}{\sqrt{2}\gamma}\bigg) 
 \,.\eeq
\eqn{Tq3pGamma} coincides, as expected, with the respective result of the
backreaction at infinitely strong coupling, presented in \eqn{tildeESL}.
As already discussed in that strong--coupling context, there is no
difficulty in interpreting this result as the average energy measured by
a probe with strongly space--like kinematics: such a probe has a poor
longitudinal resolution, hence it measures the energy integrated over the
radial profile of the decaying system, within a transverse region with
radius $\delta x_\perp\sim 1/\Delta_\perp$. This energy is correctly
given by \eqn{Tq3pGamma} or \eqref{tildeESL} for any value of the
coupling. What changes from weak to strong coupling is the radial
distribution of the energy. In particular, it is only in the zero
coupling limit that this radial distribution is correctly measured by the
three--point function \eqref{Tq3p}, as we shall explain in the next
subsection.

\subsubsection{High longitudinal resolution: $\Delta_+\gg \gamma^2/\tau$}

By choosing $\Delta_+\gg \gamma^2/\tau$, one ensures a fine longitudinal
resolution in the $x_-$ region populated by the decay. To analyze this
case, one can again rely on \eqn{Tq3p1}, which remains valid so long as
$\Delta_+\ll q_+$. Now, however, we cannot neglect the exponential
factors in \eqn{Tq3p1} anymore. Also, there is no way how the two
potentially large phases could compensate with each other, as they are
both positive definite. So, the only way to avoid strong oscillations is
that both phases separately remain of order one, or smaller. When applied to the
second phase, this condition implies that $1-z$ must be small. By using
$z\simeq 1$ together with the change of variables $u=1-z$, one can write
 \begin{align}\label{Tq3p3}
 T_q(\tau,{\bm \Delta}) &\,\simeq\,\frac{e^2Q^2}{4\pi} \int_0^\infty\rmd u
\,\rmJ_0\bigg(\sqrt{u}\,\frac{\Delta_\perp\tau}{\sqrt{2}\gamma}
\bigg)\,\exp\bigg\{i\frac{\Delta_\perp^2 \tau}{2q_+} +i {u}\,
 \frac{\Delta_+ \tau}{4\gamma^2}\bigg\}\nn
 &\,\simeq\,\frac{ie^2Q^2\gamma^2}{\pi\Delta_+\tau}
 \,\exp\bigg\{i\frac{\Delta_\perp^2 \tau}{2q_+} -i\frac{\Delta_\perp^2\tau}{2\Delta_+}
 \bigg\}
 \,,\end{align}
where the $u$ integration has been extended to $u\to\infty$ because only
the small $u$ region is important for the integral and we have
used the formula
 \beq
  \int_0^\infty\rmd u\,\rmJ_0 (a\sqrt{u})\,\rme^{iub}\,=\,\frac{i}{b}\,
  \rme^{-i\frac{a^2}{4b}}\,.\eeq
Using $\Delta_+\ll q_+$, it is clear that the dominant phase in the final
result in \eqn{Tq3p3}  is the {\em second} phase there. This phase  
constraints the values of the probe momenta such that  
$({\Delta_\perp^2}/{2\Delta_+})\tau\sim 1$ and when this happens the first phase
$i({\Delta_\perp^2\tau}/{2q_+})$ is much smaller than one and can be ignored.
For consistency with the previous manipulations, let us notice that  the integral in 
\eqn{Tq3p3}  is controlled by values of $u$ satisfying 
 \beq\label{u}
 u\, \sim\,  \frac{\gamma^2\Delta_\perp^2}{\Delta_+^2}\,\sim\,
  \frac{\gamma^2}{\tau \Delta_+}\,\ll\,1,\eeq
where the second estimate holds when $\Delta_+ /\tau\sim \Delta_\perp^2$.

After neglecting the small phase in the second line of \eqn{Tq3p3} and 
dividing the result by the decay rate
\eqref{GammaSUSY}, we finally obtain
 \beq\label{Tq3pGammaB}
 \frac{T_q(\tau,{\bm \Delta})}{\Gamma_+^{\rm SUSY}}\,\simeq\,
 \frac{8i\gamma^2 q_+}{\Delta_+\tau}\,
 \,\rme^{-i\frac{\Delta_\perp^2}{2\Delta_+}\,\tau}\,.\eeq
Once again, this coincides with the respective result at strong coupling,
\eqn{tildeELL}. As explained there, the typical value of the light--cone
energy (the quantity conjugate to the time of measurement $\tau$) is
$\Delta_-={\Delta_\perp^2}/{2\Delta_+}$, as expected for light--like kinematics. 
This is indeed consistent with our previous estimate 
$\Delta_-= p_--\bar p_-$ for this quantity, as it can be checked by 
using \eqn{pminus2} for $p_--\bar p_-$ together with $u=1-z$ from \eqn{u}.

\comment{To establish this interpretation, we return to \eqn{Tq3p}. The
phases inside the integrand there show that the light--cone energies,
$p_-$ and $\bar p_-$, of the parton prior and respectively after its
interaction with the energy--momentum tensor must obey $p_-\,,\bar
p_-\lesssim 1/\tau$. These conditions can be rewritten as (cf.
\eqn{pminus1})
 \beq
 \frac{\bmp_\perp^2}{2p_+}\,,\quad
 \frac{\bmp_\perp\cdot\bm{\Delta}_\perp}{p_+}\,,\quad
 \frac{\Delta_\perp^2}{2p_+}
 \,\lesssim\,\frac{1}{\tau}\,.\eeq
Since $\gamma^2/\tau < \Delta_+ < p_+$, the first condition above
requires the polar angle $\theta\sim p_\perp/p_+$ made by the vector
$\bmp$ with respect to the $x_3$ axis obey $\tau\theta^2\lesssim
1/\Delta_+$, or
 \beq
 \theta\,\lesssim\,\frac{1}{\sqrt{\tau\Delta_+}}
 \,\lesssim\,\frac{1}{\gamma}\,.\eeq
Similarly, the condition
${\bmp_\perp\cdot\bm{\Delta}_\perp}/{p_+}\lesssim 1/\tau$ requires the
angle $\theta_{p\Delta}$ between the vectors $\bmp$ and $\bm{\Delta}$
obey $\Delta_+ \theta_{p\Delta}^2\lesssim 1/\tau$, so that
$\theta_{p\Delta}\lesssim\theta$ and the angle that $\bm{\Delta}$ makes
with the $x_3$ axis cannot be larger than $\theta$. This in turn implies
\beq
 \frac{\Delta_\perp}{\Delta_+}\,\lesssim\,\frac{1}{\sqrt{\tau\Delta_+}}
 \quad \Longrightarrow\quad
 \frac{\Delta_\perp^2}{2\Delta_+}\,\tau\,\lesssim\,1\,,\eeq}

What is however specific to the zero--coupling limit at hand
is the fact that, in this limit, \eqn{Tq3pGammaB} is the Fourier
transform of a real measurement. This is possible since the probe
is now soft as compared to the parton (fermion or scalar)
that it interacts with: $\Delta_+\ll p_+=zq_+$ and $\Delta_\perp
\ll p_\perp$. (The second condition follows by using \eqn{perp} and \eqn{u}
to successively write $p_\perp\simeq\sqrt{u} Q \sim (q_+/\Delta_+)\Delta_\perp\gg \Delta_\perp$.)
So, for this particular problem, even a probe with relatively 
`high resolution' (which can discriminate
longitudinal and transverse structures much smaller than the maximal respective
sizes, $\Delta x_-\sim \tau/\gamma^2$ and $\Delta x_\perp\sim \tau/\gamma$, permitted
by causality) is still
soft enough to provide a coarse--grained measurement over a volume much larger than
the volume occupied by the struck parton. Since $z\simeq 1$, the struck parton
carries most of the original photon energy: $p_+=zq_+\simeq q_+$. Accordingly, we expect the result 
\eqref{Tq3pGammaB} to equal this energy $q_+$ times the probability for the trajectory of
the (small) parton to intersect the (comparatively) large area of the `detector'. And indeed,
the prefactor in \eqref{Tq3pGammaB} can be given such
a simple geometric interpretation, as we now argue.

Namely, the decay occurs over a solid angle
$1/\gamma^2$. The measurement covers a region with
transverse area $1/{\Delta}_\perp^2$ and hence a solid angle $\sim
1/({\Delta}_\perp\tau)^2$. Thus the measurement covers a fraction
$(\gamma/\Delta_\perp\tau)^2$ of the solid angle of the decay. When
$\Delta_\perp^2\sim \Delta_+/\tau$, this is the same as the prefactor in
\eqref{Tq3pGammaB} except for the factor $q_+$. Hence, \eqn{Tq3pGammaB}
is the fraction of the energy of the decaying system which
propagates within the solid angle covered by the `detector'.

\section{Jet evolution at weak coupling but late time}
\label{sect:jet}
With this section we begin the study of the four--point functions 
introduced in \eqn{2point}, first in the context of perturbation theory at weak coupling.
As anticipated in Sect.~\ref{sec:WP}, these correlations correspond
to measurements which can actually probe the space--time distribution of matter 
produced by the decaying system. This will be manifest in the subsequent discussion 
of the situation at weak coupling, where we shall see that the four--point functions reveal
the partons and their evolution. First, in Sect.~\ref{sec:DGLAP} we shall introduce
the partonic fragmentation function of a jet and discuss its evolution at weak coupling
but late times. Then in Sect.~\ref{sec:4P} we shall relate this fragmentation 
function to a specific four--point function (essentially, the function $\Pi_q$ in \eqn{2point})
that can be measured (at least in principle) via deep inelastic scattering.

\subsection{The general picture}
\label{sec:DGLAP}

We now turn to the case where the coupling constant of $\mcal{N}=4$ SYM
is small, but non--zero. As before, we are interested in the matter
distribution produced by the decay of a time--like $\mcal{R}$--current at
sufficiently large times --- much larger than the characteristic time
scale $\tau_0=2q_+/Q^2$ for the occurrence of the first decay. If one
waits for such long time, the two high--momentum partons produced by the
original decay must have evolved into a large number of softer partons.
In general if one wishes to keep track of the {\em number} of partons in
the evolution, both their small--$x$ and their DGLAP evolution are
important. However, if one only wishes to follow the time--dependence of
the {\em energy} distribution, the DGLAP evolution is sufficient. Indeed,
at weak coupling at least, the small--$x$ partons, although more
numerous, carry only a negligible fraction of the totale energy.

The DGLAP evolution characterizes the change in the parton distributions
(also known as `fragmentation functions' in the context of a time--like
evolution) with decreasing virtuality, from the original virtuality
$Q^2$ of the $\mcal{R}$--current down to the virtuality scale of interest 
$\mu^2$. This can be also viewed as an evolution in 
time, by using the relation between the lifetime of a parton generation 
and their virtuality given by the uncertainty principle (see \eqn{tform}
below). The fragmentation function $D(x, Q^2/\mu^2)$ represents
the number of partons of a given species (which for our present purposes can be 
left unspecified) per unit $x$ and with virtuality comprised between
$Q^2$ and  $\mu^2$.  As we shall explain in the next subsection, this quantity 
truly corresponds to a four--point function which can be measured via deep inelastic scattering.

In what follows, we shall assume that $\mu^2\ll Q^2$ and we 
shall limit ourselves to the `leading--logarithmic 
accuracy' (LLA), in which the DGLAP equation resums 
powers of $\lambda\ln(Q^2/\mu^2)$ to all orders. This equation 
is most conveniently solved via a Mellin transform $x\to j$. This introduces the 
`anomalous dimension' $\gamma(j)$ (the Mellin transform of the DGLAP splitting
kernel), which to the accuracy of interest reads \cite{Kotikov:2002ab}
 \beq\label{gamma}
 \gamma(j)\,=\,\frac{\lambda}{4\pi^2}\Big(\psi(1)-\psi(j-1)\Big)\,,
 \eeq
where $\psi(\gamma)\equiv \rmd \ln \Gamma(\gamma)/\rmd\gamma$ is the
di--gamma function. Then the fragmentation function $D(x, Q^2/\mu^2)$
is expressed as the inverse Mellin transform
 \beq\label{DGLAP}
 x^2 D(x, Q^2/\mu^2)\,=\int\frac{\rmd j}{2\pi i}\,\rme^{(j-2)\ln(1/x)
 +\gamma(j)\ln(Q^2/\mu^2)}\,,\eeq
where the $j$--integration goes parallel to the imaginary axis and to the
right of $j=1$. We used the initial condition
$D=\delta(x-1)$ when $Q^2=\mu^2$.
The factor $x^2$ has been introduced for convenience.
Indeed, we are mainly interested in finding what are the $x$--values of
the partons which carry most of the energy of the decay. To that aim, it
is useful to multiply $D(x, Q^2/\mu^2)$ by $x^2$, one factor of $x$ to
get an energy weighting and another one to count the number of partons.
(With the present conventions, the number of partons in an interval $\rmd x$ is $\rmd
N=D\rmd x$.)

We assume that $Q^2/\mu^2$ is very large, such that
$\lambda\ln(Q^2/\mu^2)\gg 1$, and then the integral over $j$ can be
evaluated in the saddle point approximation. The saddle point $j_s$ obeys
 \beq\label{spjs}
 \ln\frac{1}{x} + \gamma'(j_s)\ln\frac{Q^2}{\mu^2}\,=\,0,\eeq
and leads to
\beq\label{DSP}
 x^2 D(x, Q^2/\mu^2)\,\sim\,\left(\frac{Q^2}{\mu^2}\right)^{
 \gamma(j_s)-(j_s-2)\gamma'(j_s)}\,.\eeq
As already mentioned, we are interested in the values of $x$ which
maximize \eqref{DSP}. The maximum value of the function
$f(j)=\gamma(j)-(j-2)\gamma'(j)$ is $f(2)=0$, so the values of $x$ which
dominate the energy--momentum sum rule is given by \eqn{spjs} with
$j_s=2$, namely
 \beq\label{spxc}
 \ln\frac{1}{x_c} \,=\,- \gamma'(2)\ln\frac{Q^2}{\mu^2}\,=\,
 \frac{\lambda}{24}\,\ln\frac{Q^2}{\mu^2}\,
 .\eeq
We see that $x_c$ gets smaller as $\mu^2$ gets smaller, as expected in
view of our physical picture of parton branching. Now, since the
virtuality decreases along the branching process, it is convenient to
express $\mu^2$ in terms of the duration $\tau$ of the decay. To that
aim, we shall observe that, by the uncertainty principle, the partons with 
longitudinal momentum fraction $x$ and virtuality $\mu^2$ have a formation time
 \beq\label{tform}
 \tau_{\rm form}\,\simeq\,\frac{2xq_+}{\mu^2}\,.\eeq
To be able to produce the partons with a given $x$ and $\mu^2$, the evolution must
occur over a time $\tau=\tau_{\rm form}(x, \mu^2)$. (If $\tau\ll\tau_{\rm form}$, such partons
have no time to be formed, whereas if $\tau\gg\tau_{\rm form}$, then they have already
decayed by the time of measurement into partons with smaller values for $x$ and $\mu^2$.)
These considerations show that, for a given $x$, one can express the lower end $\mu^2$ 
of the virtuality evolution in terms of the evolution time $\tau$, by  equating 
$\tau_{\rm form}(x, \mu^2)$ with $\tau$. This yields the following relation
\beq\label{tauxmu}
 \frac{\tau}{\tau_0}\,=\,x\,\frac{Q^2}{\mu^2}\,\eeq
between $\tau$, $x$ and $\mu^2$. (We recall that $\tau_0=2q_+/Q^2$ is the formation time 
for the first decay of the $\mcal{R}$--current into a pair of partons.) 
Using this relation for $x=x_c$, one can finally rewrite \eqn{spxc}
as an equation for the evolution of $x_c$ with $\tau$~:
 \beq\label{xc}
 \frac{1}{x_c}
 \,\simeq\,\left(\frac{\tau}{\tau_0}\right)^{\lambda/24}\,.\eeq
This result explicitly shows which are the partons which carry most the energy of the 
decaying system at a time $\tau$ after the decay has begun.
As expected, $x_c$ decreases with time, albeit only slowly (as a small inverse power of $\tau$),
because the evolution is slow when the coupling is weak.

 \comment{
\subsection{The three--point function}

As mentioned earlier, the three--point function, as defined for example in
\eqn{Tq3p}, is not renormalized by radiative corrections. Thus the
evolution that we have just described in Sect.~\ref{sec:DGLAP} is not
visible to this three--point function, meaning that the latter cannot be a
good analyzer of the energy distribution and its evolution during the
decay. Suppose for instance we allow the decay to proceed over a long
time $\tau$ so that $x_c$, given by \eqn{xc}, is very small: $x_c\ll 1$.
If one tries to evaluate the three--point function in terms of quanta present
at $\tau$ (which is also the time of the $\hat T_{++}$ insertion in
\eqn{Tq3p}), we reach an apparent contradiction. If $\Delta_+\gg x_cq_+$,
the operator $\hat T_{++}(\tau, {\bm \Delta})$ has no elementary matrix
elements in terms of the quanta dominating the energy of the decay at
$\tau$. Yet, the result for the three--point function is non--vanishing (and
independent of the strength of the coupling), as we have seen. Hence,
there must be some high momentum parton states either in the amplitude,
or in the complex conjugate amplitude, of the decay --- but not in both
--- which allow the operator $\hat T_{++}(\tau, {\bm \Delta})$ to be
emitted or absorbed at $\tau$. These high--momentum parton modes are too
rare to show up in the analysis of Sect.~\ref{sec:DGLAP}, where it is the
{\em square} of the decay amplitude which is evaluated. But such modes
are apparently important for the transition matrix elements corresponding
to the three--point function. We have been unable to construct these modes
explicitly and hence to understand more completely this curious
phenomenon.
 }

\subsection{The four--point function and deep inelastic scattering}
\label{sec:4P}

We now turn to an analysis of the decaying state by performing a deeply
inelastic scattering, at time $\tau$, on that state. We shall use an
$\mcal{R}$--current not only to create the decaying system but also as
a probe to measure this decay via DIS. To better control the
space--time resolution and localization of the probe, we shall associate
a wave--packet to the respective $\mcal{R}$--current,
 \beq \label{jDelta}
 \hat{J}_\mu(\tau,\Delta)\,\equiv\,\int \rmd^4y\,\psi_\Delta(y;\tau)
 \,\hat{J}_\mu(x)\,,\eeq
with the probe wave--packet $\psi_\Delta(y;\tau)$ as introduced in
\eqn{PROBEWP}.  To ensure a good resolution, this wave--packet needs
to be strongly space--like (see below for the precise conditions). 
The `deep inelastic scattering' is the process where the decaying
time--like system absorbs the space--like probe current and thus evolves
into some arbitrary final state. The inclusive cross--section, also known as the
{\em DIS structure function}, is obtained by summing over all the possible final
states. Via the optical theorem, it can be related to the following
forward scattering amplitude, which is a Wightman function
 \beq\label{Pidef}
 \Pi_q(\tau,  \Delta)\,\equiv\,
 \frac{e^2}{2q_+}\,\frac{1}{2}\sum_\lambda\int \rmd^4x
 \,\rme^{-iq\cdot x}\,\left\langle \hat J_\mu(x)\,
 \hat{J}_+(\tau,\Delta)\,\hat J_+(\tau,-\Delta)\,
 \hat J_\nu(0)\right
 \rangle \,\varepsilon_\mu^{(\lambda)\,*}\,
  \varepsilon_\nu^{(\lambda)}\,.
 \eeq
As manifest from the above writing, the 4--momentum $\Delta^\mu$ transferred to the target 
by the first insertion of the probe current is then taken away by the second insertion, so
the target can be measured with high resolution without being perturbed, as anticipated 
in Sect.~\ref{sec:WP}. Strictly speaking, the above statement refers only to the {\em central} 
value $\Delta^\mu$ of the probe WP 4--momentum, but in this case one can chose the WP 
to be strongly peaked in momentum at this central value, with negligible spread.
 
For reasons to later become clear, it is now preferable to choose  $\Delta_+=0$ and use 
the other components of the probe momentum, $\Delta_\perp$ and $\Delta_-$, to control
the transverse, longitudinal and temporal resolutions of the experiment. 
These non--zero components can
be arbitrarily large and they have negligible spread, meaning that the corresponding
widths are relatively large:  $\tilde\sigma_\perp\Delta_\perp\gg 1$ and
$\tilde\sigma_+\Delta_-\gg 1$. More precisely, we shall chose these widths
large enough for the detector to cover the whole spatial region where the
decaying system can be localized at time $\tau$, in order not to miss any
parton; this requires $\tilde\sigma_\perp\gtrsim \tau/\gamma$ and $\tilde\sigma_+
\gtrsim \tau/\gamma^2$. Also, as before, we require $\tilde\sigma_+\ll \tau$
in order for the time of measurement to be well defined. As we shall see, there is indeed
no difficulty to satisfy all these conditions for the problem at hand. In particular, the probe
is (strongly) space--like, $\Delta^2=  \Delta_\perp^2 > 0$, as anticipated.
 
\begin{figure}
\begin{center}
\includegraphics[scale=0.65]{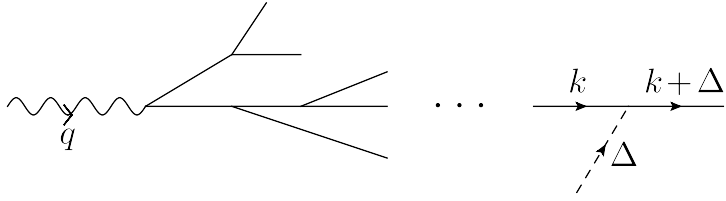}
\end{center}
\caption{The scattering of the current on a parton with momentum $k$.}
\label{fig:dis}
\end{figure}

The perturbative analysis of deep inelastic scattering at weak coupling is well developed
in the literature and will be not repeated here, especially since the corresponding result
is already known to the accuracy of interest: the structure function  \eqref{Pidef} 
is proportional to the partonic fragmentation function introduced in Sect.~\ref{sec:DGLAP}.
In what follows, we shall simply explain the relation between the kinematics of DIS and the
variables $x$ and $Q^2/\mu^2$ of the fragmentation function. To that aim, we consider
the absorption of the current with momentum $\Delta$ by a parton with momentum $k$,
as illustrated in Fig.~\ref{fig:dis}.  The parton can be assumed to be on--shell both before 
and after this interaction. This is implicit in our (Wightman) prescription for ordering the 
operators in the four--point function. It is also physically reasonable, since the virtuality 
$k^2\simeq\mu^2$ of the parton is much smaller than its longitudinal momentum squared
$k_+^2$ with $k_+=xq_+$. Hence, we can write $k^\mu\simeq(xq_+, 0, \bm{0}_\perp)$.
The on--shellness conditions $k^2=0$ and $(k+\Delta)^2=0$ then imply $\Delta^2_\perp=
2xq_+\Delta_-$, thus fixing the longitudinal momentum fraction $x$ of the struck parton. 
The corresponding virtuality $\mu^2$ can be estimated as  
in Sect.~\ref{sec:DGLAP}, by equating the time of measurement $\tau$ with the 
formation time \eqref{tform}. 

The above  considerations 
motivate the following expression for the four--point function at hand:
 \begin{align}\label{PiGamma}
 \frac{4\pi^2\Delta_-}{\Gamma_+^{\rm SUSY}}\,\Pi_q(\tau,{\Delta})\,=\,x
 D\big(x, {Q^2}/{\mu^2}, {\Delta_\perp^2}/{\mu^2}\big)\,,\end{align}
where  
 \beq
 x=\frac{{\Delta}_\perp^2}{2q_+\Delta_-}
 \quad\mbox{and}\quad \mu^2=\frac{2x q_+}{\tau}
 \,.\eeq
The fragmentation function in the r.h.s. of \eqn{PiGamma}
refers to partons with longitudinal fraction $x$, virtuality $\mu^2$ and 
transverse area $1/{\Delta}_\perp^2$. The normalization factor in the l.h.s.
can be understood as follows: the fragmentation function counts the number
of partons in one decay, whereas the four--point function \eqref{Pidef} provides
an integrated version of this quantity over the typical duration $\delta x_+ \sim 1/\Delta_-$ 
of a collision between the probe and the target. Hence, to obtain the
number of partons per decay, one needs to divide $\Pi_q(\tau,{\Delta})$ by the
typical number of decays occurring during an interval $\delta x_+$, namely $\delta x_+
\Gamma_+^{\rm SUSY}$.
 
%(in particular, much smaller than the maximal width 
%$\delta x_-\sim \gamma^2/\tau$ that would be permitted by causality and Lorentz
%contraction) 
 
By the uncertainty principle, a parton with longitudinal momentum $k_+=xq_+$ is delocalized
over a distance $\Delta x_-\sim 1/xq_+$. Since our probe can actually `see' such partons,
we deduce that it has a longitudinal resolution $\delta x_-\sim 2\Delta_-/{\Delta}_\perp^2$.
This can be made arbitrarily small 
by taking ${\Delta}_\perp$ to be sufficiently large. 
In particular, for the typical partons that carry most of the total energy
at time $\tau$, \eqn{xc} implies
 \beq
 \delta x_-\,\gtrsim\,\frac{1}{x_c q_+}\,\simeq\,\frac{1}{q_+}
 \left(\frac{\tau}{\tau_0}\right)^{\lambda/24}\,.\eeq
This grows with $\tau$, albeit only slowly: as intuitive from the picture of parton branching, 
there is a spreading of the radiation in the longitudinal direction with increasing time, but 
this spreading is slow, since so is the evolution.

In order for the scattering to give a well defined value for
$x$ as indicated above, it is important that the temporal uncertainty in
the wave--packet, $\tilde\sigma_+$, obey
 \beq\label{longit}
  \tau\,\gg\,\tilde\sigma_+\,\gg\,\frac{1}{\Delta_-}\,\simeq\,\frac{2xq_+}
 {\Delta_\perp^2}\,.\eeq
Once again, this is easy to achieve so long as $\Delta_\perp$ is not too small. (In particular,
one needs $\Delta_\perp\gg \mu$, as clear by comparing
Eqs.~\eqref{longit} and  \eqref{PiGamma}.) By the same 
token, the temporal resolution $\delta x_+ \sim {1}/{\Delta_-}$ of the scattering 
is extremely good: the  space--like current 
is absorbed over a very short time $\delta x_+ \ll \tilde\sigma_+$, hence it probes the state
of the decaying system at $\tau$.

Consider finally the dependence of the fragmentation function upon the parton
transverse size $\delta x_\perp \sim 1/\Delta_\perp$. This has not been mentioned in 
Sect~\ref{sec:DGLAP}, since it is almost trivial in the present context:
the weakly--coupled partons are {\em point--like} (they occupy a negligible area in 
the transverse space), hence the structure function is independent of the probe
transverse resolution $\Delta_\perp$ (for a
given value of the longitudinal momentum fraction $x$).
More precisely, there is a weak dependence, via powers of
$\lambda\ln({\Delta_\perp^2}/{\mu^2})$, which has been neglected in the discussion in 
Sect~\ref{sec:DGLAP} and which is generated by the quantum evolution between
the virtuality scales $\mu^2$ and ${\Delta_\perp^2}$ of the struck parton and 
of the probe, respectively. (This corresponds to radiative corrections to the interaction
vertex  in Fig.~\ref{fig:dis} and is controlled by the space--like anomalous dimension
of the probe.) But such a weak dependence, which reflects the perturbative evolution 
of the partons, 
does not alter the basic fact that partons are essentially point--like\footnote{If the
situation was different, that is, if the partons had some intrinsic,  time--dependent,
transverse size $r(\tau)$, then the fragmentation function would exhibit a strong,
geometric, dependence upon the variable $\Delta_\perp r(\tau)$ and in particular it
would rapidly die away when $\Delta_\perp r(\tau)\to\infty$ (since in that limit,
the probe would be unable to see the partons anymore).}.

To summarize, perturbation theory at weak coupling but large time predicts that the
decaying system evolves via successive parton branchings into partons which are
point--like in the transverse plane and whose longitudinal spreading (in the sense of their
deviation $\delta x_-$ from the light--cone $x_3=t$, where all the particles would be 
located in the limit of a zero coupling) is slowly increasing with time. 

\section{Witten diagrams at strong coupling}
\label{sec:Witten}

In the previous section we have seen that, at weak coupling, the four--point
function \eqref{Pidef} describes the quantum evolution of the decaying
system via parton branching, whereas the three--point function \eqref{Tq3p}
cannot do so. In what follows we would like to extend these findings to
strong coupling, by showing that the four--point function computed within
AdS/CFT from Witten diagrams shows indeed quantum broadening and no trace
of point--like substructures (`partons'). To that aim, we shall focus
directly on the space--like kinematics for the probe, since this is the
kinematics which has revealed partons at weak coupling. 
Besides, we know by now that it is only with this kinematics that we can 
access the state of the system at a time close to the measurement time.

The Witten diagrams involve interactions occurring inside 
AdS$_5$ between the bulk excitations induced by the `target' and `probe' operators 
from the boundary gauge theory. The interaction vertices are local in AdS$_5$ and 
they can be connected via bulk--to--bulk propagators.
For simplicity, we shall perform our calculations in a scalar model for SUGRA
which is a scalar theory for a massless field in AdS$_5$ with trilinear interactions. 
That is, the $\mcal{R}$--currents from
the previous discussion will be replaced (for both the decaying system
and the probe) by scalar, `dilaton', operators, denoted as $\hat{\mcal{O}}$. 
This model generates Witten diagrams with the same topology as
the relevant SUGRA diagrams \cite{Hatta:2010kt}, 
but of course there are fewer such diagrams and with simpler vertices. 
Although strictly speaking we modify the theory by doing that, we do not believe
that this could alter our conclusions. Indeed, here we are only interested
in very robust, qualitative, properties like the support of the four--point function
as a function of the probe kinematics. For a space--like probe at least,
such properties are captured (in the economy of the SUGRA calculation) 
by the various bulk--to--boundary and bulk--to--bulk propagators, 
but they are not sensitive to the detailed structure of the vertices.

To start with, in Sect.~\ref{sect:TL}
we shall present a careful construction, using wave packets,
of the bulk excitations corresponding to the target and the probe. 
This will allow us to check some approximations used in the previous manipulations,
in particular the fact that one can treat the bulk excitation associated with the decaying 
system as a `particle falling in AdS$_5$'. Then, in Sect.~\ref{sect:34pW}, we shall compute the three-- and four--point functions in the scalar model. The calculation of the three--point function is shown only for completeness, namely
to demonstrate that, even for such a scalar toy model, the Witten diagram provides a
result which is qualitatively consistent with the Fourier transform
of the backreaction computed in Sect.~\ref{sec:Delta}. The calculation of the four--point function confirms what we have been so far advocating, that jet evolution at strong coupling leads to total, 
radial and transverse, broadening, with no trace of substructures. 
This will be emphasized in the physical discussion of the results,
in Sect.~\ref{sec:phys}.

\subsection{Preliminaries: bulk excitations} \label{sect:TL}

As already explained in Sect.~\ref{sec:AdS}, the bulk excitations representing
the decaying system and the probe are obtained by propagating the respective
boundary fields towards the interior of AdS$_5$ with the help of
boundary--to--bulk propagators. From now on we shall restrict ourselves to 
scalar perturbations, corresponding
to  `dilaton' operators in the boundary field theory.

Consider first the decaying, time--like, system. The corresponding
boundary wave--packet will be taken as in  \eqn{ATLWP} . The associated
bulk excitation reads
 \beq\label{bulk}
 \Phi_q(x,z)\,=\,\int\,\rmd^4y\,D(x-y,z)\,\phi(y)\,=\,
  \int\,\frac{\rmd^4p}{(2\pi)^4}\,\rme^{ip\cdot x} D(p,z)\,\phi_q(p)\,,
  \eeq
where (with $\mcal{N}'=\mcal{N}\,(2\pi)^2 \sigma_+ \sigma_-
\sigma_\perp^2$)
 \beq\label{phimom}
 \phi_q(p)\,=\,\mcal{N}'\,\exp\left\{-\frac{\sigma_-^2(p_+-q_+)^2}{2}
-\frac{\sigma_+^2(p_--q_-)^2}{2}-\frac{\sigma_\perp^2
 p_\perp^2}{2}\right\}\,,\eeq
is the momentum--space version of \eqn{ATLWP} and (with $P^2\equiv p_\mu
p^\mu=-2p_+p_-+p_\perp^2$)
 \beq\label{Dmom}
 D(p,z) =
 \begin{cases}
 \displaystyle{\frac{z^2 P^2}{2}\,\rmK_2(P z)}
 &\quad \text{if} \quad P^2>0,
 \\*[0.25cm]
 \displaystyle{\frac{\rmi \pi z^2 |P|^2}{4}\,\rmH^{(1)}_2(|P| z)}
 &\quad \text{if} \quad P^2<0,
  %\,p^0>0,
  %\\*[0.25cm]
 %-\displaystyle{\frac{\rmi \pi z^2 |P|^2}{4}\,\rmH^{(2)}_2(|P| z)}
 %&\quad \text{if} \quad P^2<0,\, p^0<0,
 \end{cases}
 \eeq
is the time--ordered (or Feynman) boundary--to--bulk propagator in momentum space.
(For the space--like modes and also for the time--like ones with positive energy,  
this coincides with the respective retarded propagator.) Given the conditions \eqref{Asigma} 
on the widths of the WP, it is clear that the typical momenta allowed in the
integral over $p^\mu$ in \eqn{bulk} are {\em time--like}, with
$-P^2\simeq Q^2$ and $p^0\simeq q^0 >0$.

As before, we are
interested in large times $x_+\gg \sigma_+\gg 1/q_-$. Then, as we shall
shortly see, the bulk excitation is localized at relatively large values
of $z$, such that $Qz\gg 1$. Accordingly, one can use the asymptotic
form, valid for $|P|z\gg 1$, for the Hankel function within the
propagator:
 \beq\label{asymp}
 \rmH^{(1)}_2(x)\,\simeq\,\sqrt{\frac{2}{\pi
 x}}\,\rme^{ix-i({5\pi}/{4})}\qquad\mbox{when}\quad x\gg 1\,.\eeq
The momentum integral in \eqn{bulk} is then controlled by the condition
that several large and strongly oscillating phases compensate each other
within the integration domain allowed by the Gaussian WP \eqref{phimom}.
To clearly see these phases, it it convenient to change the integration
variable according to $p^\mu=q^\mu+k^\mu$ and then expand (with
$K^2\equiv -2k_+k_-+k_\perp^2$)
 \begin{align}\label{exp}
 |P| &=
 \sqrt{-(q+k)^2}=\sqrt{Q^2 - 2q\cdot k -K^2}\nn
 &\simeq Q\left(1+\frac{q_+k_-+q_-k_+}{Q^2} -\frac{K^2}{2Q^2}\right)\simeq
 Q + \sqrt{2}\gamma k_- +\,\frac{k_+}{2\sqrt{2}\gamma}\,-\,
 \frac{k_\perp^2}{2Q^2}\,.
 \end{align}
This expansion requires a few words of explanation: among the subleading
terms under the square root, the first one is of relative order $(q\cdot
k)/Q^2 \sim 1/(Q\sigma)$ and hence it is much larger than the second one,
which scales like $K^2/Q^2\sim  1/(Q\sigma)^2$. So, to be fully
consistent, one should either push the expansion to the second order or
neglect the last term, $\propto K^2/Q^2$. However, our purpose here is
merely to determine the dominant dependencies of the bulk excitation upon
$z$, $x_-$ and $x_\perp$ at large $x_+$. These are correctly encoded in
the terms proportional to $k_-$, $k_+$ and respectively $k_\perp^2$, as
kept in the last approximate equality in \eqn{exp}.

Specifically, using this approximation \eqref{exp}, we shall now
successively perform the integrations over $k_-$, $k_+$ and $k_\perp$ in
\eqn{bulk} (recall that we set $p^\mu=q^\mu+k^\mu$). To that aim, we
shall keep only the dominant, exponential, dependence upon $k^\mu$ and
replace $k^\mu\to 0$ (i.e. $p^\mu\to q^\mu$) in the prefactors. The
relevant integrals are then Gaussian and can be easily performed:
 \beq\label{kminus}
 \int\,\frac{\rmd k_-}{2\pi}\,\rme^{-ik_-(x_+- \sqrt{2}\gamma z)}\,
 \exp\left\{-\frac{\sigma_+^2k_-^2}{2}\right\}\,=\,
 \frac{1}{\sqrt{2\pi}\,\sigma_+}\,
 \exp\left\{-\frac{\big(x_+- \sqrt{2}\gamma z\big)^2}{2\sigma_+^2}\right\}\,,
 \eeq
 \beq\label{kplus}
 \int\,\frac{\rmd k_+}{2\pi}\,\rme^{-ik_+\big(x_--
 \frac{z}{2\sqrt{2}\gamma}\big)}\,
 \exp\left\{-\frac{\sigma_-^2k_+^2}{2}\right\}\,=\,
 \frac{1}{\sqrt{2\pi}\,\sigma_-}\,
 \exp\left\{-\frac{\big(x_-- z/(2\sqrt{2}\gamma)\big)^2}
 {2\sigma_-^2}\right\}\,,
 \eeq
 \beq\label{kperp}
 \int\,\frac{\rmd^2k_\perp}{(2\pi)^2}\,\rme^{ik_\perp\cdot x_\perp}
  \exp\left\{-i\frac{k_\perp^2 z}{2Q}
  -\frac{\sigma_\perp^2k_\perp^2}{2}\right\}\,=\,
  \frac{1}{2\pi\,\big(\sigma_\perp^2 + iz/Q\big)}\,
  \exp\left\{-\frac{x_\perp^2}{2\big(\sigma_\perp^2 + iz/Q\big)}\right\}.
  \eeq

\eqn{kminus} shows that the bulk excitation is itself a wave--packet
which at time $x^+\simeq \sqrt{2}t$ is localized in the radial direction
near $z=z_*$ with (recall that $\sigma_+\sim\gamma\sigma$)
 \beq\label{zstar}
 z_*\,\equiv\,\frac{x^+}{\sqrt{2}\gamma}\,=\,\frac{t}{\gamma}\,,\qquad
 |z-z_*|\,\lesssim\,\sigma\,.\eeq
\eqn{kplus} shows that for a given $z$, the bulk WP is localized near
$x^-=z/(2\sqrt{2}\gamma)$ with an uncertainty
$\sigma_-\sim\sigma/\gamma$. Since moreover $z\simeq z_*$, this implies
 \beq\label{xminus}
 x_-\,\simeq\,\frac{z_*}{2\sqrt{2}\gamma}\,=\,\frac{x_+}{4\gamma^2}
 ,\qquad\mbox{or}\qquad
 t-x_3\,\simeq\,\frac{t}{2\gamma^2}\,.\eeq
This value of $x_-$ is of course the maximal longitudinal extent
consistent with Lorentz contraction.
Finally, \eqn{kperp} together with \eqn{zstar} imply the following
condition for the average position $\langle x_\perp^2\rangle$ of the bulk
WP in the transverse plane (note that $z_*/Q=t/q^0$):
 \beq\label{xperp}
 \langle x_\perp^2\rangle \,\simeq\,\sqrt{\sigma_\perp^4+(t/q^0)^2}
 \,\simeq\,\begin{cases}
 \displaystyle{\sigma_\perp^2+\frac{t^2}{2q_0^2\sigma^2_\perp}}
 &\quad \text{if} \quad t/q^0\ll \sigma_{\perp}^2,
 \\*[0.25cm]
 \displaystyle{\frac{t}{q^0}}
 &\quad \text{if} \quad t/q^0\gg \sigma_{\perp}^2.
 \end{cases}
 \eeq
The second line shows that, for very large times, the bulk excitation
expands in the transverse plane via diffusion.

To summarize, the bulk excitation produced by the decaying system at time
$x_+\gg \sigma_+$ reads
 \begin{align}\label{bulkfin}
 \hspace*{-0.3cm}
 \Phi_q(x,z)\,=\,&\mcal{N}\,
 \rme^{-iq_+x_--iq_-x_+}\,
 \sqrt{\frac{\pi (Q z)^3}{8}}\,
 \frac{\sigma_\perp^2}
 {\sigma_\perp^2 + iz/Q}\, \nn
 &\exp\left[-\frac{\big(x_+- \sqrt{2}\gamma
 z\big)^2}{2\sigma_+^2}\right]
 \exp\left[-\frac{\big(x_-- x_+/4\gamma^2\big)^2}
 {2\sigma_-^2}\right]
 \exp\left[-\frac{x_\perp^2}{2\big(\sigma_\perp^2 + iz/Q\big)}\right].
 \end{align}

Consider now the corresponding excitation induced by the probe. The
respective boundary field is shown in \eqn{PROBEWP}, which immediately
implies
 \beq\label{SLbulk}
 \Psi_\Delta(x,z;\tau) = 
 %\int\,\rmd^4y\,D(x-y,z)\,\psi_\Delta(y)\,=\,
  \int\,\frac{\rmd^4k}{(2\pi)^4}\,
  %\frac{\rmd k_+}{2\pi}\,\frac{\rmd k_-}{2\pi}
  \rme^{ik\cdot x} \
  D(k,z)\, \psi_\Delta(k;\tau)\,,
 \eeq  
where (with $\mcal{C}'=\mcal{C}(2\pi)^2 \tilde\sigma_+
\tilde\sigma_-\tilde\sigma_\perp^2$)
 \beq \label{psimom}
 \psi_\Delta(k;\tau) \equiv 
 \mcal{C}'\,
 \exp\left[\rmi(k_--\Delta_-)\tau
 -\frac{\tilde\sigma_-^2(k_+-\Delta_+)^2}{2}
-\frac{\tilde\sigma_+^2 (k_--\Delta_-)^2}{2}
-\frac{\tilde\sigma_\perp^2(\bmk_\perp-\bm{\Delta}_\perp)^2}
{2}\right].
 \eeq
As already mentioned, the central 4--momentum $\Delta^\mu$ is taken to be
space--like, $\Delta^2 ={\Delta}_\perp^2-2\Delta_+\Delta_->0$, and the
widths of the WP are assumed to be large enough for the condition $k^2>0$
to be obeyed by the typical modes $k^\mu$ contributing to the integral in
\eqn{SLbulk}. Hence, the relevant expression for the bulk--to--boundary
propagator is that given in the first line of \eqn{Dmom}. In practice, we
shall take $\tilde\sigma_-$ and $\tilde\sigma_\perp$ to be so large that
the respective momenta have only negligible spread: $k_+\simeq\Delta_+$
and $k_\perp\simeq\Delta_\perp$. This fixes the longitudinal and transverse
resolution of the probe. As for the temporal resolution, this is controlled by
the Gaussian in $x_+$ in \eqn{PROBEWP}, which is centered at $\tau$ with
a width $\tilde\sigma_+\ll\tau$. Accordingly, the central value $\Delta_-$ of
$k_-$ is not really needed and one can choose $\Delta_-=0$
without loss of generality. Yet, the typical modes in the WP  \eqref{psimom}
will have a non--zero light--cone energy $k_-\sim 1/\tilde\sigma_+$. So, a 
typical probe mode will have a virtuality $K^2 = \Delta_{\perp}^2 - 2 \Delta_+ k_-$
with $k_-\sim 1/\tilde\sigma_+$. The condition that this virtuality $K^2$ 
remains positive (i.e. space--like) implies the constraint  
\beq\label{deltaconstraint}
 \Delta_{+} \lesssim\tilde\sigma_+ \Delta_{\perp}^2.
 \eeq
\comment{Note that, as compared to the discussion of the four--point function at weak coupling in 
Sect.~\ref{sec:4P}, where we have chosen $\Delta_+=0$ and relatively large $\Delta_-$
(and the longitudinal resolution was fixed by $xq_+ = \Delta_{\perp}^2/2\Delta_-$),
here we rather use $\Delta_+$ to control the longitudinal resolution
and take small  $k_-\lesssim 1/\tilde\sigma_+$. This is because we intend 
to discuss three-- and four--point functions on the same footing and for the three--point function
we would anyway need the strong constraint $\Delta_-\lesssim 1/\tau$, as we shall see.
With this in mind, it should be clear that} 
\eqn{deltaconstraint} plays the same role in the present context
as \eqref{longit}  in the context of  Sect.~\ref{sec:4P}: it is an upper limit on the 
longitudinal resolution of the space--like probe.
%As discussed there, this condition is not very restrictive {\em provided} one can take 
%$\Delta_\perp$ to be arbitrarily large. This was indeed possible at weak coupling, where
%the partons are point--like, but, as we shall soon discover, 
%this is not possible at strong coupling anymore.
According to this equation, the best longitudinal resolution for a given
$\Delta_\perp$ is achieved by choosing the largest possible value for
$\tilde\sigma_+$. At large time, it is convenient to let
$\tilde\sigma_+$ increase with $\tau$, like $\tilde{\sigma}_+ =\epsilon \tau$, with $\epsilon \ll1$.
Then the constraint \eqref{deltaconstraint} becomes $\Delta_+ \ll \epsilon \tau \Delta_\perp^2$.

\subsection{The three-- and four--point functions}
\label{sect:34pW}

The scalar versions of the three--point and four--point functions of interest, cf. Eqs.~\eqref{Tq3p}
and \eqref{Pidef}, read
 \beq \label{G3def}
 G^{(3)}(q;\tau,\Delta)\,\equiv\,\langle
\hat{\mcal{O}}_q^\dagger
 \,\hat{\mcal{O}}_\Delta(\tau)\,\hat{\mcal{O}}_q\rangle\,,\qquad
 G^{(4)}(q;\tau,\Delta)\,\equiv\,\langle
\hat{\mcal{O}}_q^\dagger \,\hat{\mcal{O}}_\Delta^\dagger(\tau)
 \,\hat{\mcal{O}}_\Delta(\tau)\,\hat{\mcal{O}}_q\rangle\,,
 \eeq
where $\hat{\mcal{O}}_q$ and $\hat{\mcal{O}}_\Delta$ are smeared
versions of the `dilaton' operator, as obtained after averaging over the
respective (time--like or space--like) WP:
 \beq \label{Odef}
 \hat{\mcal{O}}_q\,\equiv\,\int \rmd^4x\,\phi_q(x)\,\hat{\mcal{O}}(x)\,
,\qquad\hat{\mcal{O}}_\Delta(\tau)\,\equiv\,\int
\rmd^4x\,\psi_\Delta(x;\tau)
 \,\hat{\mcal{O}}(x)\,.\eeq
 
 \begin{figure}
\begin{center}
\begin{minipage}[b]{0.4\textwidth}
\begin{center}
\includegraphics[scale=0.55]{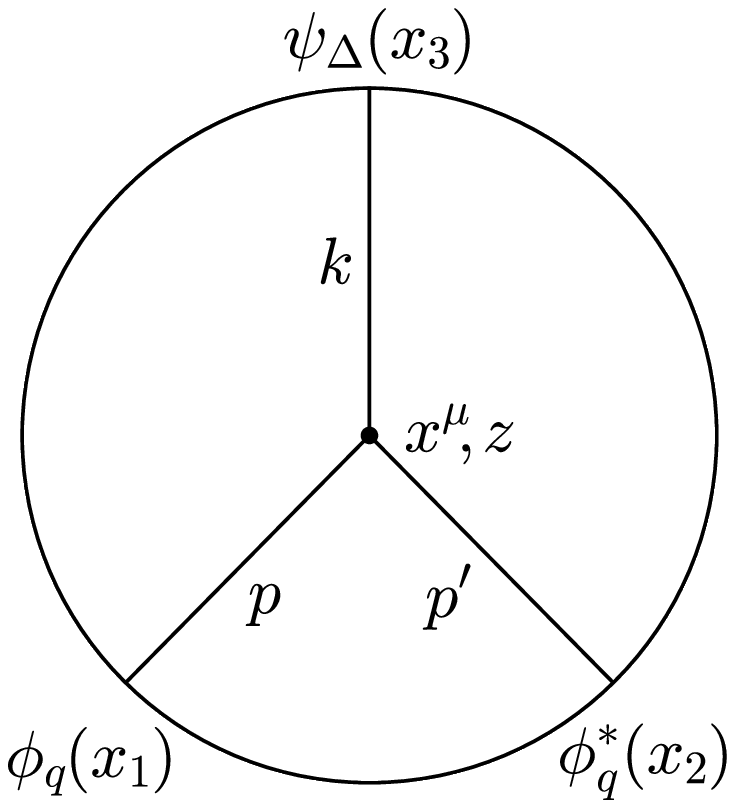}\\(a)
\end{center}
\end{minipage}
\hspace*{0.0\textwidth}
\begin{minipage}[b]{0.4\textwidth}
\begin{center}
\includegraphics[scale=0.55]{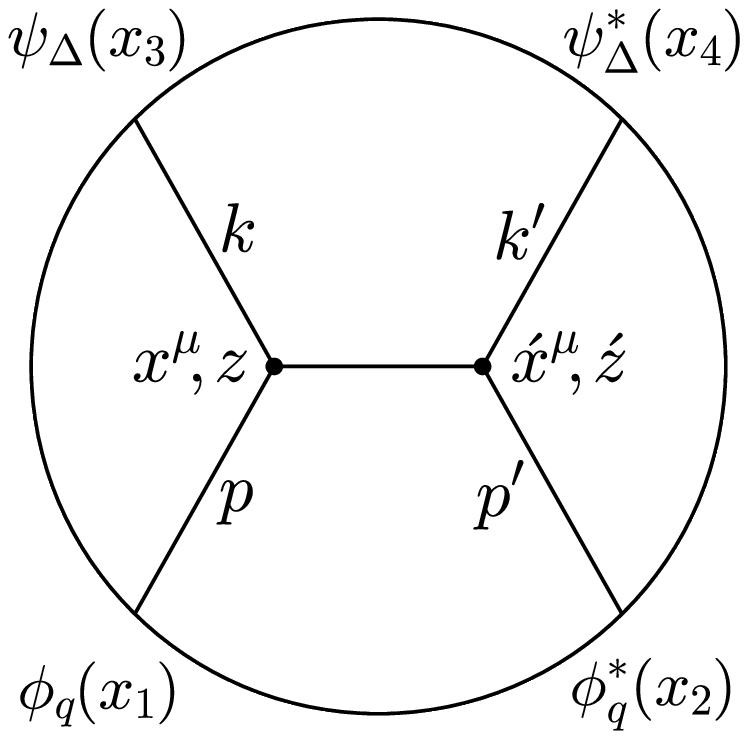}\\(b)
\end{center}
\end{minipage}
\caption{The Witten diagrams for the (a) three--point and (b) four--point functions.}
\label{fig:34point}
\end{center}
\end{figure}
As announced, the scalar toy model under consideration is characterized by
cubic self--interactions. 
At tree--level, which is the relevant approximation in the context of SUGRA, this cubic interaction contributes to the three--point function in \eqref{G3def} 
via the Witten diagram shown in Fig.~\ref{fig:34point}.a. We shall evaluate this diagram
using the SUGRA Feynman rules along the Schwinger--Keldysh contour in the complex time
plane (the `closed time path'), as appropriate for computing quantum correlations
in real time. Our use of the Schwinger--Keldysh
formalism will however be quite minimal, so we shall not describe it here in any detail. 
(See e.g. \cite{Hatta:2010kt,Arnold:2010ir,Arnold:2011hp} for recent applications of this formalism 
to Witten diagrams in SUGRA.) For the diagram in Fig.~\ref{fig:34point}.a, one finds
 \begin{align}\label{g3}
 \hspace{-0.9cm}G^{(3)}&(q;\Delta,\tau)= 
 \frac{\lambda_3}{R^5}
 \int \rmd^4 x\,\rmd z\, \sqrt{-g}\,
 \Phi_q(x,z)\,
 \Phi_q^*(x,z)\,
 \Psi_\Delta(x,z;\tau)
 \nn
 &= \lambda_3
 \int \frac{\rmd z}{z^5}\,
 \frac{\rmd^4 p}{(2\pi)^4}\, 
 \frac{\rmd^4 k}{(2\pi)^4}\,
 %D_{\rm TL}(p,z)\,
 % D_{\rm TL}^*(p+k,z)\, 
 %D_{\rm SL}(k,z)\,
 D_{\rm 11}(p,z)\,
 D_{\rm 12}(p+k,z)\, 
 D_{\rm 11}(k,z)\,
 \phi_q(p)\,
 \phi_q^*(p+k)\, 
 \psi_\Delta(k;\tau),
 \end{align}
with $\lambda_3$ denoting the strength of the cubic vertex and where we recall that $\sqrt{-g}=R^5/z^5$. Here the lower indices on the propagator refer to the branch (`1' or `2') of the 
Schwinger--Keldysh contour that the corresponding scalar field has been placed on.
$D_{\rm 11}(p,z)$ is the time--ordered bulk--to--boundary propagator, as already shown
in \eqn{Dmom}.  $D_{\rm 12}(p+k,z)$ is the Wightman  (or `cut') bulk--to--boundary propagator,
which is non--zero only for TL momenta, in which case it can be obtained by
taking the $z'\to 0$ limit of the respective bulk--to--bulk propagator shown 
in \eqn{Wightman} below. This yields an expression like the complex conjugate of  the second line in \eqn{Dmom}, but with the replacement  $\rmH^{(2)}_2\to  -2i\rmJ_2=
-i(\rmH^{(1)}_2+ \rmH^{(2)}_2)$. For the present purposes, it is only $\rmH^{(2)}_2$ 
which needs to be retained. (Indeed, we are interested in the behaviour of the integrand for 
relatively large values of $z$ and the other function $\rmH^{(1)}_2$ would lead to
strong oscillations in that regime; see below). 
To summarize, within the integrand of \eqn{g3} we can replace 
 \beq
  D_{\rm 11}(p,z)\,
 D_{\rm 12}(p+k,z)\, 
 D_{\rm 11}(k,z)\,\ \longrightarrow \ \, -\,i\, D_{\rm TL}(p,z)\,
  D_{\rm TL}^*(p+k,z)\, 
 D_{\rm SL}(k,z)\,,\eeq
with  $D_{\rm SL}$ and $D_{\rm TL}$ denoting the SL and respectively TL components 
of the Feynman propagator in \eqn{Dmom}. The boundary WPs are shown
in Eqs.~\eqref{phimom} and \eqref{psimom}.  As discussed after \eqn{psimom}
the probe (SL) WP is taken to be sharply peaked 
around $\Delta_+$ and $\bm{\Delta}_{\perp}$, so that the respective integrations over $k_+$ and $\bmk_{\perp}$ can be trivially performed. Regarding the TL propagators, expanding for large arguments and taking $k\ll p$ and $P^2 \simeq Q^2$, we have
 \beq\label{DTLexp}
 D_{\rm TL}(p,z)\,
 D_{\rm TL}^*(p+k,z)
 \simeq
 \left(\frac{\pi z^2 Q^2}{4}\right)^2
 \frac{2}{\pi Qz}\,
 \exp\left[-\rmi z\left(\sqrt{2}\gamma k_- +
 \,\frac{k_+}{2\sqrt{2}\gamma}
 \right)\right].
 \eeq 
Then the $p$--integration is simply related to the normalization of the TL WP since
 \beq\label{phiqnorm}
 \int \frac{\dif^4 p}{(2\pi)^4}\,
 \phi_q(p)\, 
 \phi^*_q(p+k)\,
 \simeq
 \int \frac{\dif^4 p}{(2\pi)^4}\, |\phi_q(p)|^2 
 = \int \dif^4 x \, 
 |\phi_q(x)|^2,
 \eeq
which is taken to be equal to unity. Putting everything together and defining $\ell_-=k_--\Delta_-$ 
and $K^2=\Delta_\perp^2 - 2\Delta_+(\ell_-+\Delta_-)\simeq \Delta_\perp^2 $ (recall the
constraint \eqref{deltaconstraint} on $\Delta_+$)
, we obtain
 \begin{align}\label{g3ell}
 \hspace{-0.25cm}
 G^{(3)}(q;\Delta,\tau) = -i
 \lambda_3\,\mcal{C}\,
 \tilde\sigma_+
 \int & \frac{\dif z}{z^5}\,
 \frac{\rmd \ell_-}{\sqrt{2\pi}}\,
 \frac{\pi (Qz)^3}{8}\,
 \frac{(K z)^2}{2}\,\rmK_2(K z)
 \nn
 &\exp\left[\rmi \ell_- (\tau - \sqrt{2}\gamma z)
 -\frac{\tilde\sigma_+^2 \ell_-^2}{2}
 -\rmi z \left(\frac{\Delta_+}{2 \sqrt{2}\gamma} + 
 \sqrt{2} \gamma \Delta_- \right)\right].
 \end{align}
Now the $\ell_-$--integration can be easily performed and we obtain
 \begin{align}\label{g3coord2}
 G^{(3)}(q;\Delta,\tau) = -i
 \lambda_3\, \mcal{C}
 \int
 \frac{\dif z}{z^5}\,
 \frac{\pi (Qz)^3}{8}\,
 \frac{(\Delta_{\perp} z)^2}{2}\,\rmK_2(\Delta_{\perp} z)\,
 \rme^{-\rmi
 \Big({\textstyle\frac{\Delta_+}{4 \gamma^2}}\,  
 +\Delta_- \Big) x_+
 -{\textstyle\frac{(x_+ - \tau)^2}{2\tilde\sigma_+^2}}},
 \end{align}
where it should be clear in the above that $x_+$ is not an independent variable, but simply determined by $x_+=\sqrt{2} \gamma z$. Also, we temporarily keep a non--zero value for $\Delta_-$,
to explicitly show that this needs to be small, $\Delta_-\lesssim 1/\tau$.
Note that the Gaussian restricts $x_+$ (the interaction time in the bulk)
to values which are relatively close to $\tau$ (the measurement time on the boundary):
$|x_+ - \tau| \lesssim \tilde\sigma_+$. Recalling our interpretation of $x_+$
as the {\em physical} emission time (the time when the signal measured
by the three--point function is actually emitted by the target, 
cf. Sect.~\ref{sec:interp}), this confirms the fact
that a space--like probe is a good measurement of the state of the system 
at the time of measurement.

Before performing the final integration over $z$, let us open a small parenthesis and notice that \eqn{g3coord2} does not carry any information on the widths of the TL WP. This remark is most easily understood in a coordinate space calculation; since we are not interested in discriminating the internal structure of the TL WP, the probe resolutions have been assumed to be low on the scales set by the 
various widths in \eqn{bulkfin}, that is, $\Delta_+\ll 1/\sigma_-$, $\Delta_\perp\ll 1/\sigma_\perp$, and
$\tilde\sigma_+\gg \sigma_+$. In view of this, we can replace (the modulus squared of) the bulk excitation \eqref{bulkfin} 
by its formal limit as obtained when all the widths approach to zero. This is a  4--dimensional 
$\delta$--function (recall that $|\mcal{N}|^2 = 1/(\pi^2 \sigma_+ \sigma_- \sigma_{\perp}^2)$)\,: 
 \beq\label{phisquare}
 |\Phi_q(x,z)|^2 = 
 \frac{\pi (Qz)^3}{8}\,
 \delta\Big(x_- - \frac{x_+}{4 \gamma^2}\Big)\,
 \delta^{(2)}(\bmx_\perp)\,
 \delta\big(x_+ - \sqrt{2}\gamma z\big).
 \eeq
As discussed in Sect.~\ref{sec:AdS}, this simplification has already been used in
the calculation of the backreaction. \eqn{phisquare} represents a particle falling into AdS$_5$  
in a highly boosted frame; indeed, the trajectory of such a particle is
 \beq
 x_-=\frac{1-\upsilon}{1+\upsilon}\,x_+ \simeq \frac{x_+}{4\gamma^2},
 \qquad
 z=\frac{\sqrt{2}}{\gamma (1+\upsilon)}\,x_+ \simeq \frac{x_+}{\sqrt{2}\gamma}
 \quad \text{and} \quad
 \bmx_{\perp} = \bm{0}_{\perp},
 \eeq
where the approximate equalities hold in the infinite momentum frame $\gamma \gg 1$. 

The presence of the $\delta$--functions in \eqn{phisquare} greatly simplifies the
calculation of the three--point function when using the coordinate space expression of \eqref{g3}. The probe excitation appears in the form
$\Psi_{\Delta}(x_+,x_+/4\gamma^2,\bm{0}_{\perp},z;\tau)$ 
with $x_+=\sqrt{2} \gamma z$. Then by using the coordinate--space version of the probe WP,
\eqn{SLbulk}, one easily recovers \eqn{g3coord2}.

Let us now close our parenthesis and return to do the $z$-integration in \eqn{g3coord2}. We would like to argue that the prefactor in the integrand, to be succinctly denoted as $f$, is slowly varying within the range of the integration
over $z$, that is, when $x_+\equiv \sqrt{2} \gamma z$ 
is changing from $\tau$ to $\tau + \tilde{\sigma}_+$. 
For definiteness, we shall do this in the two limiting cases, $\Delta_{\perp} z \gg 1$
and $\Delta_{\perp} z \ll 1$. Recalling that $z = x_+/\sqrt{2} \gamma$, we see that in
the first case, namely $\Delta_{\perp} z = \Delta_{\perp} x_+/\sqrt{2} \gamma \ll 1$, the prefactor behaves like $f \sim x_+^{\alpha}$ with $\alpha=-2$. (The actual value of $\alpha$ 
is not important for the argument, so long as this is not too large.) Then one has
 \beq\label{dff1}
 \frac{\delta f}{f} 
 \sim \frac{\tilde{\sigma}_+ f'}{f} 
 = \frac{\alpha \tilde{\sigma}_+}{x_+}
 \sim \frac{\alpha \tilde{\sigma}_+}{\tau}
 \ll 1, 
 \eeq
since we recall that $\tilde{\sigma}_+ \ll \tau$. In the other case $\Delta_{\perp} z \gg 1$, 
the dominant dependence is coming from the exponential falloff 
of the modified Bessel function and one has
  \beq\label{dff2}
  \frac{\delta f}{f}
  \sim \frac{\tilde{\sigma}_+ \Delta_{\perp}}{\gamma} \ll 1,
  \eeq
so long as $\Delta_{\perp}$ is not getting too large compared to $\gamma/\tau$.

From now on, we shall work under the assumption $\Delta_{\perp}\lesssim \order{\gamma/\tau}$,
that we shall {\em a posteriori} check to be satisfied for all the situations of interest.
Then the prefactor is slowly varying in the two limiting cases, as anticipated, and it 
keeps this property at all intermediate values. 
Hence, we can easily do the Gaussian integration to arrive at  
 \beq\label{g3coordfin}
 G^{(3)}(q;\Delta,\tau) = 
 - i \lambda_3 \sqrt{\pi}\,
 \frac{\mcal{C}\,\tilde{\sigma}_+}{\gamma z_*^5}\,
 \frac{\pi (Qz_*)^3}{8}\,
 \frac{(\Delta_{\perp} z_*)^2}{2}\,\rmK_2(\Delta_{\perp} z_*),
 \eeq 
where we have defined $z_* = \tau/\sqrt{2}\gamma$. Notice that in order to reach the 
above result we have neglected a factor originating from the phase of the space--like WP, 
namely
 \beq\label{smallfactor}
 \exp\left[-
 \rmi \left(\frac{\Delta_+}{4 \gamma^2}  
 + \Delta_- \right) \tau
 -\frac{\tilde{\sigma}_+^2}{2}
 \left(\frac{\Delta_+}{4 \gamma^2}  
 + \Delta_- \right)^2
 \right].
 \eeq
This is correct so long as we impose that phase to be small, that is,
$\Delta_+ \ll \gamma^2/\tau$ and $\Delta_- \ll 1/\tau$. 
The first condition is automatically satisfied, as it is weaker than the constraint
\eqref{deltaconstraint}.
Indeed, using $\tilde{\sigma}_+ =\epsilon \tau$ with $\epsilon\ll 1$
and $\Delta_{\perp}\lesssim {\gamma/\tau}$  within \eqn{deltaconstraint}, we deduce
   \beq\label{dpc}
 \Delta_+ \,\ll\, \epsilon\gamma^2/\tau.
 \eeq
The second condition $\Delta_- \ll 1/\tau$ is necessary to avoid strong oscillations and
motivates us to choose $\Delta_- =0$. 

The presence of the modified Bessel function $\rmK_2(\Delta_{\perp} z_*)$ in 
\eqn{g3coordfin}, which originates from the boundary--to--bulk propagator
for the space--like probe, effectively restricts the transverse momenta to 
$\Delta_{\perp} \lesssim \gamma/\tau$, as anticipated. The result
\eqref{g3coordfin} is formally independent of $\Delta_+$, but clearly this is valid
only for longitudinal momenta $\Delta_+$ obeying the constraint \eqref{dpc},
which is the condition that the probe be space--like.
The physical consequences of these constraints will be  
discussed in the next section. Also, notice that, in so far as the dominant 
behaviour upon $\Delta_{\perp} z_*$ is concerned, this result, \eqn{g3coordfin},
of the scalar toy theory is in fact consistent with the respective result of the backreaction,
for the same type of bulk excitation (a falling particle) and the same, space--like, kinematics
(compare to \eqn{tildeESL}).

The Witten diagram contributing to the four--point function in the toy--model, scalar
theory under consideration is shown in Fig.~\ref{fig:34point}.b. In analogy to the three--point function, it can be estimated as 
 \begin{align}\label{g4}
 \hspace{-0.9cm}
 G^{(4)}(q;\Delta,\tau) = 
 \lambda_3^2
 \int& \rmd^4 x\,\frac{\rmd z}{z^5}
 \int \rmd^4 x'\,\frac{\rmd z'}{z'^5}\,
 \Phi_q(x,z)\,
 \Psi_\Delta(x,z;\tau)\,
 G(x-x',z,z')\,
 \Phi_q^*(x',z')\,
 \Psi_\Delta^*(x',z';\tau)
 \nn
 = \lambda_3^2
 \int& \frac{\rmd z}{z^5}\,
 \frac{\rmd z'}{z'^5}\,
 \frac{\rmd^4 p}{(2\pi)^4}\,
 \frac{\rmd^4 p'}{(2\pi)^4}\,
 \frac{\rmd^4 k}{(2\pi)^4}\,
 \frac{\rmd^4 k'}{(2\pi)^4}\,
 (2\pi)^4 \delta^{(4)}(p+k-p'-k')
 \nn
 &
 G_{12}(p+k,z,z')\,
 D_{\rm TL}(p,z)\,
 D_{\rm TL}^*(p',z')\, 
 D_{\rm SL}(k,z)\,
 D_{\rm SL}^*(k',z')\,
 \nn
 &
 \phi_q(p)\,
 \phi_q^*(p')\, 
 \psi_\Delta(k;\tau)\,
 \psi_\Delta^*(k';\tau) 
 \end{align}
where the only new ingredient is the Wightman bulk--to--bulk propagator:
 \beq\label{Wightman}
 G_{12}(\ell,z,z') = 
 \pi \Theta(-\ell^2)\, z^2 z'^2
 \rmJ_2(|\ell| z)\,
 \rmJ_2(|\ell| z').
 \eeq
The various bulk--to--boundary propagators visible in the integrand of \eqn{g4} have
entered the calculation as time--ordered ($D_{11}$) or anti--time--ordered ($D_{22}$) 
propagators in real time.

Since $k,k'\ll p,p'$, we shall shortly see that the dominant exponential dependence on $p$ and $p'$ in the propagators product cancels. Regarding the remaining dependence upon $p$ and $p'$
in the prefactors (as arising from the large argument expansion of the propagators), these are weak and hence we can simply replace $p$ and $p'$ by their central value $q$. 
Then one can integrate over $p$ and $p'$ to recover the
normalization conditions for the TL wave--packets:
 \beq
 \int\frac{\rmd^4 p}{(2\pi)^4}\,
 \frac{\rmd^4 p'}{(2\pi)^4}\,
 (2\pi)^4 \delta^{(4)}(p+k-p'-k')\,
 \phi_q(p)\,
 \phi_q^*(p')
 \simeq
 \int\frac{\rmd^4 p}{(2\pi)^4}\,
 |\phi_q(p)|^2
 =1\,.
 \eeq
At this stage, we managed to bring  \eqn{g4} into a {\em factorized} form:
 \beq\label{fact}
 G^{(4)}(q;\Delta,\tau) = G_{\rm L}(q;\Delta,\tau)\,G_{\rm R}(q;\Delta,\tau)
 = |G_{\rm L}(q;\Delta,\tau)|^2,
 \eeq
with the ``left'' part given by
 \beq\label{gleft}
 G_{\rm L}(q;\Delta,\tau) = 
 \lambda_3 \sqrt{\pi}
 \int \frac{\dif z}{z^3}\, 
 \frac{\rmd^4 k}{(2\pi)^4}\,
 \rmJ_2(|q+k| z)\,
 D_{\rm TL}(q,z)\,
 D_{\rm SL}(k,z)\,
 \psi_\Delta(k;\tau).
 \eeq
From this point on, the calculation is very similar to the one of the three--point function. For large arguments we can approximate
 \beq
 \rmJ_2(|q+k| z)\,\rmH_2^{(1)}(Q z) \simeq \frac{1}{2} \frac{2}{\pi Q z}\,
 \exp\left[-\rmi z\left(\sqrt{2}\gamma k_- +
 \,\frac{k_+}{2\sqrt{2}\gamma}
 \right)\right] + \dots,
 \eeq
where we have used \eqn{exp} and with the dots standing for a term proportional to $\exp(2 \rmi Q z)$ which is neglected since it is varying rapidly. Taking again the SL WP to be sharply peaked we find (cf.~the similarity with \eqn{g3ell})
 \begin{align}\label{g3leftell}
 \hspace*{-0.5cm}
 G_{\rm L}(q;\Delta,\tau) = 
 \rmi\,\sqrt{\pi}\,\lambda_3\,\mcal{C}\,
 \tilde\sigma_+
 \int & \frac{\dif z}{z^3}\,
 \frac{\rmd \ell_-}{\sqrt{2\pi}}\,
 \frac{Qz}{4}\,
 \frac{(K z)^2}{2}\,\rmK_2(K z)
 \nn
 &\exp\left[\rmi \ell_- (\tau - \sqrt{2}\gamma z)
 -\frac{\tilde\sigma_+^2 \ell_-^2}{2}
 -\rmi z \left(\frac{\Delta_+}{2 \sqrt{2}\gamma} + 
 \sqrt{2} \gamma \Delta_- \right)\right].
 \end{align}
Performing the integrations over $\ell_-$ and $z$ as usual we get
 \beq
 G_{\rm L}(q;\Delta,\tau) = \rmi \lambda_3 \pi\,
 \frac{\mcal{C} \tilde\sigma_+}{\gamma z_*^3}\,
 \frac{Qz_*}{4}\,
 \frac{(\Delta_{\perp} z_*)^2}{2}\,\rmK_2(\Delta_{\perp} z_*).
 \eeq
One should also include in the above a factor equal to that in \eqn{smallfactor}, that is,
the product of a phase factor times a Gaussian. However, when we construct
the modulus squared according to \eqn{fact}, the respective phase factors mutually 
compensate\footnote{At this level, there is a small difference compared 
to the case of the three--point function: the two phases $ \rme^{\pm\rmi \Delta \cdot  x}$ 
(recall \eqn{psimom}) automatically cancel out between the two insertions, 
$\Psi_\Delta$ and $\Psi_\Delta^*$, of the space--like WP, so the condition 
$\Delta_-\lesssim 1/\tau$ is not necessary anymore.},
while the Gaussian factors can be safely set to unity.  We thus finally arrive at  
 \beq\label{g4fin}
 G^{(4)}(q;\Delta,\tau) = \lambda_3^2\,\pi^2\,
 \frac{\mcal{C}^2 \tilde\sigma_+^2}{\gamma^2 z_*^6}\,
 \frac{(Qz_*)^2}{16}\,
 \frac{(\Delta_{\perp} z_*)^4}{4}\,[\rmK_2(\Delta_{\perp} z_*)]^2.
 \eeq
There is some model--dependence inherent in this formula, but this is harmless for
our present purposes: the only information that we actually need is the
dependence of the four--point function upon the dimensionless variable  $\Delta_{\perp} z_*
= \Delta_{\perp} \tau/\sqrt{2}\gamma$. This dependence is robustly predicted by 
\eqn{g4fin} and could have been anticipated without explicitly performing the 
calculation, as we explain now. The four--point function defined in \eqn{G3def} 
represents the 
imaginary part of a forward scattering amplitude, which at the level of the SUGRA
calculation is obtained by taking the `cut' of the 4--leg Witten diagram depicted in 
Fig.~\ref{fig:34point}.b. (This cut is manifest in our use of the Wightman 
prescription for the bulk--to--bulk propagator in \eqn{Wightman}.) In turn, this
cut diagram  is proportional to the modulus squared of the 3--leg diagram
shown in Fig.~\ref{fig:34point}.a and evaluated in  \eqn{g3coordfin}.  
(This is merely the optical theorem adapted to the SUGRA context at hand.)
We conclude that, at strong coupling, the four--point function $G^{(4)}(q;\Delta,\tau)$ 
must depend upon $\Delta_{\perp} z_*$ in the same way as the square 
$|G^{(3)}(q;\Delta,\tau)|^2$ of the three--point function.
This conclusion is indeed consistent with our previous results for 
$G^{(4)}(q;\Delta,\tau)$ (see \eqn{g4fin}) and, respectively, for  
$G^{(3)}(q;\Delta,\tau)$ (cf. \eqn{tildeESL} or \eqref{g3coordfin}). 
Yet, this formal similarity between the three--point and the 
four--point functions should not be misleading: the physical content of these two 
quantities is very different, as it will be further discussed in the next subsection.
 
%as repeatedly emphasized, the physical content of these two 
%quantities is different in general and this difference becomes particularly important 
%at strong coupling. This will

\subsection{Physical discussion}
\label{sec:phys}

The physical discussion to follow will only exploit those aspects of our above
result, \eqn{g4fin}, for the four--point function at strong coupling which are firmly
under control: its dependence upon the transverse momentum of the probe,
which enters via the dimensionless variable
$\Delta_{\perp} z_*= \Delta_{\perp} \tau/\sqrt{2}\gamma$, and the
upper limit \eqref{dpc} on the longitudinal momentum $\Delta_+$ of the probe.
It is useful to recall that, in this boosted frame, the decaying system has an overall
transverse size $\Delta x_{\perp} \sim \tau/\gamma$ and that the maximal longitudinal
broadening permitted by causality and Lorentz contraction is 
$\Delta x_- \sim \tau/\gamma^2$ (cf Fig.~\ref{fig:lorentz}). We would like to check
whether the system involves some substructures like partons with sizes much
smaller than this maximal sizes. To that aim, one needs to estimate the four--point 
function for relatively high momenta $\Delta_{\perp} \gg \gamma/\tau$ and
$\Delta_- \gg \gamma^2/\tau$. (Such values are compatible with the space--likeness
condition  \eqref{dpc} provided $\Delta_{\perp}$ is chosen to be high enough.) 
But for such large values of the momenta, the four--point function \eqref{g4fin} 
is exponentially suppressed, because $\rmK_2(\Delta_{\perp} z_*)\propto
\exp\{-\Delta_{\perp} z_*\}$ when $\Delta_{\perp} z_*\gg 1$. This simply means
that there are no substructures in the decaying system at large times
$\tau\gg\sigma_+$: the matter is
uniformly distributed within a region with transverse size $\Delta x_{\perp} \sim \tau/\gamma$ 
and  longitudinal size $\Delta x_- \sim \tau/\gamma^2$. In particular, it exhibits
maximal longitudinal broadening.

It is amusing to notice that, at strong coupling, the four--point function
and the three--point function are formally similar to each other (compare
Eqs.~\eqref{g3coordfin} and \eqref{g4fin}) --- they both predict the
exponential suppression of the respective correlation for transverse momenta 
$\Delta_{\perp} \gg \gamma/\tau$. However, as it should be clear from 
our previous analysis, this similarity is deceiving.  The three--point function is 
independent of the coupling, so it looks quasi--homogeneous in the transverse 
plane (when probed with a low longitudinal resolution) even at weak coupling,
where point--like partons are well known to exist. This is so since, by construction, 
the three--point function integrated over $x_-$ represents the 
average energy per unit transverse area, which in this problem is homogeneous
by symmetry.

On the other hand, the four--point function has the potential to reveal small fluctuations
in the transverse plane, as manifest from the corresponding discussion at weak coupling.
So the corresponding exponential suppression for transverse momenta 
$\Delta_{\perp} \gg \gamma/\tau$ is a unambiguous proof of the absence of partons.

It is furthermore interesting to re-express our results in terms of the typical momenta
and virtualities of the quanta composing the decaying system at time $\tau$. 
By the uncertainty principle, a quantum distribution of matter which looks homogeneous 
(in a given event) over transverse distances $\delta x_{\perp} \lesssim \tau/\gamma$ 
involves Fourier modes with transverse momenta and virtualities 
$k_\perp\sim\mu \sim \gamma/\tau$ and hence with longitudinal momenta $k_+\sim \gamma^2/\tau$. 
For comparison with the weak coupling discussion in Sect.~\ref{sect:jet}, we note that
the longitudinal momentum fraction $x\equiv {k_+}/{q_+}$ 
carried by a typical mode at strong coupling is $x\simeq x_c(\tau) $ with
 \beq\label{stc}
 x_c(\tau) \,\simeq\,\frac{\gamma^2}{\tau q_+} \,\sim\,
 \frac{q_+}{\tau Q^2}\,\sim\,\frac{\tau_0}{\tau}\,.
 \eeq
This is independent of the coupling, unlike the corresponding weak--coupling result in
\eqn{xc}. In fact, \eqn{stc} looks more like the extrapolation of \eqn{xc}
to values of the coupling of order one, rather than to $\lambda\to\infty$. 
This is consistent with the fact that the time dependence of $x$ shown in \eqn{stc}
is the fastest one to be allowed by causality. 
Interestingly, the above estimate for $x_c$ can also be written as $x_c
\sim \mu/Q$, which shows that, at strong coupling\footnote{For comparison,
note that at weak coupling the virtuality $\mu$ is decreasing with time, via successive
branchings, much faster than the longitudinal momentum fraction $x_c$; this
can be checked using the general relation \eqn{tauxmu} together with \eqn{xc}
for $x_c$ at weak coupling.}, the fraction of longitudinal
momentum carried by the typical quanta in the decaying system is commensurable
with the respective fraction of {\em virtuality}. This is in agreement with the
picture of {\em democratic parton branching}, as put forward in Ref.~\cite{HIM3},
in which the energy and the virtuality are quasi--democratically split among 
the daughter partons at any branching.
In Appendix \ref{sec:fragstrong}, we shall present an alternative derivation 
of \eqn{stc}, which is in the spirit of the perturbative calculation for  
the fragmentation function in Sect.~\ref{sec:DGLAP} --- that is,
it relies on the expression for the time--like anomalous 
dimension at strong coupling, as obtained in Ref.~\cite{Hatta:2008tn}.

To summarize, at strong coupling and for sufficiently large times after the
decay has been initiated, the decaying system occupies
the maximal region in space and time which is allowed by causality and special relativity,
that is, $\Delta x_{\perp} \sim \tau/\gamma$ and $\Delta x_- \sim \tau/\gamma^2$.
In the center of mass frame of the virtual photon, the matter produced by its 
decay at time $t\gg \sigma$ is spread over the 
whole ball with radius $r\le t$ and its distribution within this ball is
(quasi)homogeneous.
The strongly coupled matter looks like a soft, smooth, jelly.

\section*{Acknowledgments}
We would like to thank Simon Caron-Huot for an observation which triggered part of this work.
One of us (E.I.) would like to thank the Physics Department at the Universidad Federal de 
Rio de Janeiro for hospitality and Prof. Nelson Braga for inciting discussions during the
completion of this work. The work of E.I. is supported in part by the European Research 
Council under the Advanced Investigator Grant ERC-AD-267258. The work
of A.H.~M. is supported in part by the US Department of Energy.
All figures were created with Jaxodraw \cite{Binosi:2003yf}.

\appendix

\section{The $\mathcal{E}_B$ term for the falling particle}
\label{sec:EB}

In Sect.~3.1 of \cite{Hatta:2010dz} we have calculated the boundary `energy density' $\mathcal{E}$ generated by a particle falling in AdS$_5$ and with velocity $\upsilon$ along the $x_3$ axis. To this end, following \cite{Athanasiou:2010pv} we have separated $\mathcal{E}$ in two pieces $\mathcal{E}_A$ and $\mathcal{E}_B$. We have erroneously stated that $\mathcal{E}_B$ vanishes; this is true only when $\upsilon=0$ as we shall shortly see. Notice that, since in \cite{Hatta:2010dz} we were in fact interested only in this particular case where $\upsilon=0$, all subsequent calculations there were correct.

We shall follow the notation of \cite{Hatta:2010dz} with the only exception being the replacement $E_0 \mapsto q_0$. In general, for an arbitrary 5D stress energy tensor $t_{MN}$
the term $\mathcal{E}_B$ reads
 \beq \label{EB}
 \mathcal{E}_B = 
 \frac{2 L^3}{3 \pi}
 \int \frac{\dif^4\rp \,\dif z}{z}\, \Theta(t-\tp)
 \delta'''(\mcal{W})
 \left[(\bm{r} - \acute{\bm{r}})^2 (2 t_{00} -2 t_{55} + t_{ii}) -
 3 (x - \xp)^i (x - \xp)^j t_{ij} \right],
 \eeq
where $\mathcal{W} = - (t-\tp)^2 + (\bm{r} - \acute{\bm{r}})^2 +z^2$. For the falling particle under consideration we have
 \beq
 t^{MN} = q_0 \left( \frac{z}{L} \right)^7
 \delta(\xp^3 - \upsilon \tp)\,
 \delta^{(2)}(\acute{\bm{x}}_{\perp})\,
 \delta(z - \tp/\gamma)
 \frac{\xp^M \xp^N}{\tp^2},
 \eeq
and by lowering indices with the metric in \eqn{metric} we find that the non-zero components of interest are
 \beq\label{tmnb}
 t_{00} = q_0 \left( \frac{z}{L} \right)^3
 \delta(\xp^3 - \upsilon \tp)\,
 \delta^{(2)}(\acute{\bm{x}}_{\perp})\,
 \delta(z - \tp/\gamma),
 \quad t_{33} = \upsilon^2 t_{00},
 \quad t_{55} = t_{00}/\gamma^2.
 \eeq
Thus, the square bracket in the integrand of \eqn{EB} becomes
 \beq
 (\bm{r} - \acute{\bm{r}})^2 (2 t_{00} -2 t_{55} + t_{ii}) -
 3 (x - \xp)^i (x - \xp)^j t_{ij}= 
 3 \upsilon^2 x^2_{\perp} \left( \frac{z}{L} \right)^3
 \delta(\xp^3 - \upsilon \tp)\,
 \delta^{(2)}(\acute{\bm{x}}_{\perp})\,
 \delta(z - \tp/\gamma),
 \eeq
and using these $\delta$-functions it is straightforward to integrate over $\xp$, $\acute{\bmx}_{\perp}$ and $z$ to obtain
 \beq\label{EBapp}
 \mathcal{E}_{B} = 
 \frac{2q_0}{\pi}\,\frac{v^2 x_{\perp}^2}{\gamma^2}\
 \del^3_{r^2} \int_0^\infty \dif \tp\, \tp^{\,2} \
 \delta\big(t^2 -r^2 - 2(t-v x_3)\tp\big).
 \eeq
As announced earlier, this contribution vanishes only when $\upsilon=0$. \eqn{EBapp}  is the second term in \eqn{EApart} and the remaining part of the calculation leading to the final expression in \eqn{Eparticle} is given in the main body of the present paper.

\section{Fragmentation function at strong coupling}
\label{sec:fragstrong}

The time--like anomalous dimension at strong coupling reads \cite{Hatta:2008tn}
 \beq\label{sad}
 \gamma(j)= 
 -\frac{1}{2}\left(j-j_0-\frac{j^2}{2\sqrt{\lambda}}\right), 
 \eeq
when $j\ll \sqrt{\lambda}$ and with $j_0=2-2/\sqrt{\lambda}$. 
The counterpart of \eqn{DGLAP} at strong coupling is
 \begin{align}\label{frags}
 x^2D(x,Q^2/\mu^2) = & 
 \int \frac{\rmd j}{2\pi i} \, D(j,1)\,  
 \rme^{(j-2)\ln (1/x) + \gamma(j) \ln  (Q^2/\mu^2) }
 \nonumber \\  
 = &
 \int \frac{\rmd j}{2\pi i} \, D(j,1) \, 
 \exp{\left(\frac{j^2}{2\sqrt{\lambda}}
 \ln \frac{Q}{\mu} -(j-j_0)\ln \frac{Q}{\mu} 
 + (j-2)\ln \frac{1}{x} \right)}, 
 \end{align}
where the fragmentation function $D(x,Q^2/\mu^2)$ is the special case of the four--point
function computed in Sect.~\ref{sec:Witten} for ${\Delta_\perp^2}={\mu^2}$. Yet, it turns
out that one cannot rely on \eqn{frags} to recover the results of Sect.~\ref{sec:Witten}
in this special limit because of non--commutativity issues to be later explained.
As in Sect.~\ref{sec:DGLAP}, we shall evaluate \eqn{frags} using the saddle point method.
The saddle point is located at
 \beq\label{saddlepoint}
 j_s \,=\,\sqrt{\lambda}\ \frac{\ln (xQ/\mu)}{\ln (Q/\mu)}\,.
 \eeq
For consistency this has to be much smaller than $\sqrt{\lambda}$, so that
 \beq
 \ln \frac{Qx}{\mu} \,=\,\ln \frac{Q}{\mu}-\ln \frac{1}{x} \ll \ln \frac{Q}{\mu}\,,
 \eeq
which requires $x$ to be relatively close to $\mu/Q$; this condition is indeed
satisfied, as we shall {\em a posteriori} check. 
Evaluating \eqn{frags} at the saddle point, we get
 \beq\label{go}
 x^2D(x,\mu^2) \sim \exp \left(-\frac{1}{4\sqrt{\lambda}}(j_s-2)^2
 \ln \frac{Q^2}{\mu^2} \right) = 
 x^2 \left(\frac{Q}{\mu}\right)^{j_0}
 \exp{ \left(-\frac{\sqrt{\lambda}(\ln xQ/\mu)^2}{2\ln Q/\mu}\right)}.  
 \eeq
This has a maximum at $j_s=2$ for which
 \beq
 x_c= \left(\frac{\mu}{Q}\right)^{1-\frac{2}{\sqrt{\lambda}}} 
 \simeq \frac{\mu}{Q}\,.
 \eeq
In terms of the formation time $\tau = {2x_c q_+}/{\mu^2}$, we equivalently have
(with $\tau_0\equiv {2 q_+}/{Q^2}$)
 \beq
 {x_c} \simeq   \frac{\tau_0}{\tau} \,,
 \eeq
in agreement with \eqn{stc}.
As $\lambda$ gets larger, the distribution \eqn{go} becomes more strongly peaked at $x=x_c$, but the limit $\lambda \to \infty$ is subtle. \eqn{go} becomes the delta function $\delta(x-x_c)$, however, the limit $\lambda \to \infty$ and the $j$--integral in \eqn{frag} do not commute because the saddle point in \eqn{saddlepoint} depends on $\lambda$. Indeed, if we set $\lambda= \infty$ first,  the integrand of \eqn{frag} becomes 
 \beq
 x^2D(x,Q^2/\mu^2) =
 \int \frac{\rmd j}{2\pi i} \, D(j,1) \, \left(\frac{xQ}{\mu}\right)^{2-j},
 \label{frag}
 \eeq
for which there is no saddle point and the dependence of $D(j,1)$ on $j$ cannot be neglected. Still, one can see that the right-hand-side depends only on $xQ/\mu$ and decays rapidly as a function of this variable. In order to determine its functional form,  one needs a direct evaluation of the fragmentation function in the context of SUGRA, as we did in Sect.~\ref{sect:34pW} (see also
Ref.~\cite{Hatta:2008qx}) .

\providecommand{\href}[2]{#2}\begingroup\raggedright\endgroup

%\bibliographystyle{utcaps}
%\bibliography{ADSref}

\end{document}